# Traffic generation for benchmarking data centre networks

Christopher W.F. Parsonson *, Joshua L. Benjamin, Georgios Zervas

*Optical Networks Group, Department of Electronic and Electrical Engineering, University College London, Roberts Building, WC1E 7JE, London, United Kingdom*

ABSTRACT

Benchmarking is commonly used in research fields, such as computer architecture design and machine learning, as a powerful paradigm for rigorously assessing, comparing, and developing novel technologies. However, the data centre network (DCN) community lacks a standard open-access and reproducible traffic generation framework for benchmark workload generation. Driving factors behind this include the proprietary nature of traffic traces, the limited detail and quantity of open-access network-level data sets, the high cost of real world experimentation, and the poor reproducibility and fidelity of synthetically generated traffic. This is curtailing the community's understanding of existing systems and hindering the ability with which novel technologies, such as optical DCNs, can be developed, compared, and tested.

We present TrafPy; an open-access framework for generating both realistic and custom DCN traffic traces. TrafPy is compatible with any simulation, emulation, or experimentation environment, and can be used for standardised benchmarking and for investigating the properties and limitations of network systems such as schedulers, switches, routers, and resource managers. We give an overview of the TrafPy traffic generation framework, and provide a brief demonstration of its efficacy through an investigation into the sensitivity of some canonical scheduling algorithms to varying traffic trace characteristics in the context of optical DCNs. TrafPy is open-sourced via GitHub and all data associated with this manuscript via RDR.

## 1. Introduction

A benchmark is a series of experiments performed within some standard framework to measure the performance of an object. Researching data centre network (DCN) systems and objects such as networks, resource managers, and topologies involves understanding which types of mechanisms, principles or architectures are generalisable, scalable and performant when deployed in real world environments. Benchmarking is a powerful paradigm for investigating such questions, and has proved to be a strong driving force behind innovation in a variety of fields [3]. A famous example of a successful benchmark is the ImageNet project [4], which has facilitated a range of significant discoveries in the field of deep learning over the last decade.

In order to benchmark a DCN system, a traffic trace with which to load the network is required. This presents several challenges: (i) Data related to DCNs are often considered privacy-sensitive and proprietary to the owner, therefore few DCN traffic traces are openly available; (ii) when a real DCN trace is made available, it is often specific to a particular DCN and possibly not representative of current and future systems, too limited for cutting-edge data-hungry applications such as reinforcement learning, and not sufficient for stress-testing different loads in networks with arbitrary capacities to understand system limitations and vulnerabilities to future workloads; (iii) even if an attempt is made to make a real DCN available for live testing, deploying experimental systems in such large-scale production environments is often too expensive and time consuming; and (iv) reducing or approximating DCN traffic down to small-scale experiments is often unfruitful since many DCN application traffic patterns only emerge at large scales.

For these reasons, most DCN researchers revert to simulating DCN traffic in order to conduct their experiments. However, synthetic DCN traffic generation is often plagued by numerous inadequacies. A common simplification approach is to assume uniform or 'named' (Gaussian, Pareto, log-normal, etc.) distributions from which to sample DCN traffic characteristics. However, such distributions often ignore fluctuations caused by the short bursty nature of real DCN traffic, rendering the simulation unrealistically simple. Sometimes researchers will try to implement their own unique distributions to better describe real DCN traffic, however this brings difficulties with trying to reproduce and benchmark against literature reports since there is no standard framework for doing so. Another common approach is to only focus on the temporal (arrival time) dependence of DCN traffic characteristics and assume uniform spatial (server-to-server) dependencies. However, this fails to capture the spatial variations in server-to-server communication which are needed to accurately mimic real traffic. Works by Alizadeh et al. [5,6] and Bai et al. [7] introduced important DCN systems, but the traffic generators released with their papers fall short of addressing the issues of fidelity, reproducibility, and compatibility with generic network architectures (see Section 2).

* Corresponding author.
*E-mail address:* zciccwf@ucl.ac.uk (C.W.F. Parsonson).

These difficulties with simulating DCN traffic have meant that no traffic generation framework, and subsequently no universal DCN system benchmark, has emerged as the networking research field's tool-of-choice. The lack of a rigorous benchmarking framework has been a major issue in DCN literature since individual researchers have often used their own tests without adhering to the aforementioned requirements. This has limited reproducibility, stifled network object prototype benchmarking, and hindered training data supply for novel machine learning systems. Without benchmarking, it is difficult to systematically and consistently test and validate new heuristics for specific tasks such as flow scheduling. Furthermore, without sufficient training data, state-of-the-art machine learning models are less able to replace existing heuristics.

To address the lack of openly available traffic data sets, the aforementioned problems with simulation, and the absence of a system benchmark, a common DCN traffic generation framework is needed. We introduce TrafPy: An open-source Python API for realistic and custom DCN traffic generation for any network under arbitrary loads, which can in turn be used for investigating a variety of network objects such as networks, schedulers, buffer managers, switch/route architectures, and topologies. TrafPy is open-access via GitHub [1] and all data associated with this manuscript via RDR [2]. TrafPy contributes two key novel ideas to traffic generation, which we detail in this paper:

1. **Reproducibility guarantee** A novel method for providing a distribution reproducibility guarantee when generating traffic based on the Jensen-Shannon distance metric (see Section 3.3).
2. **Traffic generation algorithm**: A novel method for efficiently creating reproducible flow-level traffic with granular control over both spatial and temporal characteristics (see Section 3.5).

In addition to the above, TrafPy also contains the following features which, when combined with these novel aspects, make TrafPy a useful tool for benchmark workload generation:

- **Interactivity**: A distribution shaping tool for rapid creation of complex distributions which accurately mimic realistic workloads given only high-level characteristic descriptions (see Appendix C).
- **Compatibility**: Compatibility with any simulation, emulation, or experimentation environment by exporting traffic into universally compatible file formats; and
- **Accessibility**: Open-source code and documentation with a low barrier to entry.

## 2. Related work

While there is limited literature on DCN traffic generation, data sets, and benchmarking for the reasons outlined in Section 1, there have been notable works striving towards their creation.

**Real workloads** There are a collection of publicly available DCN workload traces and job computation graph data sets [8–29]. However, almost all of these stem from Hadoop clusters and are limited to data mining applications [14], therefore their use is primarily suited to application-specific testing and evaluation rather than as a generic tool for generating arbitrary loads and testing and designing DCN systems as TrafPy is proposed for. Additionally, many of them lack flow-level data, which is needed to accurately benchmark network systems.

**Real workload characteristics** There is a limited body of work, primarily from private corporations, aiming to characterise real DCN workloads without open-accessing the underlying proprietary raw data. Benson et al. [30] built on work done by Kandula et al. [31] and Benson et al. [32] by characterising DCN traffic into one of three categories; university, private enterprise, and commercial cloud DCNs. They identified that each of these categories serviced different applications and therefore had different traffic patterns. University DCNs serviced applications such as database backups, distributed file system hosting (e.g. email servers, web services for faculty portals, etc.), and multicast video streams. Private enterprise hosted the same applications as university DCNs but additionally serviced a significant number of custom applications and development test beds. Commercial cloud DCNs focused more on internet-facing applications (e.g. search indexing, webmail, video, etc.), and intensive data mining and MapReduce-style jobs. They also went further than prior works by quantifying the number of hot spots and characterising the flow-level properties of DCN traffic.

The above cloud DCN studies came almost exclusively from Microsoft, who primarily service MapReduce-style applications. Roy et al. [33] broke this homogeneous view of cloud traffic by reporting the traffic characteristics of Facebook's DCNs, thereby introducing a fourth DCN category; social media cloud DCNs. Social media cloud applications include generating responses to web requests (email, messenger, etc.), MySQL database storage and cache querying, and newsfeed assembly. This results in network traffic being more uniform and, in contrast to Microsoft's commercial cloud DCNs, having a much lower proportion (12.9%) of traffic being intra-rack.

Note that the above examples did not open-access the full data sets, but rather provided quantitative characterisations of their nature for other researchers to inform their own traffic generators.

**Traffic generators** In their seminal pFabric work, Alizadeh et al. [6] provided open-access traffic generation code which loosely replicated web search and data mining DCN workloads by following a Poisson flow inter-arrival time distribution whose arrival rate was adjusted to meet a required target load and with a mix of small and large characteristically heavy-tailed flow sizes. Additionally, the same authors [5] released a simple generator which used a server application to create many-to-one flow requests from 9 servers, again following a load-adjustable Poisson arrival time distribution with 80% of flows having a size of 1 kB (a single packet) and 20% being 10 MB. As the authors noted themselves, these workloads were not intended to be realistic, but rather were designed to demonstrate clear impact comparisons between different DCN design schemes and the small latency-sensitive and large bandwidth-sensitive flows. TrafPy, on the other hand, can facilitate the shaping of complex inter-arrival and flow size distributions with one-to-one, many-to-one, and one-to-many non-uniform server-server distributions with ease. Furthermore, TrafPy enables the generation of traffic with the same characteristics as Alizadeh et al. [5,6], but for any network topology with an arbitrary number of servers and link capacities, allowing for the straightforward comparison of novel DCN fabrics with pre-established benchmark workloads.

Similarly, Bai et al. [7] conducted an extensive experiment into the trade-off between throughput, latency, and weighted fair sharing in scenarios where each switch port had multiple queues. Alongside their study they released an open-access traffic generator which could take a configuration file as input and generate both uniform and non-uniform server-server flow requests from pre-defined discrete probability distributions. However, to produce traffic, users had to manually enter numbers into a configuration file, which made the code difficult to use. Furthermore, Bai et al.'s generator had no mechanism for ensuring distribution reproducibility when sampling from a pre-defined probability distribution; a feat achieved by TrafPy via the Jensen-Shannon distance method (see Section 3.3).

The key objective of TrafPy is to augment DCN research projects such as those cited above [5–7]. Unlike our work, the primary focus of such projects was not on the traffic generator itself, but rather on using traffic generation as a means of benchmarking innovative ideas. We posit that the fidelity, generality, reproducibility, and compatibility of TrafPy, achieved by generating custom server-level flow traffic, would make such works easier to conduct and to compare against as baselines in future projects.

## 3. Proposed framework

### 3.1. Design objectives

Designing successful network object benchmarks requires a flexible, modular, and reproducible traffic generation framework. The framework should enable fair comparisons between different systems whilst maintaining a rigorous experimental setting. In light of the issues highlighted in Section 1, the following criteria are required of such a framework:

1 *Fidelity*: Generate demands which represent realistic DCN traffic.
2 *Generality*: Generate traffic for arbitrary DCN applications and topologies.
3 *Scalability*: Efficiently scale to large networks.
4. *Reproducibility*: Reliably reproduce traffic traces to run multiple test repeats or to reproduce other researchers' traffic conditions.
5. *Repeatability*: Summarise traffic distributions such that, given just a few parameters, other researchers can repeat the demand data set for cross-validation and comparison.
6. *Replicability*: Interactively shape characteristic distributions visually to replicate realistic data given only a plot or written description (i.e. in the absence of raw data).
7. *Compatibility*: Export generated demands into universally compatible data formats such that they can be imported into any simulation, emulation, or experimentation test bed.
8. *Comparability*: Compare a set of standardised performance metrics across different studies.

### 3.2. TrafPy overview

An overview of the TrafPy API user experience is given in Fig. 1 and further elaborated on throughout this manuscript, with Table A.1 summarising the notation used and some API examples given in Appendix C. The core component of TrafPy is the *Generator*, which can be used for generating custom, literature, or standard benchmark network traffic traces. These traces can be saved in standard formats (e.g. JSON, CSV, pickle, etc.) and imported into any script or network simulator. Researchers can therefore design their systems and experiments independently of TrafPy and use their own programming languages, making TrafPy compatible with already-developed research projects and future network objects. This also means that TrafPy can be used with any simulation, emulation, or experimentation test bed. The Generator has an optional interactive visual tool for shaping and reproducing distributions, therefore little to no programming experience is required to use it to generate and save traffic data in standard formats. As the nature of DCN traffic changes, new traffic distributions can be generated with TrafPy and state-of-the-art benchmarks established.

#### 3.2.1. Flow traffic

The flow-centric paradigm considers a single demand as a *flow*, which is a task demanding some information be sent from a source node to a destination node in the network. Flow characteristics include *size* (how much information to send), *arrival time* (the time the flow arrives ready to be transported through the network, as derived from the network-level *inter-arrival time* which is the time between a flow's time of arrival and its predecessor's), and *source-destination node pair* (which machine the flow is queued at and where it is requesting to be sent). Together, these characteristics form a network-level *source-destination node pair distribution* ('how much' (as measured by either probability or load) each machine tends to be requested by arriving flows). When a new flow arrives at a source and requests to be sent to a destination, it can be stored in a buffer until completed (all information fully transferred) or, if the buffer is full, dropped. Once dropped or completed, the flow is not re-used.

#### 3.2.2. TrafPy distributions

At the heart of TrafPy are two key notions; that no raw data should be required to produce network traffic, and that every aspect of the API should be parameterised for reproducibility. To achieve the first, rather than using clustering and autoregressive models to fit distributions to data [34,35], TrafPy provides an interactive tool for visually shaping distributions. This way, researchers need only have either a written (e.g. 'the data followed a Pareto distribution with 90% of the values less than 1') or visual description of a traffic trace's characteristics in order to produce it. To achieve the second, all distributions are parameterised by a handful of parameters (termed $D'$; see Appendix B for an example of the parameters used in this paper), and a third party need only see $D'$ in order to reproduce the original distribution. As such, TrafPy traces are discrete distributions in the form of hash tables, which can be sampled at run-time to generate flows. These tables map each possible value taken by all flow characteristics to fractional values which represent either the 'probability of occurring' for size and time distributions, or the 'fraction of the overall traffic load requested' for node distributions. This enables traffic traces to be generated from common TrafPy benchmarks for custom network systems in a reproducible manner without needing to reformat the original data in order to make it compatible with new systems and topologies, as would be needed if the benchmarks were hard-coded request data sets instead of distributions.

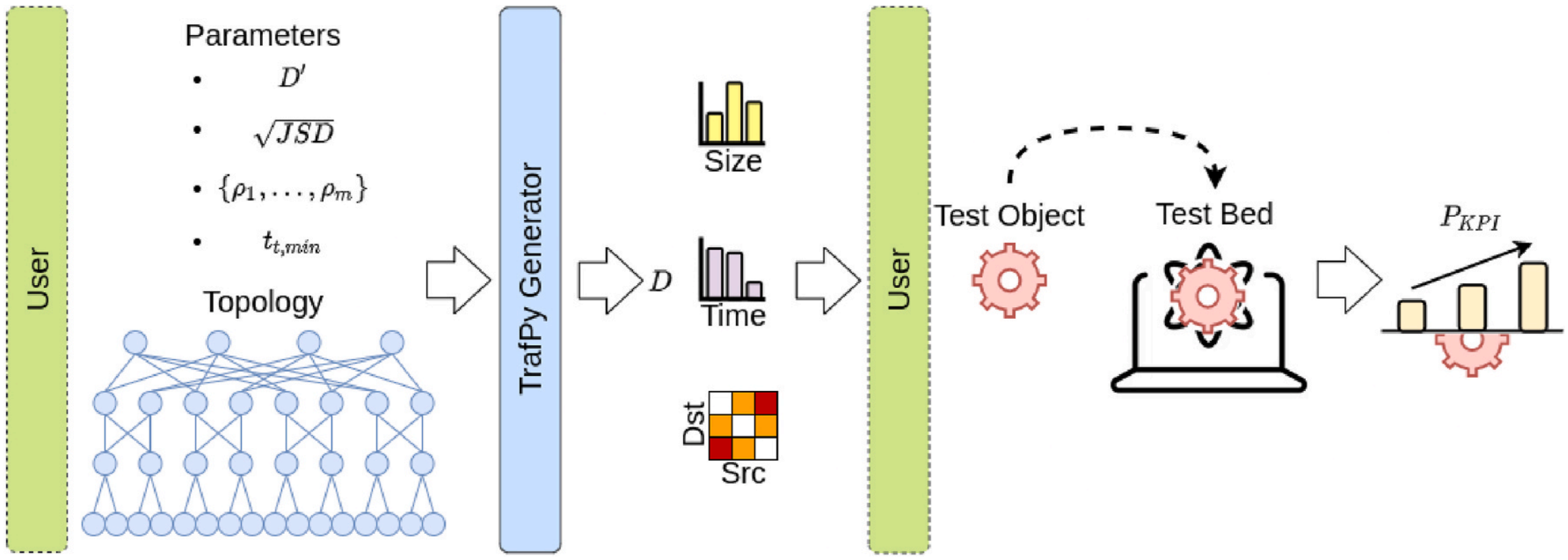

**Fig. 1.** TrafPy API user experience for using custom or benchmark TrafPy parameters $D'$ to make flow traffic trace $D$ with maximum Jensen-Shannon distance threshold $\sqrt{JSD}$ and minimum flow arrival duration $t_{t,min}$ for $m$ loads $\{\rho_1, \ldots, \rho_m\}$. The generated trace $D$ can then be used to benchmark a DCN system test object (e.g. a scheduler) in a test bed (a simulation, emulation, or experimentation environment) to measure the key performance indicators $P_{KPI}$. The user need only use TrafPy to generate the traffic; all other tasks can be done externally to TrafPy in any programming language.

### 3.3. Accuracy and reproducibility of distributions

All TrafPy distributions are summarised by a set of parameters $D'$. Once $D'$ has been established (by e.g. the community as a benchmark or a researcher as a custom stress-test or future workload trace), TrafPy must be able to reliably and accurately reproduce (via sampling) the ‘original’ distribution parameterised by $D'$ each time a new set of traffic data is generated. Therefore, a guarantee that the sampled distribution will be close to the original is required to ensure reproducibility. TrafPy utilises the *Jensen-Shannon Divergence* (JSD) [36,37] to quantify how *distinguishable* discrete probability distributions are from one another. Given a set of $n$ probability distributions $\{\mathbb{P}_1, \dots, \mathbb{P}_n\}$, a corresponding set of weights $\{\pi_1, \dots, \pi_n\}$ to quantify the contribution of each distribution’s entropy to the overall similarity metric, and the entropy $H(\mathbb{P}_i)$ of a discrete distribution with $m$ random variables $X_i = \{x_1^i, \dots, x_m^i\}$ which occur with probability $\mathbb{P}_i = \{\mathbb{P}_i(x_1^i), \dots, \mathbb{P}_i(x_m^i)\}$ where $H(X_i) = -\sum_{j=1}^{m} \mathbb{P}_i(x_j^i) \log \mathbb{P}_i(x_j^i)$, the JSD between the distributions can be calculated as in Equation (1). In the context of TrafPy, the $\mathbb{P}_i$ distributions are the hash tables of variable value-fraction pairs and the weights are simply set to 1.

$$\mathrm{JSD}_{\pi_1, \dots, \pi_n}(\mathbb{P}_1, \dots, \mathbb{P}_n) = H\left(\sum_{i=1}^{n} \pi_i \mathbb{P}_i\right) - \sum_{i=1}^{n} \pi_i H(\mathbb{P}_i) \quad (1)$$

The square root of the Jensen-Shannon Divergence gives the *Jensen-Shannon distance* [37], which is a metric between 0 and 1 used to describe the similarity between distributions (0 being exactly the same, 1 being completely different). The TrafPy API enables users to specify their own maximum $\sqrt{JSD}$ threshold, $\sqrt{JSD_{\text{threshold}}}$, when sampling data from a set of original distributions to create their own data set(s). A lower distance requires that the sampled distributions be more similar to the original distributions. TrafPy will automatically sample more demands until, by the law of large numbers, the user-specified $\sqrt{JSD}$ threshold is met.

Fig. 2 shows how, for an example benchmark’s flow size and inter-arrival time distribution, the $\sqrt{JSD}$ between the original and the sampled distributions changes with the number of samples (number of demands). As shown, most characteristic parameters (mean, minimum, maximum, and standard deviation) of the sampled distributions converge at $\sqrt{JSD} \approx 0.1$; a threshold reached after 137,435 demands for the flow size distribution and 27,194 for the inter-arrival times. The greater the number of possible random variable values and complexity in the original distribution, the more demands which will be needed to lower the $\sqrt{JSD}$. The distribution which requires the most demands to meet the $\sqrt{JSD}$ threshold will determine the minimum number of demands needed for the generated flow data set to accurately reproduce the original set from which it is sampled.

### 3.4. Node distributions

‘Node distributions’ are a mapping of how much each machine (network node) pair tends to be requested by arriving flows, as measured by the pair’s load (flow information arriving per unit time), to form a source-destination pair matrix. These distributions can be defined *explicitly* on a per-node basis. However, explicit mappings would result in $D'$ being defined for a specific topology (since each topology might have a different number of machines and/or a different machine labelling convention). Therefore, TrafPy node distributions can also be *implicitly* defined by high-level parameters. These parameters are the fraction of the nodes and/or node pairs which account for some proportion of the overall traffic load and, optionally, the fraction of the traffic which is intra- vs. inter-cluster (where ‘clusters’ are usually considered as ‘racks’ in the context of DCNs). In this way, node distributions can be defined independently of the network topology, enabling greater generality and the use of custom topologies with traffic traces and benchmarks parameterised by $D'$, even if $D'$ was originally defined for a different topology. Furthermore, this allows individual or groups of network nodes to be set as ‘hot’, ‘cold’, or any combination of hot and cold as desired by the user. Note that this formalism also enables both in-cast (many-to-one) and out-cast (one-to-many) traffic patterns, since any node(s) can have multiple out-cast and in-cast flow demands generated at a given point in time when sampling from the node distribution.

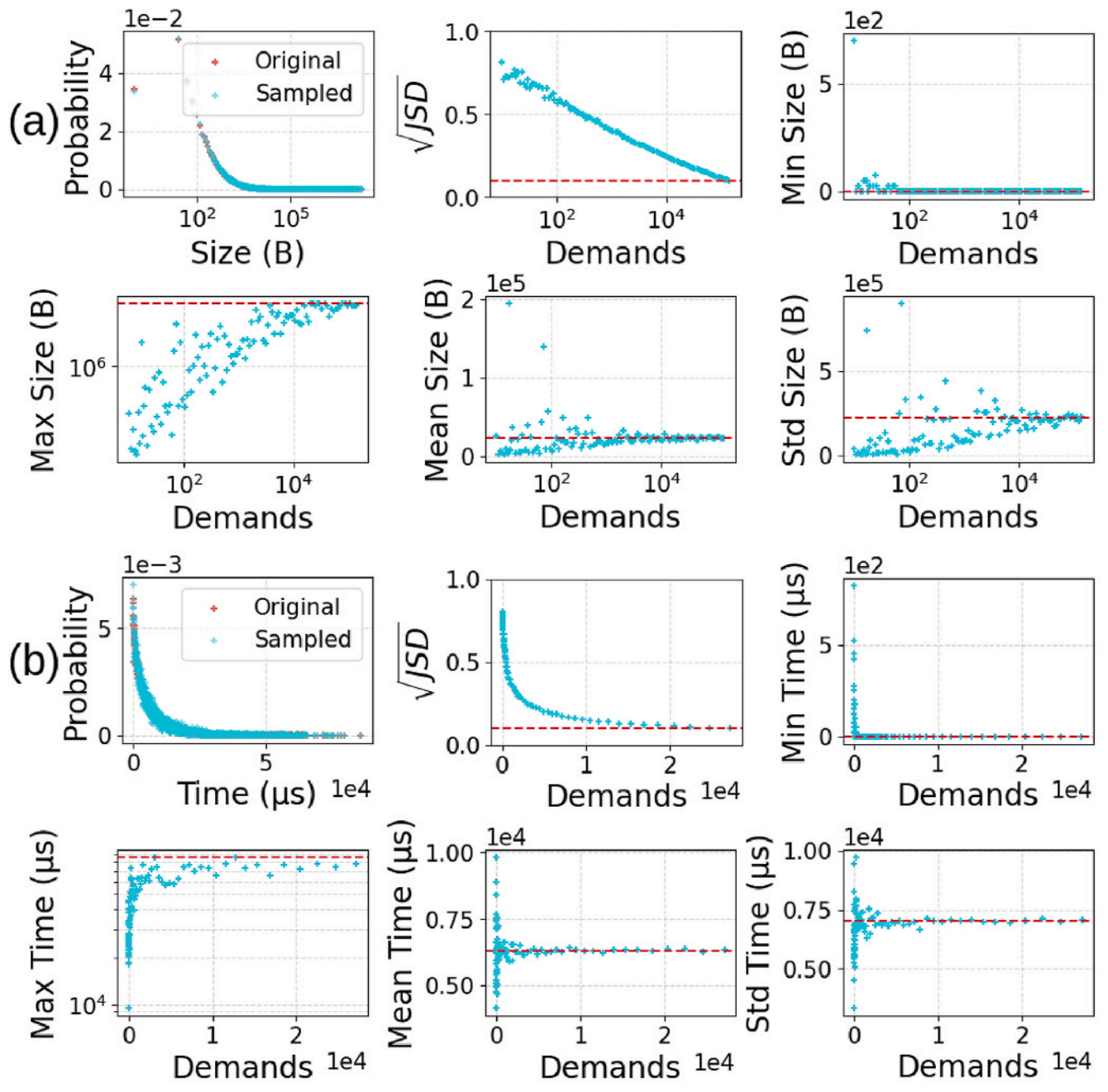

**Fig. 2.** How the Jensen-Shannon distances between the original (red) and sampled (cyan) distributions and the sampled distributions’ characteristic parameters (target from original distribution plotted as red dotted line) vary with the number of demands for (a) flow size and (b) inter-arrival time. Note that the first sub-plots of (a) and (b) are plotting the probability distribution of the flow characteristic in question, whereas the other sub-plots are plotting various metrics ($\sqrt{JSD}$, minimum value, maximum value, etc.) of the generated traffic as a function of the number of demands (flows) generated.

### 3.5. Traffic generation methodology

**Algorithm 1**
TrafPy traffic generation process.

Input: $\mathbb{P}(B^s)$, $\mathbb{P}(B^t)$, $\mathbb{P}(B^n)$, $\sqrt{JSD_{\text{threshold}}}$ $\rho_{\text{target}}$, $\langle n_n, n_c, C_c \rangle$, $t_{t,min}$
**Output:** $\{\boldsymbol{b}^s, \boldsymbol{b}^a, \boldsymbol{b}^p\}$
**Initialise:** $n_f$, $\{\boldsymbol{b}^s, \boldsymbol{b}^t\}$ empty arrays

**Step 1:** Partially initialise $n_f$ flows $\{\boldsymbol{b}^s, \boldsymbol{b}^a\}$
**while** $\sqrt{JSD(\mathbb{P}(B^s), \mathbb{P}(\boldsymbol{b}^s))} \leq \sqrt{JSD_{\text{threshold}}}$ **do**
  $\mathbf{b}^s \leftarrow$ Sample $b^s$ from $\mathbb{P}(B^s)$ $n_f$ times
  $n_f := \lceil 1.1 \times n_f \rceil$
**end while**
**while** $\sqrt{JSD(\mathbb{P}(B^t), \mathbb{P}(\boldsymbol{b}^t))} \leq \sqrt{JSD_{\text{threshold}}}$ **do**
  $\mathbf{b}^t \leftarrow$ Sample $b^t$ from $\mathbb{P}(B^t)$ $n_f$ times
  $n_f := \lceil 1.1 \times n_f \rceil$
**end while**

(*continued on next page*)

**Algorithm 1** (*continued*)

$n_f = \mathbf{max}(\{\mathbf{length}(\boldsymbol{b}^s), \mathbf{length}(\boldsymbol{b}^t)\})$
Resample so that $\mathbf{length}(\boldsymbol{b}^s) = \mathbf{length}(\boldsymbol{b}^t) = n_f$
Initialise $\boldsymbol{b}^a$ zero array of length $n_f$
**for** $i$ in $[2, \ldots, n_f]$ **do**
  $\boldsymbol{b}^a_i := \boldsymbol{b}^a_{i-1} + \boldsymbol{b}^t_{i-1}$
**end for**
$\varrho = \frac{\sum_{i=1}^{n_f} \boldsymbol{b}^s_i}{\boldsymbol{b}^a_{n_f} - \boldsymbol{b}^a_0} \rightarrow \rho = \frac{\varrho}{\frac{n_n \cdot C_c \cdot n_c}{2}} \rightarrow \alpha_t = \frac{\rho}{\rho_{\text{target}}}$
**for** $i$ in $[1, \ldots, n_f]$ **do**
  $\boldsymbol{b}^a_i := \alpha_t \times \boldsymbol{b}^a_i$
**end for**
$\varrho := \frac{\sum_{i=1}^{n_f} \boldsymbol{b}^s_i}{\boldsymbol{b}^a_{n_f} - \boldsymbol{b}^a_0} \rightarrow \rho := \frac{\varrho}{\frac{n_n \cdot C_c \cdot n_c}{2}}$

**Step 2:** 'Pack the flows' → fully initialise $n_f$ flows $\{\boldsymbol{b}^s, \boldsymbol{b}^a, \boldsymbol{b}^p\}$
Initialise $\boldsymbol{b}^p$ and $\boldsymbol{b}^n$ from $\mathbb{P}(B^n)$ with $n_n^2 - n_n$ elements
$\boldsymbol{d} = \varrho \cdot \boldsymbol{b}^n \cdot (\boldsymbol{b}^a_{n_f} - \boldsymbol{b}^a_0)$
**for** $i$ in $[1, \ldots, n_f]$ **do**
  Sort pairs in descending $\boldsymbol{d}_p$ order and randomly self-shuffle equal $\boldsymbol{d}_p$ pairs
  **First pass:** Attempt $\boldsymbol{d}_p \approx 0 \forall p \in [1, \ldots, n_n^2 - n_n]$
  **for** $p$ in $[1, \ldots, n_n^2 - n_n]$ **do**
    **if** $\boldsymbol{d}_p - \boldsymbol{b}^s_i \geq 0$ **then**
      $\boldsymbol{b}^p_i = p$
      $\boldsymbol{d}_p := \boldsymbol{d}_p - \boldsymbol{b}^s_i$
      **break**
    **end if**
**end for**
**if** first pass unsuccessful **then**
    **Second pass:** Ensure no link capacity exceeds $\frac{C_c}{2}$
    **for** $p$ in $[1, \ldots, n_n^2 - n_n]$ **do**
      **if** capacity not exceeded **then**
        $\boldsymbol{b}^p_i = p$
        $\boldsymbol{d}_p := \boldsymbol{d}_p - \boldsymbol{b}^s_i$
        **break**
      **end if**
    **end for**
  **end if**
**end for**

**Step 3:** Ensure $\boldsymbol{b}^a_{n_f} - \boldsymbol{b}^a_0 \geq t_{t,min}$
**if** $\boldsymbol{b}^a_{n_f} - \boldsymbol{b}^a_0 < t_{t,min}$ **then**
  $\beta = \left\lceil \frac{\boldsymbol{b}^a_{n_f} - \boldsymbol{b}^a_0}{t_{t,min}} \right\rceil$
  $\{\boldsymbol{b}^s, \boldsymbol{b}^a, \boldsymbol{b}^p\} := \mathbf{double}(\{\boldsymbol{b}^s, \boldsymbol{b}^a, \boldsymbol{b}^p\})$ $\beta$ times
**end if**

Given the distributions of flow sizes, inter-arrival times, and node pairs $\mathbb{P}(B^s)$, $\mathbb{P}(B^t)$, and $\mathbb{P}(B^n)$ of a benchmark $B$, TrafPy can generate traffic at a (optionally) specified target load fraction (fraction of overall network capacity being requested for a given time period) $\rho_{\text{target}} \in [0, 1]$ with maximum Jensen-Shannon distance threshold $\sqrt{JSD_{\text{threshold}}}$ for an arbitrary topology $T$ with $n_n$ server nodes, $n_c$ channels (light paths) per communication link, and $C_c$ capacity per server node link channel (divided equally between the source and destination ports such that each machine may simultaneously transmit and receive data), forming tuple $\langle n_n, n_c, C_c \rangle$ with total network capacity per direction (maximum information units transported per unit time) $C_t = \frac{n_n \cdot C_c \cdot n_c}{2}$. Since load rate is defined as information arriving per unit time, in order to generate traffic at arbitrary loads, either the amount of information (flow sizes) or the rate of arrival (flow inter-arrival times) must be adjusted in order to change the load rate. Since DCNs tend to handle particular types of applications and jobs which result in particular flow sizes, we posit that a reasonable assumption is that changing loads are the result of changing rates of demand arrivals rather than changing flow sizes (which remain fixed for a given application type). Therefore, if a target load is specified, TrafPy automatically adjusts the scale of the inter-arrival time distribution values in $\mathbb{P}(B^t)$ by a constant factor to meet the target load whilst keeping the same general shape of the $\mathbb{P}(B^t)$ distribution that was initially input to the generator. The following 3-step traffic generation process (summarised in Algorithm 1) is used to achieve the above:

**Step 1** (generate $n_f$ flows with size and arrival time characteristics $\{\boldsymbol{b}^s, \boldsymbol{b}^a\}$): First, $n_{b^s}$ flow sizes and $n_{b^t}$ inter-arrival times are independently sampled from $\mathbb{P}(B^s)$ and $\mathbb{P}(B^t)$ to form vectors $\boldsymbol{b}^s$ and $\boldsymbol{b}^t$ respectively, where $n_{b^s}$ and $n_{b^t}$ are incrementally increased by a constant factor until $\sqrt{JSD(\mathbb{P}(B^s), \mathbb{P}(\boldsymbol{b}^s))} \leq \sqrt{JSD_{\text{threshold}}}$ and $\sqrt{JSD(\mathbb{P}(B^t), \mathbb{P}(\boldsymbol{b}^t))} \leq \sqrt{JSD_{\text{threshold}}}$ by the law of large numbers. Whichever distribution needed fewer samples to meet $\sqrt{JSD} \leq \sqrt{JSD_{\text{threshold}}}$ is then continually sampled such that there are $n_f$ flow sizes and inter-arrival times, where $n_f = \mathbf{max}(\{n_{b^s}, n_{b^t}\})$. Then, $\boldsymbol{b}^t$ (whose order is arbitrary from the previous random sampling process) can be converted to an equivalent arrival time vector $\boldsymbol{b}^a$ by initialising a zero array of length $n_f$ and setting $\boldsymbol{b}^a_i := \boldsymbol{b}^a_{i-1} + \boldsymbol{b}^t_{i-1} \forall i \in [2, \ldots, n_f]$, resulting in a total time duration of $t_t = \boldsymbol{b}^a_{n_f} - \boldsymbol{b}^a_0$ over which the flows arrive. Next, the load rate $\varrho$ is evaluated with $\varrho = \frac{\sum_{i=1}^{n_f} \boldsymbol{b}^s_i}{t_t}$, converted to a load fraction $\rho = \frac{\varrho}{C_t}$, and adjusted to meet $\rho_{\text{target}}$ by multiplying the elements of $\boldsymbol{b}^t$ by a constant factor $\alpha_t = \frac{\rho}{\rho_{\text{target}}}$. Then, $\boldsymbol{b}^a$ can be re-initialised with the updated $\boldsymbol{b}^t$ as before, and a set $\{\boldsymbol{b}^s, \boldsymbol{b}^a\}$ of $n_f$ flows can be partially initialised each with size $b^s$ and arrival time $b^a$ and an overall load $\rho = \rho_{\text{target}}$ on network $T$.

**Step 2** ('pack the flows' → generate $n_f$ flows with size, arrival time, and source-destination node pair characteristics $\{\boldsymbol{b}^s, \boldsymbol{b}^a, \boldsymbol{b}^p\}$): Next, to meet the source-destination node pair load fractions specified by $\mathbb{P}(B^n)$, the flows are packed into node pairs with a simple packing algorithm. First, a vector of $n_n^2 - n_n$ node pairs $\boldsymbol{b}^p$ (which do not include self-similar pairs) and their corresponding load pair fractions $\boldsymbol{b}^n$ are extracted from $\mathbb{P}(B^n)$. Next, these 'target' load pair fractions $\boldsymbol{b}^n$ are converted into a hash table mapping each pair $p$ of the $[1, \ldots, n_n^2 - n_n]$ pairs to their current 'distance' from their respective target total information request magnitudes $\boldsymbol{d} = \varrho \cdot \boldsymbol{b}^n \cdot t_t$. In other words, we take the load fractions (fraction of overall information requested) of each node pair $\boldsymbol{b}^n$ and multiply them by the total simulation load rate (information units arriving per unit time) $\varrho$ and the total simulation time $t_t$ to create a vector $\boldsymbol{d}$ which, when first initialised, represents the total amount of information which is requested by each source-destination pair across the whole simulation. The task of the packer is therefore to assign source-destination pairs to each flow such that $\boldsymbol{d}_p \approx 0 \forall p \in [1, \ldots, n_n^2 - n_n]$. For each sequential $i$th flow $\forall i \in [1, \ldots, n_f]$, after sorting the pairs in descending $\boldsymbol{d}_p$ order (with any pairs with equal $\boldsymbol{d}_p$ randomly shuffled amongst one-another), the packer will try to 'pack the flow' (given its size $\boldsymbol{b}^s_i$) into a source-destination pair in two passes. For the first pass the packer loops through each sorted $p$th pair $\forall p \in [1, \ldots, n_n^2 - n_n]$ and checks that assigning the flow to this pair would not result in $\boldsymbol{d}_p < 0$. If this condition is met, the packer sets $\boldsymbol{b}^p_i = p$ and $\boldsymbol{d}_p := \boldsymbol{d}_p - \boldsymbol{b}^s_i$ before moving to the next flow. However, if the condition is violated for all pairs, the packer moves to the second pass, where it again loops through each sorted pair $p$ but now, rather than ensuring $\boldsymbol{d}_p \geq 0$, only ensures that assigning the pair would not exceed the maximum server link's source/destination port capacity $\frac{C_c}{2}$ before setting $\boldsymbol{b}^p_i = p$ and $\boldsymbol{d}_p := \boldsymbol{d}_p - \boldsymbol{b}^s_i$. In other words, the first pass attempts to achieve $\boldsymbol{d}_p \approx 0 \forall p \in [1, \ldots, n_n^2 - n_n]$ to try to match $\mathbb{P}(B^n)$ but, failing that, the second pass ensures that no server link load exceeds 1.0 of the link capacity. Consequently, as $\rho_{\text{target}}$ approaches 1.0, so too will the resultant packed node distribution's server links, thereby converging on a uniform distribution no matter what the original skewness was of $\mathbb{P}(B^n)$ as shown in Fig. 3 and further elaborated on in Appendix E. Once this packing process is complete, a set $\{\boldsymbol{b}^s, \boldsymbol{b}^a, \boldsymbol{b}^p\}$ of $n_f$ flows each with size $b^s$, arrival time $b^a$, and source-destination node pair $b^p$, an overall load $\rho_{\text{target}}$ on network $T$, and a flow size, inter-arrival time, and node distribution of approximately $\mathbb{P}(B^s)$, $\mathbb{P}(B^t)$, and $\mathbb{P}(B^n)$ will have been fully initialised.

**Step 3** (ensure $\boldsymbol{b}^a_{n_f} - \boldsymbol{b}^a_0 \geq t_{t,min}$): The final stage of the flow generation process is then to ensure that the flow arrival duration $t_t$ is greater than or equal to some minimum duration $t_{t,min}$ (a parameter often required for test bed measurement reliability) specified by either the user. This is done by simply doubling the set $\{\boldsymbol{b}^s, \boldsymbol{b}^a, \boldsymbol{b}^p\}$ of flows $\beta = \left\lceil \frac{t_t}{t_{t,min}} \right\rceil$ times to make an updated set of $n_f := \beta \cdot n_f$ flows with $t_t \geq t_{t,min}$ and the same distribution and load statistics as before.

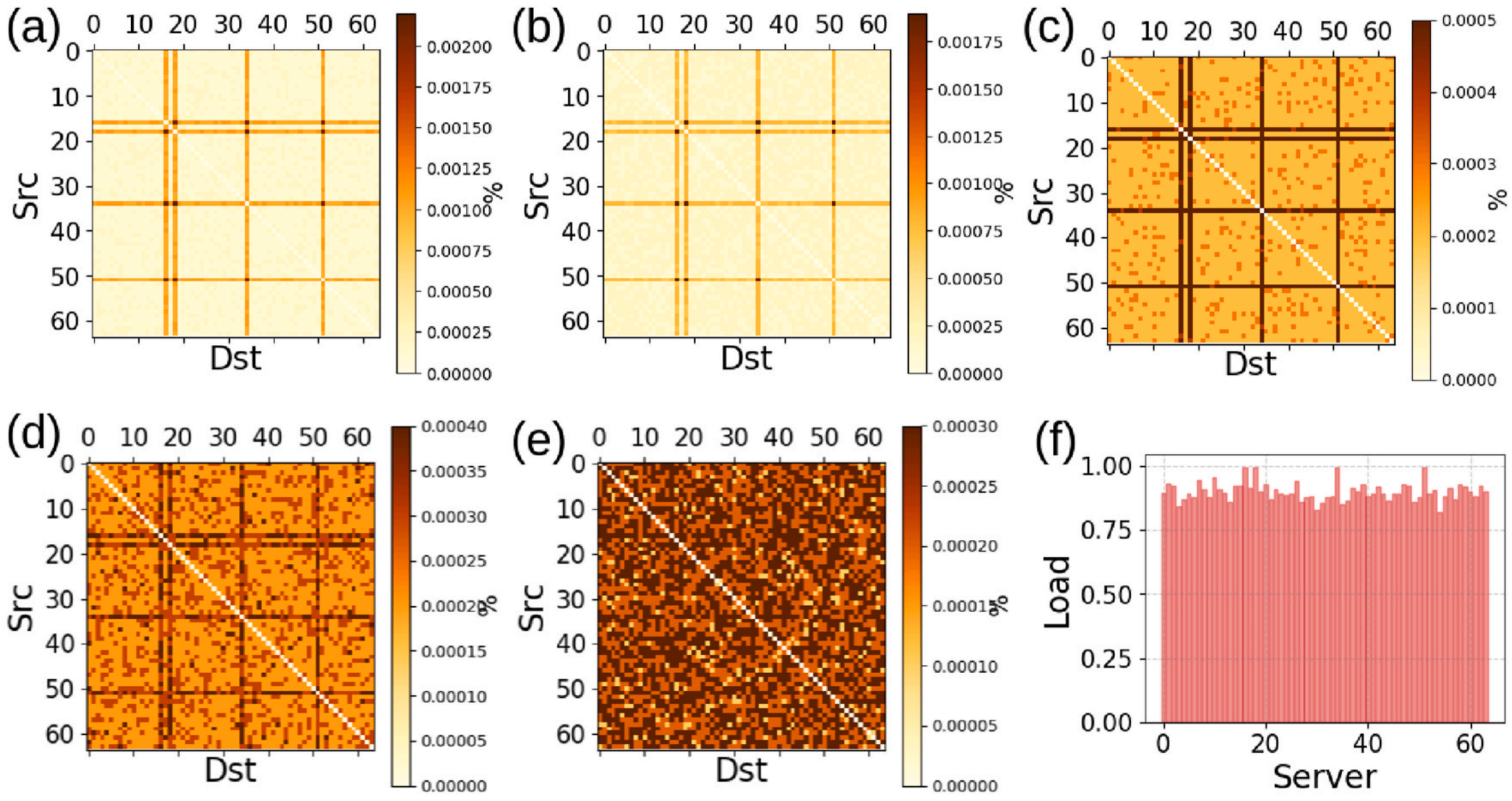

**Fig. 3.** Visualisation of the packed flow nodes converging on uniform distributions as the total network load approaches 1.0 regardless of how skewed the original target node distribution is. The plotted distributions are for overall network loads (a) 0.1, (b) 0.3, (c) 0.5, (d) 0.7, and (e) 0.9, and (f) the final demonstrably uniform endpoint loads on each server at 0.9 overall load.

### 3.5.1. *Traffic generation guidelines*

Given a user- or benchmark-specified set of distribution parameters $D'$, TrafPy generates traffic trace $D$. As such, whenever using TrafPy to generate $D$, $D'$ should always be reported to help others reproduce the same trace (as done in Table B.2 of Appendix B for this manuscript). For the same reason, all traffic traces $D$ generated from $D'$ should have a maximum $\sqrt{JSD_{\text{threshold}}}$ of 0.1 as outlined in Section 3.3. Enough demands should be generated so as to have a last demand arrival time $t_t$ larger than the time needed to complete the largest demands in the user-defined network $T$ under the test conditions used; not doing so would result in all large flows being dropped regardless of what decisions were made. This would unfairly punish systems optimised for large demands, since such systems would allocate network resources to requests which ultimately could never be completed during the experiment. TrafPy conveniently generates and saves traffic data sets in a range of formats including JSON, CSV, and pickle. Therefore if desired, users may generate traffic in TrafPy and then use their own custom test bed and analysis scripts written in any programming language thereafter by simply importing the TrafPy-generated traffic. For result reliability, each trace $D$ should be generated $R$ times from $D'$ and used to test the network object, where $R$ should be sufficiently large enough so as to have a satisfactory confidence interval (which might vary from project-to-project but should be reported regardless).

## 4. Optical networks

The key purpose of TrafPy is for it to be used as a tool to explore novel areas of DCN research. One such area of particular importance is that of optical DCNs, which strive to replace electronically interconnected networks with optical systems in order to improve performance whilst reducing power consumption.

### 4.1. *Limitations of current electronic packet switched networks*

The servers of traditional multi-tier data centre and high performance computing (HPC) systems are interconnected by electronic packet switched (EPS) networks. Such ‘electronic DCNs’ have poor scalability, bandwidth, latency, and power consumption. Data centres now consume 2% of the World’s electricity; more than the entire aviation industry and estimated to increase to 15% by 2030, with the network sometimes accounting for >50% of total power consumption [38]. Furthermore, the sensitivity of electronic switches to workloads limits their computational and application performance. Compounding this, the slowing of Moore’s Law coinciding with new data-hungry demands means that electronic switches are unable to keep up with emerging applications (internet-of-things, artificial intelligence, genome processing, etc.) which follow data-heavy trends [39,40]. Although the compute power of DCN server nodes, as measured by flops per second, has increased by a factor of 65 over the last 18 years, the bandwidth of the DCN network facilitating communication between these nodes has only increased by a factor of 4.8, resulting in an 8-factor decrease in bytes communicated per flop. This has created a performance bottleneck not in the server nodes themselves, but rather in the network connecting them. As a result, management systems such as machine placers, schedulers and topology controllers are being forced to minimise data movement and constrain applications to operate locally, which would otherwise benefit from utilising more distributed architectures. Further degrading system and application performance, these systems also suffer from high median and 99th percentile network latencies on the order of 100 μs and 100 ms respectively.

### 4.2. *Optical circuit switched networks*

DCNs with optical interconnects have the potential to offer orders-of-magnitude improvements in performance and energy efficiency and thereby address the limitations of EPS networks [48,64–66]. Optical circuit switched (OCS) networks offer a promising avenue with which to realise optical DCNs, and have been used in many DCN solutions as they offer stable non-blocking circuit configurations with high-capacity and scalability [41]. In contrast to optical packet switching, they are simpler to implement and they eliminate the need for in-switch buffering or queuing and addressing. They establish single-hop connections with a wide range of circuit establishment time, lasting from orders of magnitude less than a second to hours or days. Leveraging stable circuit establishments, they can employ wavelength division multiplexing (WDM) and modulation formats to reach higher capacity. OCS switches are readily available [41] and are being used as part of many existing networks. They are mainly employed as part of a hybrid network, as in Ref. [42], in order to cater to specific types of traffic. However, they cannot be used on their own as they suffer from two key limitations: the long reconfiguration time (time taken to switch) and the long circuit computation time (time taken to compute the schedule), as shown by Fig. 4.

Fig. 4 shows the circuit computation and the reconfiguration time of the key state-of-the-art OCS technologies. In summary, slow beam steering and light guiding technologies (millisecond OCS) were assisted with slow software-based circuit computation to provide reconfiguration, also in milliseconds (HELIOS, Firefly and OSA) [42–44]. More recent work has shown microsecond speed WSS-based OCS reliant on FPGA-based control (REACToR, Mordia) [45,46]. Rotor switches and fast SOA-based switches with schedule-less control were also explored for fast OCS in RotorNet [47] and Sirius [48] respectively. Although schedule-less architectures simplified the control plane, they result in performance-inefficient networks as network resources are allocated uniformly even in dynamic and skewed traffic environments.

However, with transceivers growing at a staggering rate, already reaching 100 Gbps [49] (trending towards 400G and 800G) and switch bandwidth increasing beyond 6.4 Tbps [50], the increased data-rate makes OCS 5-6 orders of magnitude too slow. This ever increasing gap between OCS switching/control speed and transceiver data rate makes OCS unsuitable as standalone solutions. Hence, PULSE (indicated by a star in Fig. 4) [51] proposed a two-fold solution: The first is the use of SOA-aided widely tunable-switching methods to minimize the reconfiguration time to sub-nanoseconds [52]. The second is a custom-made ASIC controller or scheduler that reduces reconfiguration computation time to nanoseconds. PULSE matches OCS switching times to packet-level granularity, making them suitable and adaptable to modern high capacity, bandwidth and speed switching data centre networks.

However, the performance of PULSE is heavily reliant on the performance of the scheduling heuristic employed. TrafPy can therefore be used as a tool with which to evaluate the performance of different design choices and resource management systems in novel OCS networks, such as PULSE (an OCS DCN system which was developed with the help of TrafPy [53]), and thereby help to realise future all-optical DCNs.

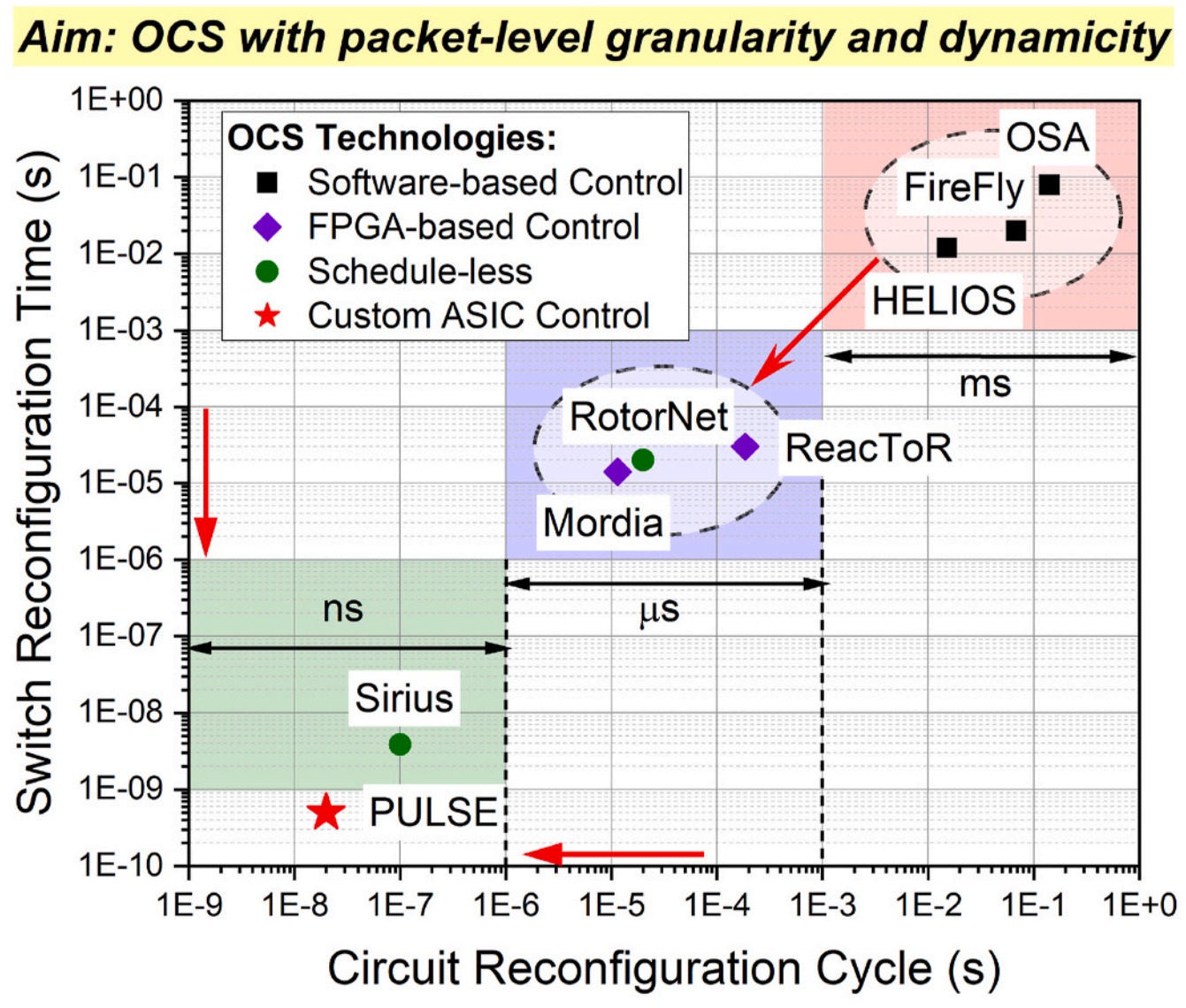

**Fig. 4.** Reconfiguration and computation times of various OCS architectures.

## 5. Experiment

Here we conduct a brief experiment into the sensitivity of 4 schedulers to different traffic traces. Specifically, we look at shortest remaining processing time (SRPT) [6,54,55], fair share [54], first fit (FF) [56], and random DCN flow scheduling.

### 5.1. Network

All experiments assume an optical TDM-based circuit switched network architecture with a 64-server folded clos (spine-leaf) topology made up of 2 core switches, 4 top-of-the-rack (ToR) switches, and 64 servers (16 servers per rack) with bidirectional links, as shown in Fig. 5. The server-to-rack and ToR-to-core links each have 1 channel with 10 Gbps and 80 Gbps capacity respectively, leading to a 1:1 subscription ratio and a total network capacity of 640 Gbps (320 Gbps bisection bandwidth). Flows are mapped to TDM circuits, and we assume ideal server-level time multiplexing of the flows' packets such that the bandwidth of each channel can be fully utilised. The core switch performs link/fiber switching. There are various ways to perform packet/TDM aggregation of flows at the server and to realise such networks, but neither are the focus of this paper.

### 5.2. Traffic traces

We use TrafPy to generate 2 categories of traffic with which to investigate our schedulers; DCN traces based on real-world application data, and custom skewed node and rack data for testing system performance under extreme conditions. We use a maximum $\sqrt{JSD_{\text{threshold}}}$ of 0.1, setting $t_{t,min} = 3.20 \times 10^5 \mu s$ (≈10 times larger than the time taken to complete the largest ≈$20 \times 10^6$ B flow amongst our benchmarks), and generating traffic of loads 0.1–0.9 for each data set. We generate each set $R = 5$ times to run 5 repeats of our experiments and therefore ensure reliability. All TrafPy parameters $D'$ used to generate the traffic are reported in Table B.2 of Appendix B for reproducibility.

#### 5.2.1. 'Realistic' DCN traces

Four types of DCN and their network flow demand distributions are explored; *University* [30], *Private Enterprise* [57], *Commercial Cloud* [31], and *Social Media Cloud* [33]. Each DCN type services different applications and therefore has a different traffic pattern. Using TrafPy, flow distributions for each of these categories were generated to established a set of open-source traffic traces for the *DCN benchmark*. The tuned TrafPy parameters $D'$ of each flow characteristic have been summarised in Table B.2. The resultant distributions are shown in Fig. 6, and the subsequent quantitative summary of each distribution's characteristics is given in Table B.3 of Appendix B.

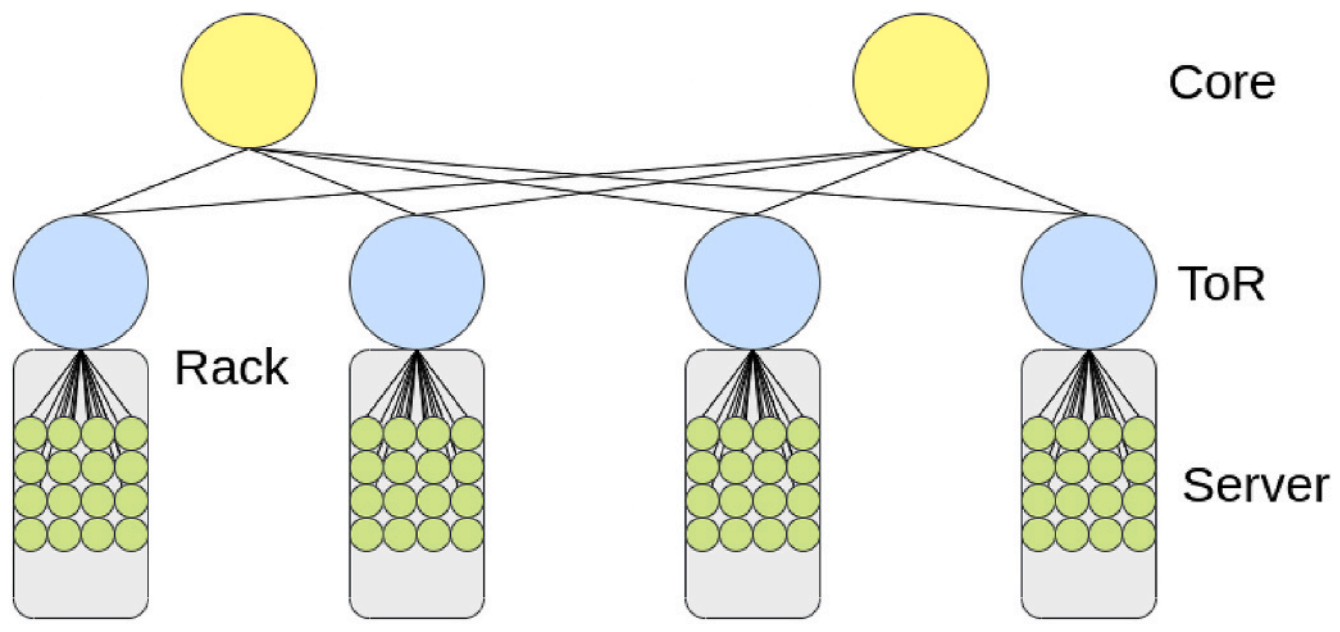

**Fig. 5.** 2-layer spine-leaf topology used with 64 end point (server) nodes, 10 Gbps server-to-ToR links, and 80 Gbps ToR-to-core links (1:1 subscription ratio, 640 Gbps total network capacity).

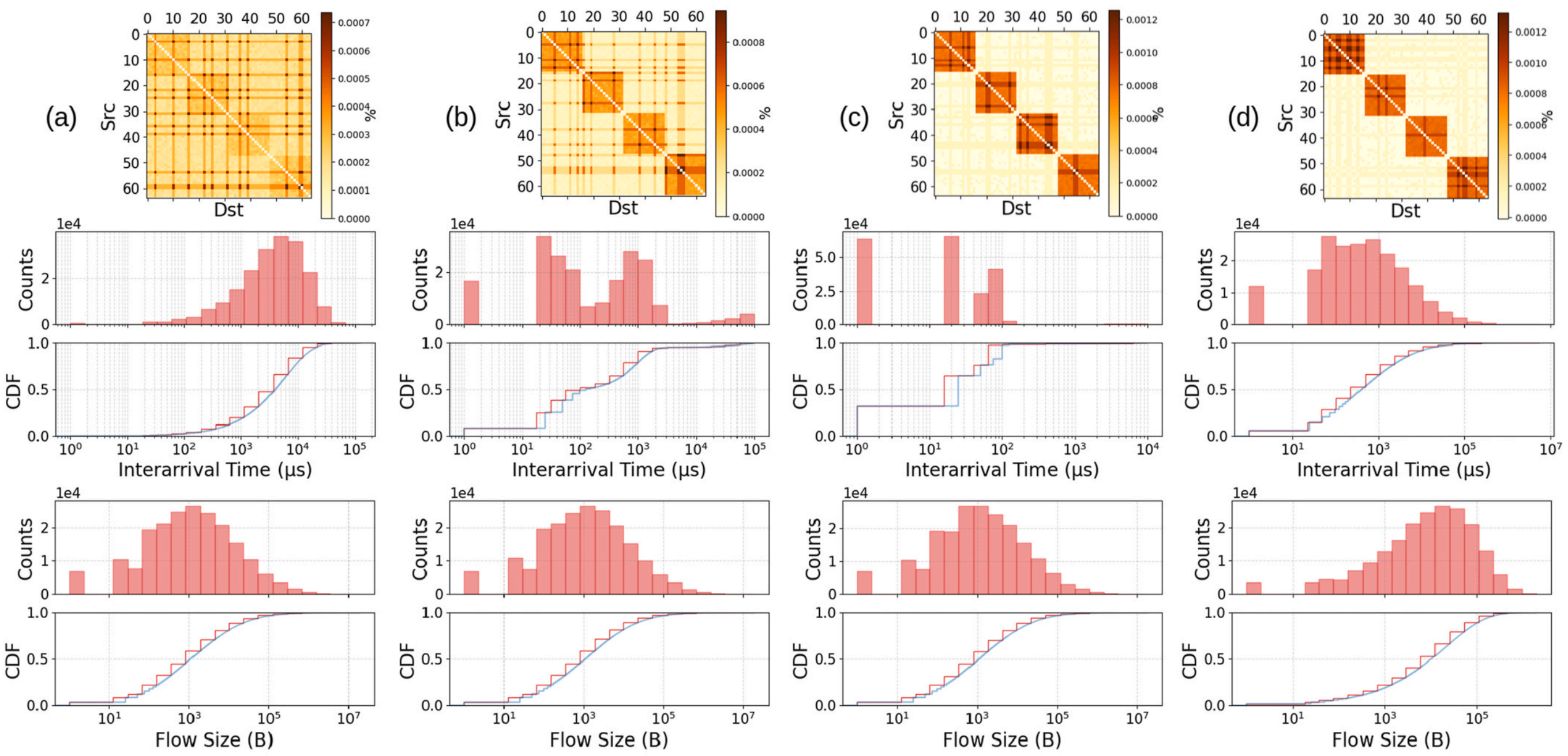

**Fig. 6.** TrafPy distribution plots for the *DCN benchmark* containing the (a) University [30], (b) Private Enterprise [57], (c) Commercial Cloud [31], and (d) Social Media Cloud [33] data sets. Each plot contains (i) the end point node load distribution matrix and (ii) the flow size and inter-arrival time histogram and CDF distributions.

#### 5.2.2. 'Extreme' skewed node and rack sensitivity traces

We generated two additional traces; the *skewed nodes sensitivity* benchmark and the *rack sensitivity* benchmark. These were not based on realistic data, but rather designed to test and better understand our systems under extreme conditions. Both use the same flow size and inter-arrival time distributions as the commercial cloud data set in Fig. 6, however the node distribution is adjusted. Specifically, the skewed nodes benchmark is made up of 5 sets with uniform, 5%, 10%, 20%, and 40% of the server nodes being 'skewed' by accounting for 55% of the total overall traffic load, named *skewed_nodes_sensitivity_uniform*, *0.05*, *0.1*, *0.2*, and *0.4* respectively (see Appendix E for further justification and analysis of these values). Similarly, the rack distribution benchmark is made up of 5 sets with uniform, 20%, 40%, 60%, and 80% of the traffic being intra-rack (and the rest inter-rack) named *rack_sensitivity_uniform*, *0.2*, *0.4*, *0.6*, and *0.8* respectively. Therefore, these distributions allow for investigations into DCN system sensitivity to i) the number of skewed nodes and ii) the ratio of intra- vs. inter-rack traffic. They have been plotted in Fig. 7.

### 5.3. Simulation details

We use a time-driven simulator where scheduling decisions are made at fixed intervals. The time between decisions is the 'slot size'; smaller slot sizes result in greater scheduling decision and measurement metric granularity, but at the cost of longer simulation times and the need for scheduler and switch hardware optimisation [52,53,58–60,63]. We use a slot size of 1 ms. We assume perfect packet time-multiplexing whereby the scheduler is allowed to schedule as many flow packets for the next time slot as the channel bandwidth of its rate-limiting link in its chosen path will allow. We run 9 simulations (loads 0.1–0.9) for each benchmark data set, terminating the simulation when the last demand arrives at $t = t_t$ (which is $\geq t_{t,min} = 3.20 \times 10^5 \mu s$). We set the warm-up time as being 10% of the simulation time $t_t$ before which no collected data contribute to the final performance metrics. Similarly, since the simulation is terminated at $t_t$, we exclude any cool-down period from measurement. For each experiment, we then record: (1) mean flow completion time (FCT); (2) 99th percentile (p99) FCT; (3) maximum (max) FCT; (4) absolute throughput (total number of information units transported per unit time); (5) relative throughput (fraction of arrived information successfully transported); and (6) fraction of arrived flows accepted. We report each of these metrics' mean across the $R = 5$ runs and their corresponding 95% confidence intervals.

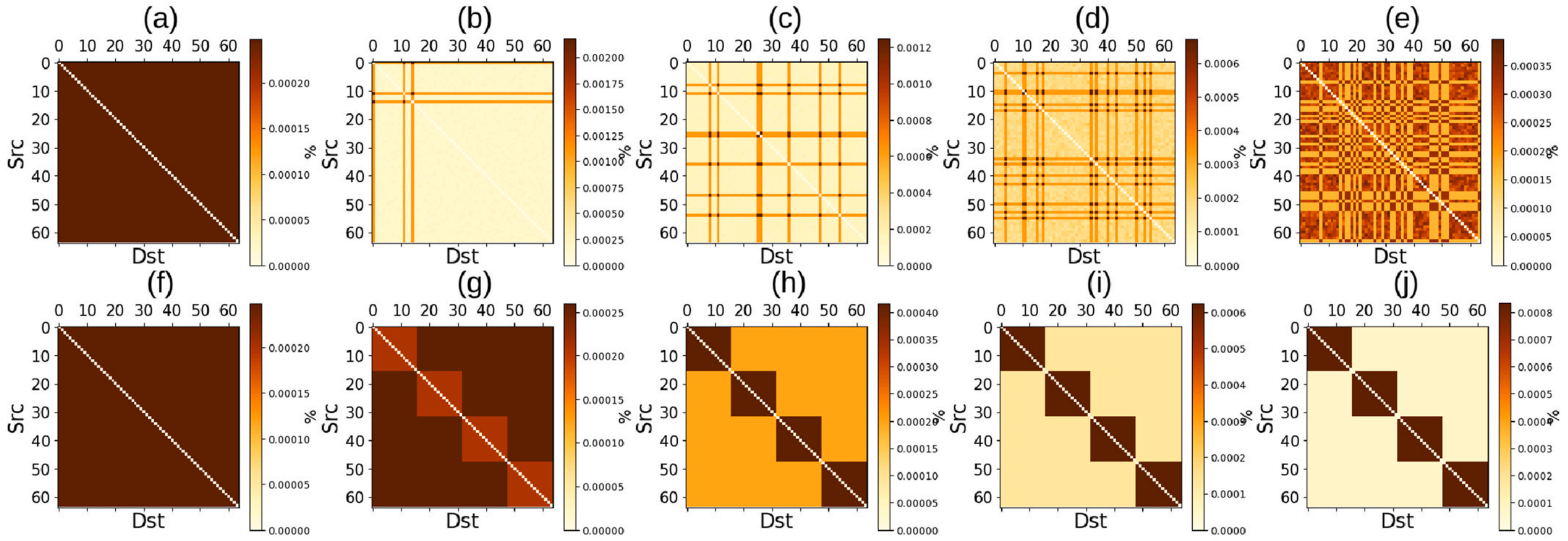

**Fig. 7.** TrafPy node distribution plots for the *skewed nodes sensitivity* benchmark with (a) uniform, (b) 5%, (c) 10%, (d) 20%, and (e) 40% of nodes accounting for 55% of the overall traffic load, and for the *rack sensitivity* benchmark with (f) uniform, (g) 20%, (h) 40%, (i) 60%, and (j) 80% traffic being intra-rack and the rest inter-rack.

## 6. Results

To begin the investigation into the sensitivity of different schedulers, we first input TrafPy-generated traffic with heavily skewed nodes and racks (see Section 5.2.2) into our simulator to understand how the four schedulers considered behave at the extremes. We then test the same schedulers under traces for different DCN types to see how the results from the 'extreme' condition investigation translate into more realistic scenarios. For brevity, we provide the full results in Appendix G and a summary in this section.

**Extreme Rack Conditions** As shown in Table F.17, as the rack distribution becomes heavily skewed to intra-rack, the completion time metrics of FS become increasingly superior to SRPT. This suggests that real DCNs which have heavy intra-rack traffic (e.g. social media cloud DCNs) would benefit from deploying pure FS scheduling policies, at least at higher loads, whereas DCNs with heavy inter-rack traffic (e.g. university DCNs) would benefit from deploying FS at medium loads and SRPT at low and high loads.

In terms of throughput and demands accepted, FF is competitive with SRPT and FS at low intra-rack traffic levels, but as the DCN becomes more heavily intra-rack (e.g. social media cloud DCNs), SRPT and FS are preferable, with FS achieving the best performances at higher loads. Again, a preferable strategy would likely be to utilise SRPT strategies at low loads before switching to FS at loads about 0.3–0.5 (depending on the level of intra-rack traffic).

**Extreme Node Conditions** As shown in Table F.18, at the two extremes of heavily skewed and uniform traffic, scheduler completion time performances are similar in that SRPT outperforms FS at low and high loads, but FS performs well at medium loads. However, in between these two extremes (around 40% of nodes requesting 55% of overall traffic), there is a point where FS becomes the dominant scheduler in terms of completion time.

In terms of throughput and demands accepted, under heavily skewed conditions (5% nodes requesting 55% of traffic), FF and/or Rand beat SRPT and FS across all 0.1–0.9 loads in terms of throughput and fraction of information accepted. This suggests that FF and SRPT are strained under high skews with respect to these two metrics. However, as observed with the uniform distribution, this comes at the cost of the fraction of arrived flows accepted, where SRPT and FS outperform FF and Rand across all loads. As the proportion of nodes requesting 55% of traffic is increased to 10%, 20%, and 40%, relative scheduler performances converge to those seen with the uniform distribution, with FS and SRPT being mostly dominant except at high 0.8 and 0.9 loads, where FF often has the better throughput and fraction of information accepted.

**Realistic Conditions** Table F.19 summarises the results for the four schedulers on each of the four 'realistic' DCN benchmarks considered. As shown, the SRPT scheduler tends to achieve the best completion time metrics when loads are low ($\leq 0.7$) and where traffic is primarily inter-rack (the University and Private Enterprise DCNs). This is to be expected, since a policy which prioritises completion of the smallest flows as soon as possible will keep its completion time averages low. However, as traffic reaches higher loads ($> 0.7$), the fair share policy achieves the best completion time metrics. This indicates that networks would benefit from scheduling policies which can dynamically adapt to changing traffic loads. Moreover, for networks with characteristically intra-rack traffic (the Commercial Cloud and Social Media Cloud DCNs), the fair share policy attains the best completion time and throughput metrics. These results therefore validate the predictions made by the rack distribution sensitivity analysis study; namely that completion time metrics in real DCN traces with heavily intra-rack (e.g. Commercial Cloud and Social Media Cloud) traffic benefit from FS scheduling strategies. On the other hand, at least for low loads, low intra-rack DCN traces (e.g. University and Private Enterprise) benefit from SRPT scheduling strategies.

These results suggest that not only should scheduling policies be adapted to changing traffic loads, but also to changing characteristics such as the level of inter- vs. intra-rack communication. Note that, as expected, the fair share policy provides the best worst-case completion time (max FCT), the greatest network utilisation (throughput), and the strongest service guarantee (number of flow requests satisfied) across most loads and DCN types.

## 7. Conclusion & further work

In conclusion, we have introduced TrafPy; an API for generating custom and realistic DCN traffic and a standardised protocol for benchmarking DCN systems which is compatible with any simulation, emulation, or experimentation test bed. These systems can be any combination of networked devices or methods such as schedulers, switches, routers, admission control policies, management protocols, topologies, buffering methods, and so on. TrafPy has been developed with a focus on having a high level of *fidelity*, *generality*, *scalability*, *reproducibility*, *repeatability*, *replicability*, *compatibility*, and *comparability* in the context of DCN research, which in turn will aid in accelerating innovation.

We have demonstrated the efficacy of TrafPy by briefly investigating the sensitivity of four canonical schedulers to varying traffic loads and characteristics. The scheduler performances were heavily dependent on the level of intra-rack traffic and overall network load. We found that SRPT was generally the dominant scheduler for low intra-rack traffic (particularly at low loads), but that FS became superior across all loads at high intra-rack levels. These insights were then found to translate into realistic DCN traces, with low intra-rack users such as University and Private Enterprise DCNs benefiting from SRPT policies at low and medium loads and high intra-rack traces such as Commercial Cloud and Social Media Cloud being more suited to FS strategies. This shows that there is no 'one size fits all' strategy for scheduling different types of DCNs, and that there would be great value in the development of traffic-informed and dynamic DCN systems. With its standardised traffic generation and benchmark protocol, TrafPy is an ideal tool for developing such systems via the benchmark paradigm described throughout this manuscript.

The space of potential research areas from this work is vast. We hope presently unforeseeable avenues will be pursued with the support of TrafPy's standardised traffic generation and rigorous benchmarking framework. For our own work, based on the preliminary results of scheduler sensitivity to varying load conditions and traffic trace characteristics, we expect to develop new scheduling heuristics and learning algorithms which can dynamically adapt to network traffic states and outperform literature baselines in open-source TrafPy benchmarks. The 2.5 TB of open-access simulation data from this manuscript open some interesting offline reinforcement learning opportunities. We also anticipate adding more sensitivity-testing and realistic DCN traffic traces to the suite of TrafPy benchmarks. Furthermore, there are some exciting features which could enhance TrafPy. For example, although TrafPy can generate traces without any raw data given whatever characteristic distributions the user provides, it would be useful to be able to input real data (e.g. Ref. [7]) and have TrafPy automatically characterise the traffic in order to generate realistic data. Additionally, we plan to include a computation graph view of DCN network traffic in the TrafPy

API, unifying the flow-centric paradigm from the networking community with the job-centric perspective from computer science. This could lead to exciting novel research, such as network- and job-aware DCN scheduling.

## Author statement

We declare that all authors made notable contributions to this manuscript, and that this work is our own original work. This work was completed with the support of the following funders: EPSRC Distributed Quantum Computing and Applications EP/W032643/1; the Innovate UK Project on Quantum Data Centres and the Future 10004793; OptoCloud EP/T026081/1; TRANSNET EP/R035342/1; the Engineering and Physical Sciences Research Council EP/R041792/1 and EP/L015455/1; and the Alan Turing Institute.

## Declaration of competing interest

The authors declare the following financial interests/personal relationships which may be considered as potential competing interests: Christopher Parsonson reports financial support was provided by the Engineering and Physical Sciences Research Council, OptoCloud, TRANSNET, and the Innovate UK Project on Quantum Data Centres for the Future. Christopher Parsonson reports a relationship with The Alan Turing Institute that includes: funding grants.

## Appendix A. Table of Notation

**Table A.1**
Table summarising the symbol notation used throughout the paper.

| Symbol | Definition |
|---|---|
| $D'$ | Set of parameters defining the TrafPy distributions |
| $D$ | Traffic trace generated using the $D'$ TrafPy parameters |
| $\mathbb{P}$ | Probability distribution |
| $X$ | Discrete random variables |
| $H$ | Entropy |
| JSD | Jensen-Shannon divergence |
| $\sqrt{\text{JSD}}$ | Jensen-Shannon distance |
| $\{\pi_1, \ldots, \pi_n\}$ | Weightings for the JSD of $n$ distributions |
| $B^s$, $B^t$, $B^n$ | Flow size, inter-arrival time, and node pair random variables for benchmark workload $B$ |
| $\boldsymbol{b}^s$, $\boldsymbol{b}^t$, $\boldsymbol{b}^n$ | Flow sizes, inter-arrival times, and node pairs sampled from benchmark workload $B$ |
| $\boldsymbol{b}^a$ | Flow arrival times derived from inter-arrival times $\boldsymbol{b}$t |
| $T$ | DCN network topology |
| $\rho$ | Load fraction (fraction of overall network capacity requested) |
| $n_n$ | Number of server nodes |
| $n_c$ | Number of channels per communication link |
| $C_c$ | Capacity per server node link channel |
| $C_t$ | Total network capacity per direction |
| $n_f$ | Number of flows generated |
| $t_t$ | Total time duration of simulation |
| $\varrho$ | Load rate (information arriving per unit time) |
| $\alpha_t$ | Inter-arrival time adjustment factor |
| $d_p$ | Difference between a node pair's current and target information request magnitude |
| $\beta$ | Number of flows adjustment factor |
| $R$ | Number of traffic traces to generate and simulate for a suitable confidence interval |

## Appendix B. TrafPy Distribution Parameters

**Table B.2**
Benchmark categories with their real traffic characteristics reported in the literature (where appropriate) and the corresponding TrafPy parameters $D'$ needed to reproduce the distributions. DCN $_{<i,ii,iii,iv>}$ → <*university, private_enterprise, commercial_cloud, social_media_cloud* > Skewed $_{<i,ii,iii,iv,v>}$ → skewed_nodes_sensitivity_-<*uniform, 0.05, 0.1, 0.2, 0.4*> Rack $_{<i,ii,iii,iv,v>}$ → rack_sensitivity_<*uniform, 0.2, 0.4, 0.6, 0.8*>.

| Benchmark Category | Applications | Size, Bytes | Inter-arrival Time, $\mu$s | Inter- \| Intra-Rack Traffic, % | Hot Nodes \| Load Requested, % |
|---|---|---|---|---|---|
| **DCN$_i$** [30,57] | Database backups, hosting distributed file systems (email, servers, web services for faculty portals etc.), multi-cast video streams | [a] 80% < 10, 000 [b] 'lognormal', $\{\mu$: 7, $\sigma$: 2.5}, min_val = 1, max_val = 2$e$7, round = 25 | [a] 10% < 400, 80% < 10, 000 [b] 'weibull', $\{\alpha$: 0.9, $\lambda$: 6000}, min_val = 1, round = 25 | [a]70\|30 [b] $r = \{r_d$: $c$, $p$: 0.7 } | [a]20\| 55[b] 'multimodal', $n_s = d(0.2)$, $n_p = e$ (0.2, 0.55) |
| **DCN$_{ii}$** [30] | University + 'custom' applications and development test beds | [a] 80% < 10, 000 [b] 'lognormal', $\{\mu$: 7, $\sigma$: 2.5}, min_val = 1, max_val = 2$e$7, round = 25 | [a] 80% < 1, 000 [b] 'multimodal', min_val = 1, max_val = 100,000, locations = [40,1], skews = [-1,4], scales = [60,1000], num_skew_samples = [10]$e$3, round = 25, bg_factor = 0.05 | [a] 50 \| 50 [b] $r = \{r_d$: $c$, $p$: 0.5} | [a]20\| 55[b] 'multimodal', $n_s = d(0.2)$, $n_p = e$ (0.2, 0.55) |
| **DCN$_{iii}$** [30,31] | | [a] 80% < 10, 000 [b] 'lognormal', | [a] Median 10 [b] 'multimodal', | [a] 20 \| 80 [b] $r = \{r_d$: $c$, $p$: 0.2} | [a]20\| 55[b] 'multimodal', |

(continued on next page)

**Table B.2** (*continued*)

| Benchmark Category | Applications | Size, Bytes | Inter-arrival Time, $\mu s$ | Inter- \| Intra-Rack Traffic, % | Hot Nodes \| Load Requested, % |
|---|---|---|---|---|---|
| | Internet-facing applications (search indexing, webmail, video, etc.), data mining and MapReduce-style applications | {$\mu$: 7, $\sigma$: 2.5}, min_val = 1, max_val = 2$e$7, round = 25 | min_val = 1, max_val = 100,000, locations = [10,20,100,1], skews = [0,0,0,100], scales = [1,3,4,50], num_skew_samples = [10,7,5,20]$e$3,round = 25, bg_factor = 0.01 | | $n_s = d(0.2)$, $n_p = e(0.2, 0.55)$ |
| **DCN$_{iv}$** [33] | Web request response generation (mail, messenger, etc.), MySQL database storage & cache querying, newsfeed assembly | [a] 10% < 300, 90% < 100, 000 [b] 'weibull', {$\alpha$: 0.5, $\lambda$: 21,000}, min_val = 1, max_val = 2$e$6, round = 25 | [a] 10% < 20, 90% < 10, 000 [b] 'lognormal', {$\mu$: 6, $\sigma$: 2.3}, min_val = 1, round = 25 | [a] 12.9 \| 87.1 [b] $r = \{r_d: c, p: 0.129\}$ | [a]20\| 55[b] 'multimodal', $n_s = d(0.2)$, $n_p = e(0.2, 0.55)$ |
| **Skewed$_i$, Rack$_i$** | – | [b]DCN$_{iii}$ | [b]DCN$_{iii}$ | [b] 'uniform', $r$ = None | [b] 'uniform' $n_s = n_p$ = None |
| **Skewed$_{ii}$** | – | [b]DCN$_{iii}$ | [b]DCN$_{iii}$ | [b] 'uniform', $r$ = None | 5\| 55[b] 'uniform' $n_s = d(0.05)$ $n_p = e(0.05, 0.55)$ |
| **Skewed$_{iii}$** | – | [b]DCN$_{iii}$ | [b]DCN$_{iii}$ | [b] 'uniform', $r$ = None | 5\| 55[b] 'uniform' $n_s = d(0.1)$ $n_p = e(0.1, 0.55)$ |
| **Skewed$_{iv}$** | – | [b]DCN$_{iii}$ | [b]DCN$_{iii}$ | [b] 'uniform', $r$ = None | 5\| 55[b] 'uniform' $n_s = d(0.2)$ $n_p = e(0.2, 0.55)$ |
| **Skewed$_v$** | – | [b]DCN$_{iii}$ | [b]DCN$_{iii}$ | [b] 'uniform', $r$ = None | 5\| 55[b] 'uniform' $n_s = d(0.4)$ $n_p = e(0.4, 0.55)$ |
| **Rack$_{ii}$** | – | [b]DCN$_{iii}$ | [b]DCN$_{iii}$ | 80\| 20[b] 'uniform', $r = \{r_d: c, p: 0.8\}$ | [b] 'uniform' $n_s = n_p$ = None |
| **Rack$_{iii}$** | – | [b]DCN$_{iii}$ | [b]DCN$_{iii}$ | 60\| 40[b] 'uniform', $r = \{r_d: c, p: 0.6\}$ | [b] 'uniform' $n_s = n_p$ = None |
| **Rack$_{iv}$** | – | [b]DCN$_{iii}$ | [b]DCN$_{iii}$ | 40\| 60[b] 'uniform', $r = \{r_d: c, p: 0.4\}$ | [b] 'uniform' $n_s = n_p$ = None |
| **Rack$_v$** | – | [b]DCN$_{iii}$ | [b]DCN$_{iii}$ | 20\| 80[b] 'uniform', $r = \{r_d: c, p: 0.2\}$ | [b] 'uniform' $n_s = n_p$ = None |

[a] Real traffic characteristics reported in the literature.

[b] Corresponding TrafPy parameters $D'$. $c$ = net.graph['rack_to_ep_dict'] → Network cluster (i.e. rack) configuration. $d(u)$ = int($u$ * len(net.graph['endpoints'])) → Number of nodes to skew. $e(u, v) = [v/d(u)$ for _in range($d(u)$)] → Fraction of overall traffic load to distribute amongst the skewed nodes. $r|r_d|p \mid n_s \mid n_p$ = rack_prob_config | 'racks_dict' | 'prob_inter_rack' | num_skewed_nodes | skewed_node_probs.

**Table B.3**
Flow size, inter-arrival time, and node load distribution characteristics for the University (U), Private Enterprise (PE), Commercial Cloud (CC), and Social Media Cloud (SMC) data sets of the DCN benchmark after generating the distributions from TrafPy parameters $D'$.

| Variable | DCN | # Modes | Min. | Max. | Mean | Variance | Skewness | Kurtosis |
|---|---|---|---|---|---|---|---|---|
| Size (B) | U | 1 | 1 | $19.80 \times 10^6$ | $22.90 \times 10^3$ | $42 \times 10^9$ | 39.4 | $2.41 \times 10^3$ |
| | PE | 1 | 1 | $19 \times 10^6$ | $23.30 \times 10^3$ | $53.50 \times 10^9$ | 44.1 | $2.79 \times 10^3$ |
| | CC | 1 | 1 | $19.20 \times 10^6$ | $22.30 \times 10^3$ | $38.40 \times 10^9$ | 36.9 | $2.08 \times 10^3$ |
| | SMC | 1 | 1 | $3.17 \times 10^6$ | $42 \times 10^3$ | $8.87 \times 10^9$ | 6.20 | 66.4 |
| Inter-arrival time ($\mu s$) | U | 1 | 1 | $126 \times 10^3$ | $6.30 \times 10^3$ | $49.90 \times 10^6$ | 2.44 | 9.92 |
| | PE | 2 | 1 | $100 \times 10^3$ | $2.83 \times 10^3$ | $154 \times 10^6$ | 5.7 | 33.1 |
| | CC | 4 | 1 | $10 \times 10^3$ | 84.5 | $0.32 \times 10^6$ | 13 | 179 |
| | SMC | 1 | 1 | $54.60 \times 10^5$ | $5.51 \times 10^3$ | $2.11 \times 10^9$ | 47.8 | $3.75 \times 10^3$ |

| Variable | DCN | % Hot Nodes | % Hot Node Traffic | % Inter-Rack |
|---|---|---|---|---|
| Node load distribution (%) | U | 20 | 55 | 70 |
| | PE | 20 | 55 | 50 |
| | CC | 20 | 55 | 20 |
| | SMC | 20 | 55 | 12.9 |

## Appendix C. TrafPy API Examples

### *Appendix C.1. Custom Distribution Shaping*

#### *Appendix C.1.1. Interactively & Visually Shaping a Custom 'Named' Distribution in a Jupyter Notebook*

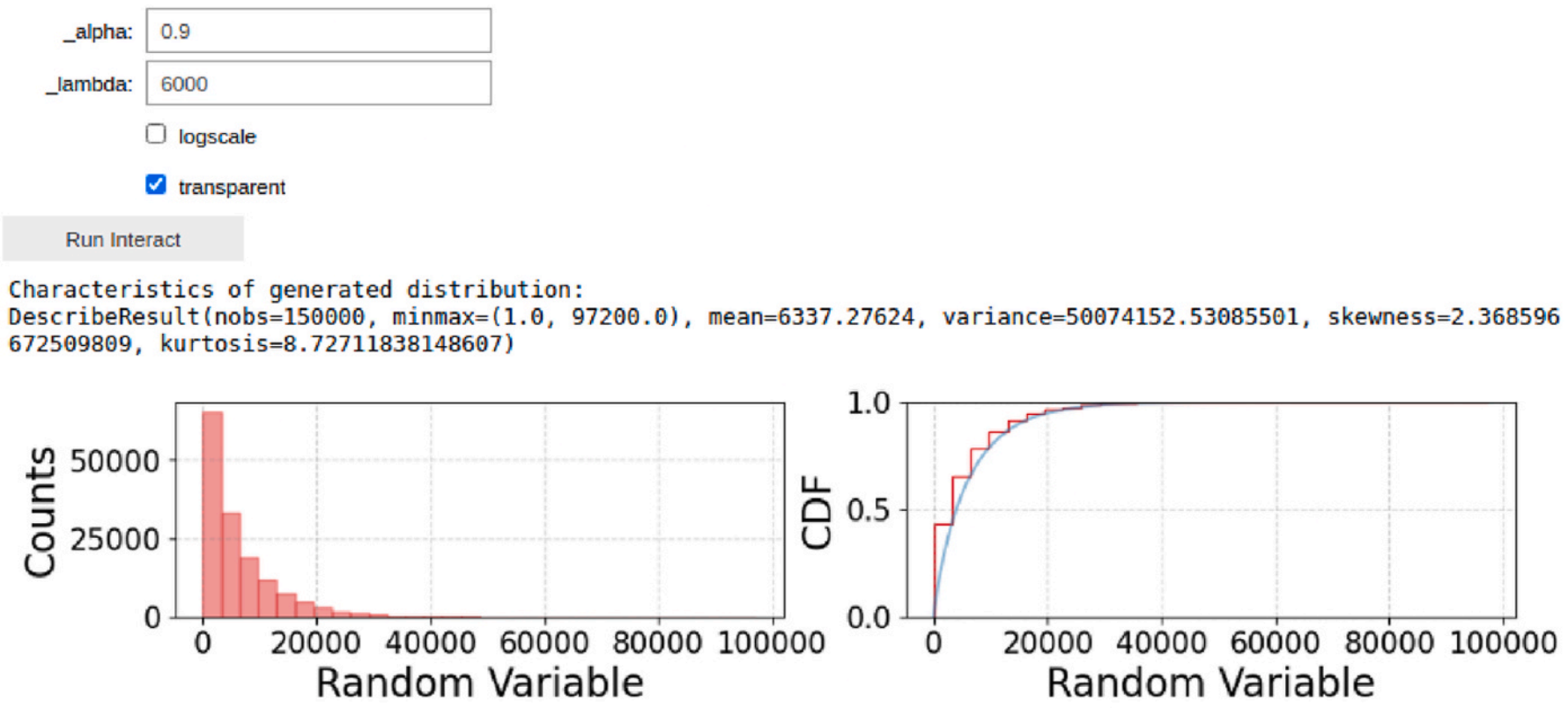

**Fig. C.8.** Output of example code for interactively and visually shaping a 'named' distribution in a Jupyter Notebook.

Example of interactively and visually shaping a weibull distribution's parameters to achieve a target distribution for some random variable in Jupyter Notebook (output in Fig. C.8):

```
import trafpy.generator as tpg

dist = tpg.gen_named_val_dist(dist='weibull',
                              interactive_plot=True,
                              min_val=1,
                              max_val=None,
                              size=15e4)
```

This same distribution can then be reproduced by using the same parameters:

```
dist = tpg.gen_named_val_dist(dist='weibull',
                              params={'_alpha': 0.9, '_lambda': 6000}
                              min_val=1,
                              max_val=None)
```

#### *Appendix C.1.2. Interactively & Visually Shaping a Custom 'Multimodal' Distribution in a Jupyter Notebook*

To generate a multimodal distribution, first shape each mode individually (output in Fig. C.9):

```
import trafpy.generator as tpg

data_dict = tpg.gen_skew_dists(min_val=1,
                               max_val=1e5,
                               num_modes=2)
```

Then combine the distributions, filling the distribution with a tuneable amount of 'background noise' (output in Fig. C.10):

```
multimodal_dist = tpg.combine_multiple_mode_dists(data_dict,
                                                  min_val=1,
                                                  max_val=1e5)
```

This same distribution can be reproduced using the same parameters:

```
multimodal_dist = tpg.gen_multimodal_val_dist(min_val=1,
                                              max_val=1e5,
                                              locations=[40, 1],
                                              skews=[-1, 4],
                                              scales=[60, 1000],
                                              num_skew_samples=[1e4, 1e4],
                                              bg_factor=0.05)
```

N.B. An equivalent function can be used for generating custom skew distributions with a single mode which also do not fall under one of the canonical ‘named’ distributions.

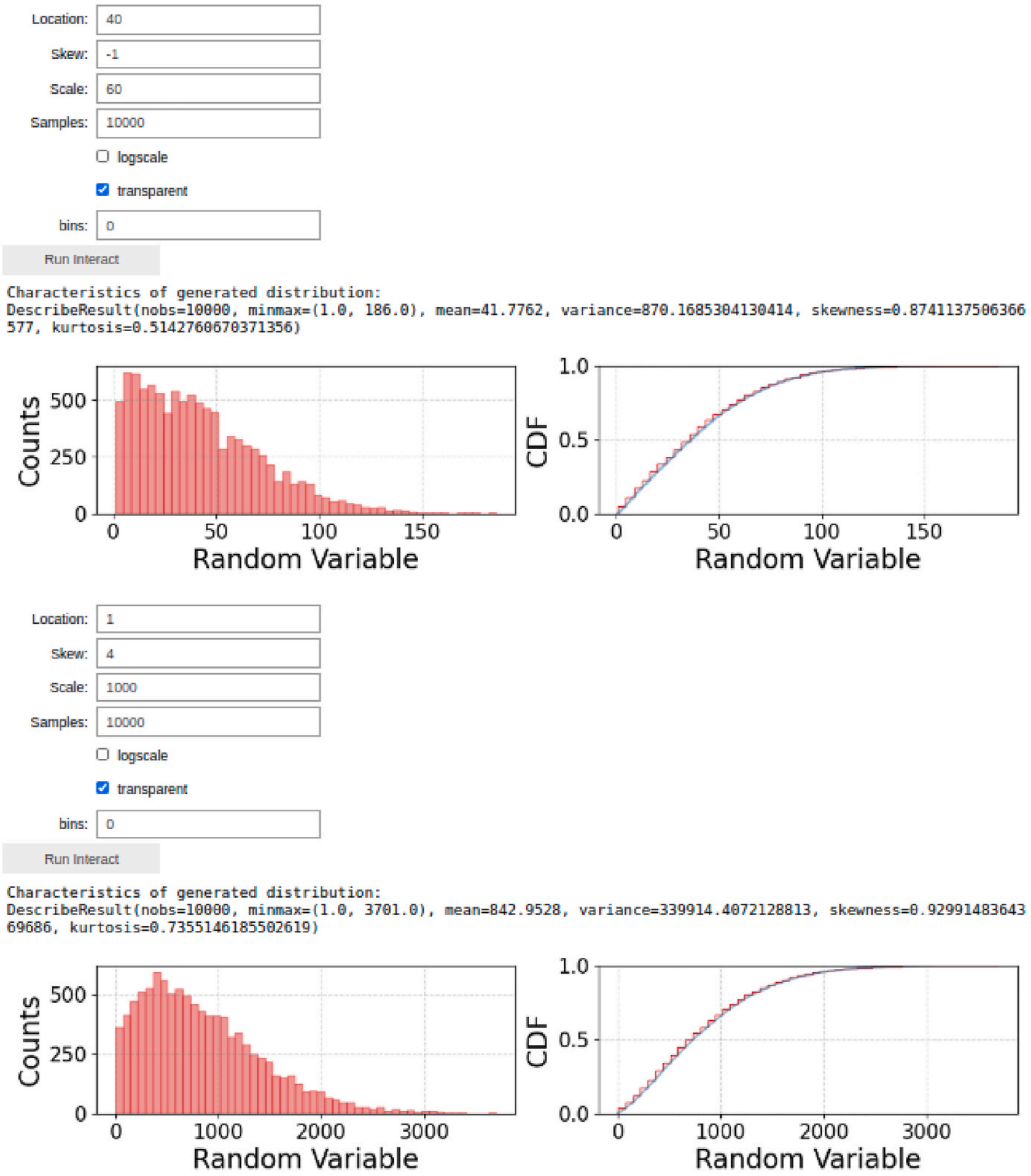

**Fig. C.9.** Output for step 1 of example code for interactively and visually shaping a ‘multimodal’ distribution in a Jupyter Notebook, where you must first shape each mode individually.

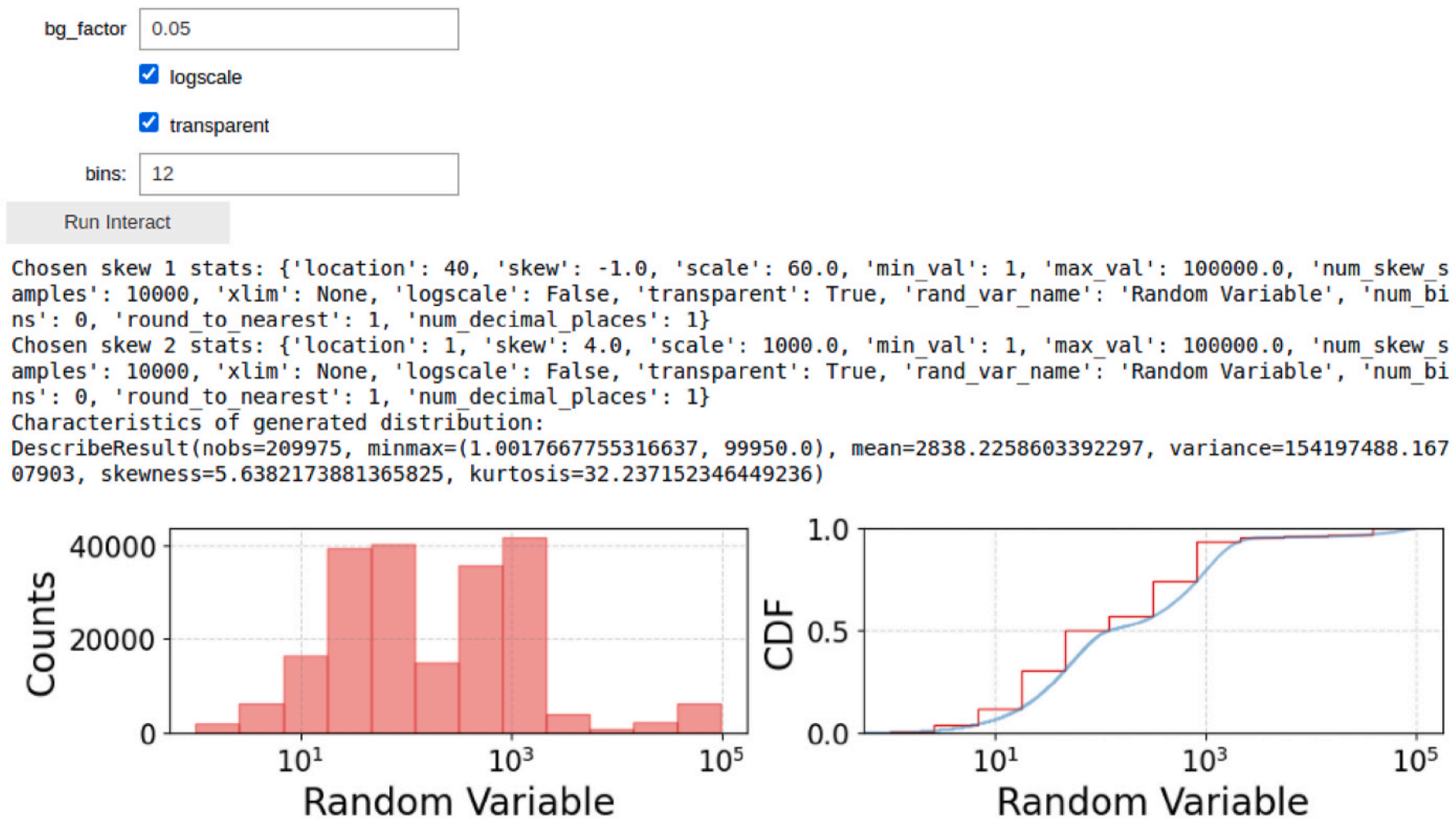

**Fig. C.10.** Output for step 2 of example code for interactively and visually shaping a 'multimodal' distribution in a Jupyter Notebook, where you must combine your individually shaped modes into a single distribution.

### *Appendix C.2. Benchmark Importing & Flow Generation*

Example code for generating and visualising a load 0.1 University benchmark data set of flows for a custom topology (output in Fig. C.11):

```
import trafpy.generator as tpg
from trafpy.benchmarker import BenchmarkImporter
from trafpy.generator import Demand, DemandsAnalyser, DemandPlotter

# set variables
min_duration = 1000
jsd_threshold = 0.1

# initialise network
net = tpg.gen_arbitrary_network(num_eps=64, ep_channel_capacity=1250)

# initialise benchmark distributions
importer = BenchmarkImporter(benchmark_version='0.01')
dists = importer.get_benchmark_dists(benchmark='university', eps=net.graph['endpoints'])

# generate flow-centric demand data set
network_load_config = {'network_rate_capacity': net.graph['max_nw_capacity'],
                       'ep_channel_capacity': net.graph['ep_channel_capacity'],
                       'target_load_fraction': 0.1}
flow_centric_demand_data = tpg.create_demand_data(eps=net.graph['endpoints'],
                                                  node_dist=dists['node_dist'],
                                                  flow_size_dist=dists['flow_size_dist'],
                                                  interarrival_time_dist=dists['interarrival_time_dist'],
                                                  network_load_config=network_load_config,
                                                  jsd_threshold=jsd_threshold,
                                                  min_duration=min_duration)

# print summary table
demand = Demand(flow_centric_demand_data, net.graph['endpoints'])
DemandsAnalyser([demand], net).compute_metrics(print_summary=True)

# visualise distributions
plotter = DemandPlotter(demand)
plotter.plot_flow_size_dist()
plotter.plot_interarrival_time_dist()
plotter.plot_node_dist()
```

| | Name | Flows | 1st | Last | Duration | Info | Load | Smallest | Largest |
|---|---|---|---|---|---|---|---|---|---|
| 0 | demand | 135916 | 0.0 | 788907.126500 | 788907.126500 | 3.155629e+09 | 4000.0 | 1.0 | 17236750.0 |

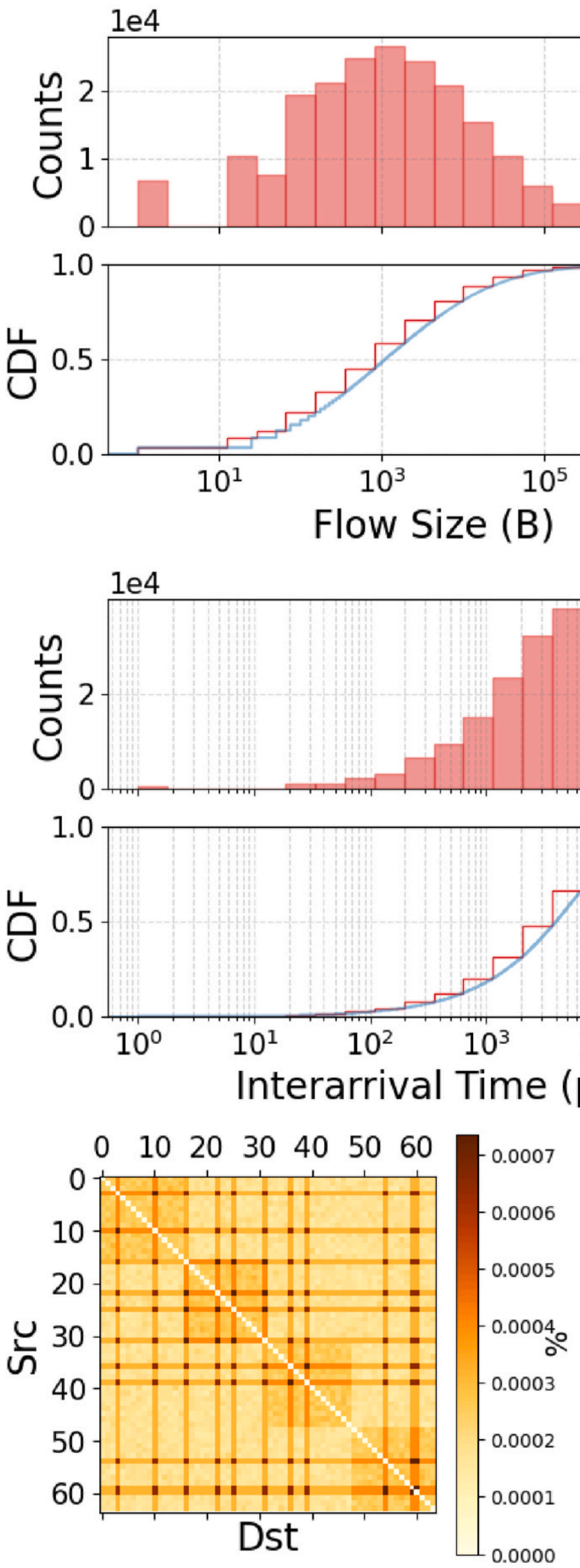

**Fig. C.11.** Output of example code for generating a benchmark.

## Appendix D. Pseudocode

### *Appendix D.1. Scheduling*

The flow scheduling pseudocode is shown in Algorithm 2. First, information about the queued flows such as their characteristics (packets left, time of arrival, flow queue, destination node, etc.), the network links requested in the source-destination path, and the bandwidth requested, is collected. If the scheduler uses cost-based scheduling (e.g. SRPT uses flow completion time cost), a cost is also assigned to each flow. Next, for each link being requested by the flows, while the link in question has some available bandwidth left to allocate for the current time slot, the scheduler chooses flows until either there is no bandwidth left or there are no flows demanding the link which have not been chosen. Finally, for each flow in the set of these provisionally chosen flows, the smallest number of packets scheduled for the flow in question across all links is chosen as the flow's number of packets to schedule. Note that this simulation methodology considers bandwidth bottlenecks throughout all layers of the network. The pseudocode in Algorithm 3 is used to resolve any contentions and attempt to set up the flow, thus adding the flow to the ultimate set of flows chosen by the scheduler for the given time slot. The parts which are **scheduler**-specific have been marked in bold.

**Algorithm 2**
Flow scheduling process.

```
Collect flow information
link_allocations = []
for link in links do
  while link bandwidth ≠ 0 do
    link_allocations.append(scheduler choose flow)
  end while
end for
chosen_flows = []
for flow in flows do
  if flow in link_allocations then
    flow_packets = min(packets allocated for flow in link_allocations)
    establish, removed_flows = scheduler resolve_contentions(flow, chosen_flows)
    if establish then
      chosen_flows.append(flow)
      chosen_flows.remove(removed_flows)
    end if
  end if
end for
```

**Algorithm 3**
Flow contention resolution process.

```
Require: flow, chosen_flows
    removed_flows = []
  while True do
    if no_contention(flow) then
      establish = True
      return establish, removed_flows
    else
      contending_flow = find_contending_flow()
      establish = scheduler resolve_contention(flow, contending_flow)
      if not establish then
        chosen_flows.append(removed_flows)
        return establish
      else
        chosen_flows.remove(contending_flow)
        continue
      end if
    end if
end while
```

*Appendix D.2. TrafPy Benchmark Protocol*

**Algorithm 4**
TrafPy benchmark protocol.

for $r$ in range($R$) do
  **for** $d$ in $D$ **do**
    **for** $\rho \leftarrow 0.1$ to 0.9 step 0.1 **do**
      $P_{\text{KPI}} = \Upsilon(\chi, d, \rho)$
    **end for**
  **end for**
**end for**

## Appendix E. Traffic Skew Convergence

A constraint of any traffic matrix is that the load on each end point (the fraction of the end point's capacity being requested) cannot exceed 1.0. Consequently, certain traffic skews become infeasible at higher loads (for example, it is impossible for an $n > 1$ network to have 1 node requesting 100% of the traffic if the overall network is under a 1.0 load). As shown in Fig. 3, this results in all traffic matrices tending towards uniform (i.e. having no skew) as the overall network load tends to 1.0.

A question traffic trace generators may ask is: for a given load, what combination of i) number of skewed nodes, ii) corresponding fraction of the arriving network traffic the skewed nodes request, and iii) overall network load results in the traffic matrix being skewed or not skewed? To answer this question, we make the following assumptions:

- All network end points have equal bandwidth capacities.
- All end points are either 'skewed' or 'not skewed' by the same amount.

- 'Skew' is defined by a skew *factor*, which is the fractional difference between the load rate per skewed node and the load rate per non-skewed node (the highest being the numerator, and the lowest being the denominator).
- For a given combination of skewed nodes and the load rate they request of some overall network load, any excess load (exceeding 1.0) on a given end point is distributed equally amongst all other end points whose loads are $< 1.0$.

With the above assumptions, we can calculate the skew factor for each combination of skewed nodes, corresponding traffic requested, and overall network load. Doing this for 0–100% of the network nodes being skewed and requesting 0–100% of the overall network load under network loads 0.1–0.9, we can construct a look-up table of skew factors for each of these combinations before generating any actual traffic. Fig. E.12 shows a high resolution (0.1%) heat map of these combinations, with any skew factors $\geq 2.0$ set to the same colour for visual clarity. Fig. E.13 shows the corresponding plots with lower resolution (5%) but with the skew factors labelled. As expected, above 0.6 network loads, certain combinations of number of skewed nodes and traffic requested become restricted as to how much skew there can be in the matrix, with many combinations tending towards uniform (skew factor 1.0) at 0.9 loads.

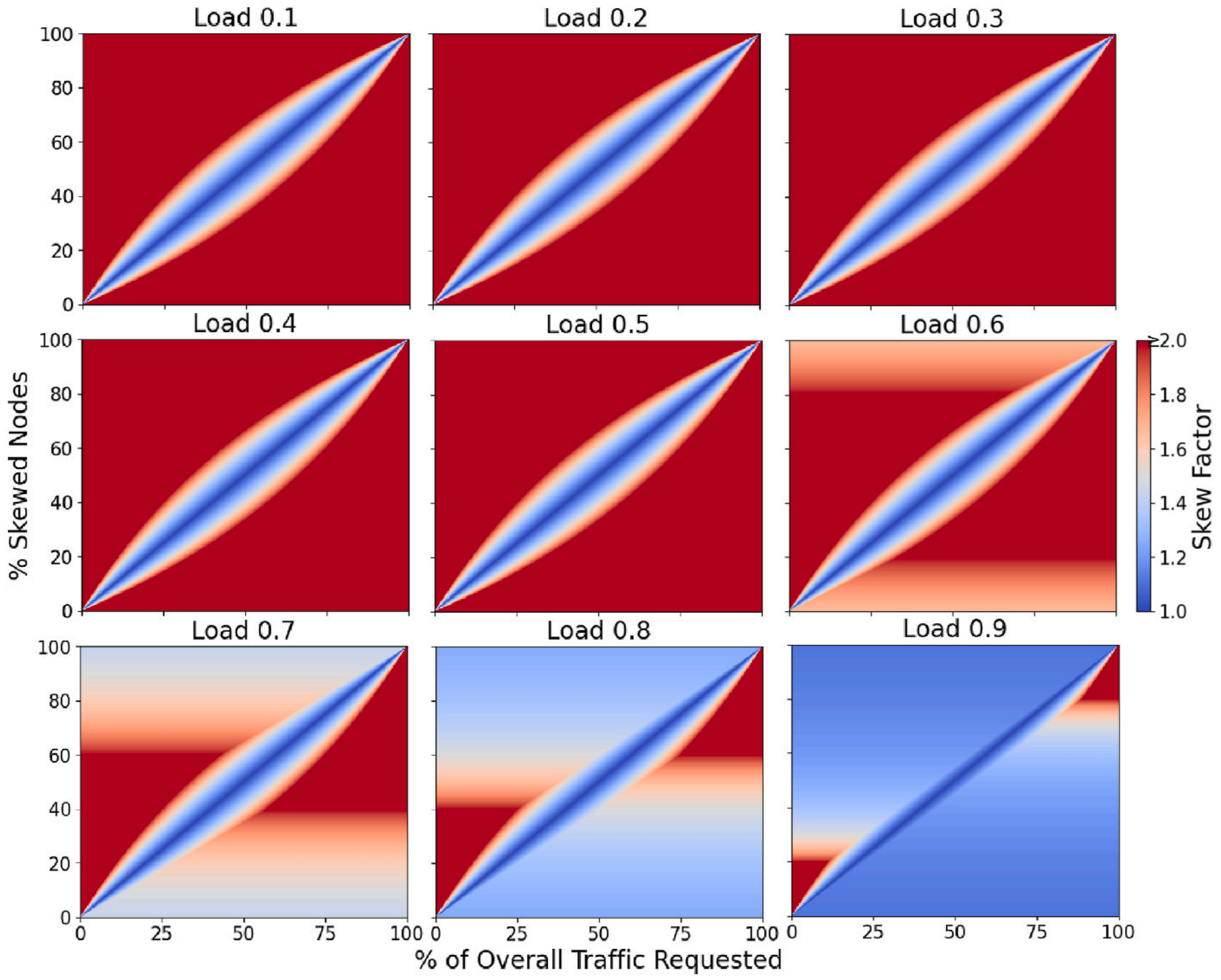

**Fig. E.12.** Skew factor heat maps for 0–100% of network nodes requesting 0–100% of the overall network traffic across loads 0.1–0.9 plotted at 0.1% resolution. For clarity, combinations with skew factors $\geq 2$ have been assigned the same colour.

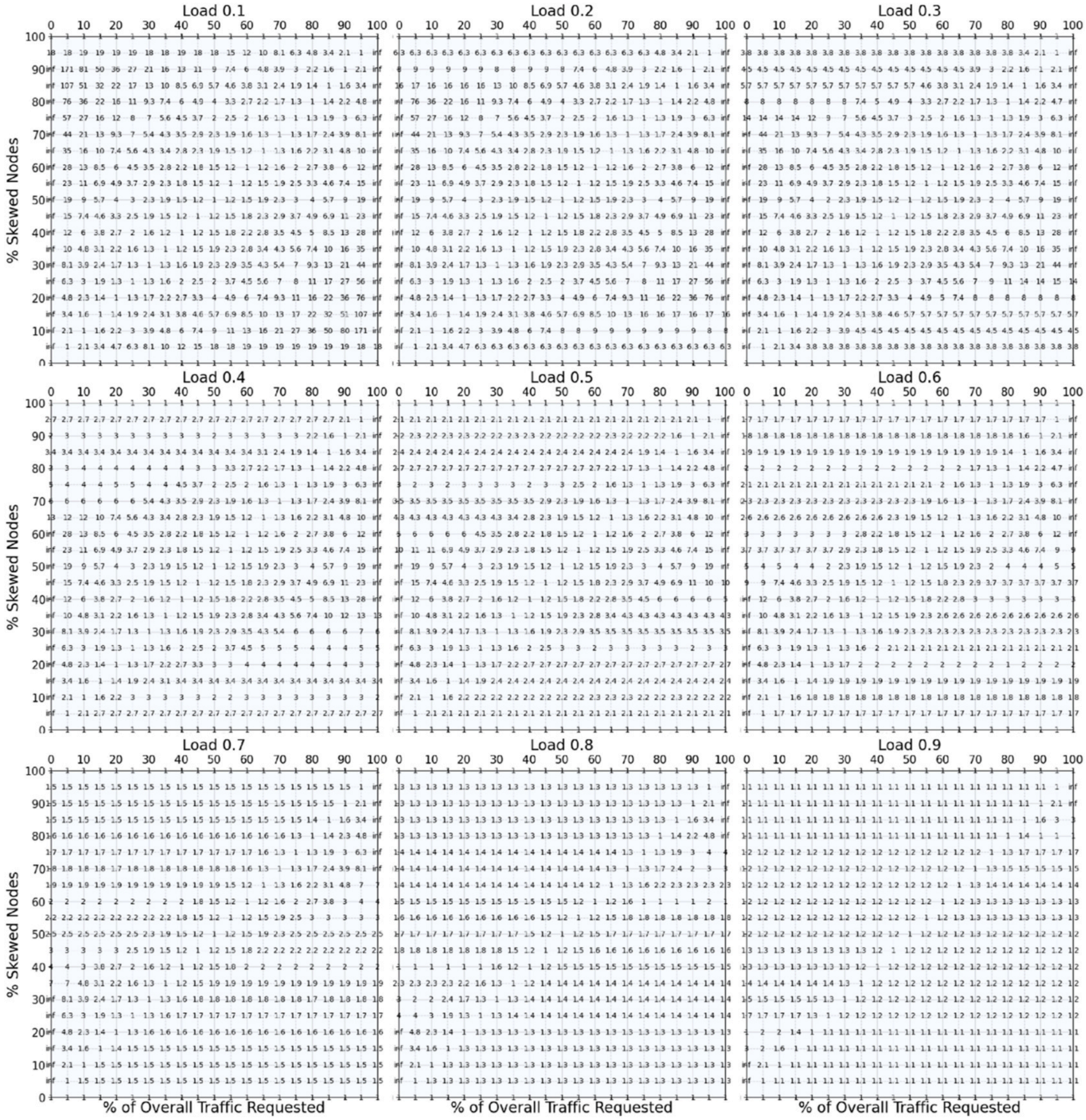

**Fig. E.13.** Labelled skew factor tables for 0–100% of network nodes requesting 0–100% of the overall network traffic across loads 0.1–0.9 plotted at 5% resolution.

Using the skew factor data from Figs. E.12 and E.13, we can be confident at 5%, 10%, 20%, and 40% of the network nodes requesting 55% of the overall network traffic that the skew factor will be $> 1.0$ across loads 0.1–0.9. Fig. E.14 shows the skew factor as a function of load for these combinations. Therefore, these were the combinations chosen for the skewed nodes sensitivity benchmark defined in Section 5 of this manuscript.

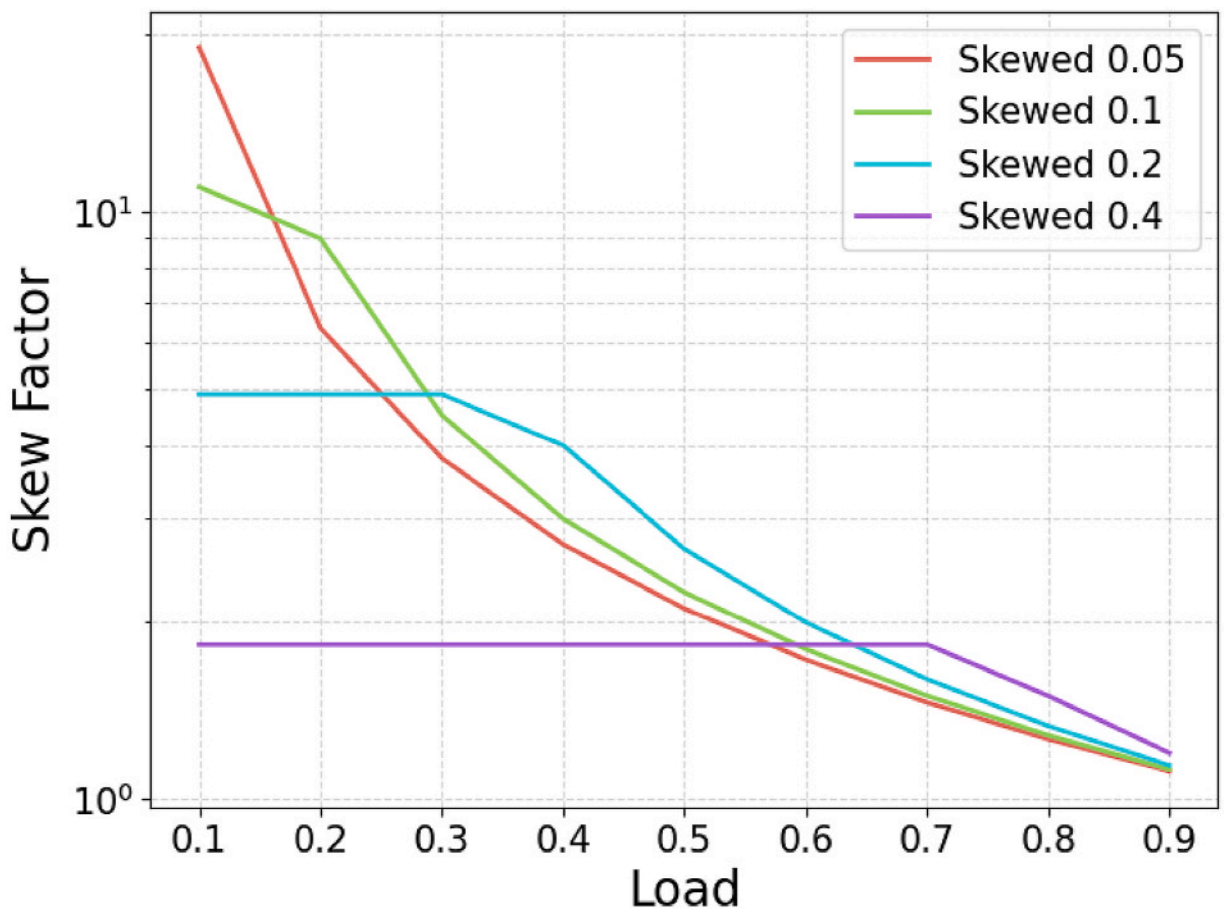

**Fig. E.14.** Skew factor as a function of load for 5%, 10%, 20%, and 40% of the network nodes requesting 55% of the overall network traffic.

## Appendix F. Scheduler Performance Summary

### *Appendix F.1. Completion Time Performance Plots*

Plots showing the schedulers' completion performances are provided for the realistic DCN (Fig. F.15) uniform (Fig. F.16), extreme rack (Fig. F.17), and extreme nodes (Fig. F.18) traffic traces.

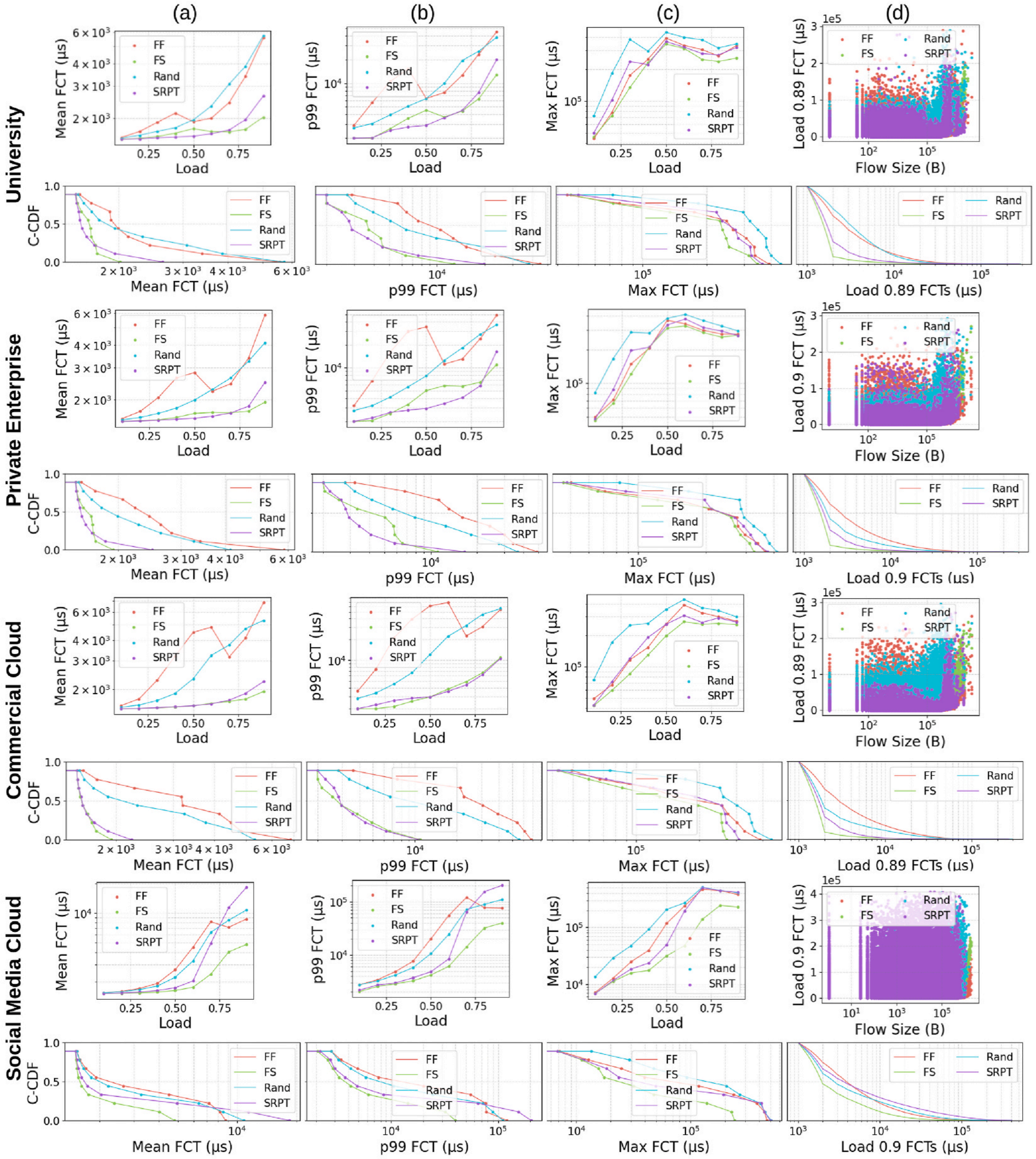

**Fig. F.15.** The schedulers' (a) mean, (b) 99th percentile, and (c) maximum flow completion time metrics for the **DCN benchmark distributions** across loads 0.1–0.9, and (d) a scatter plot of flow completion time as a function of flow size for the same distribution at load 0.9.

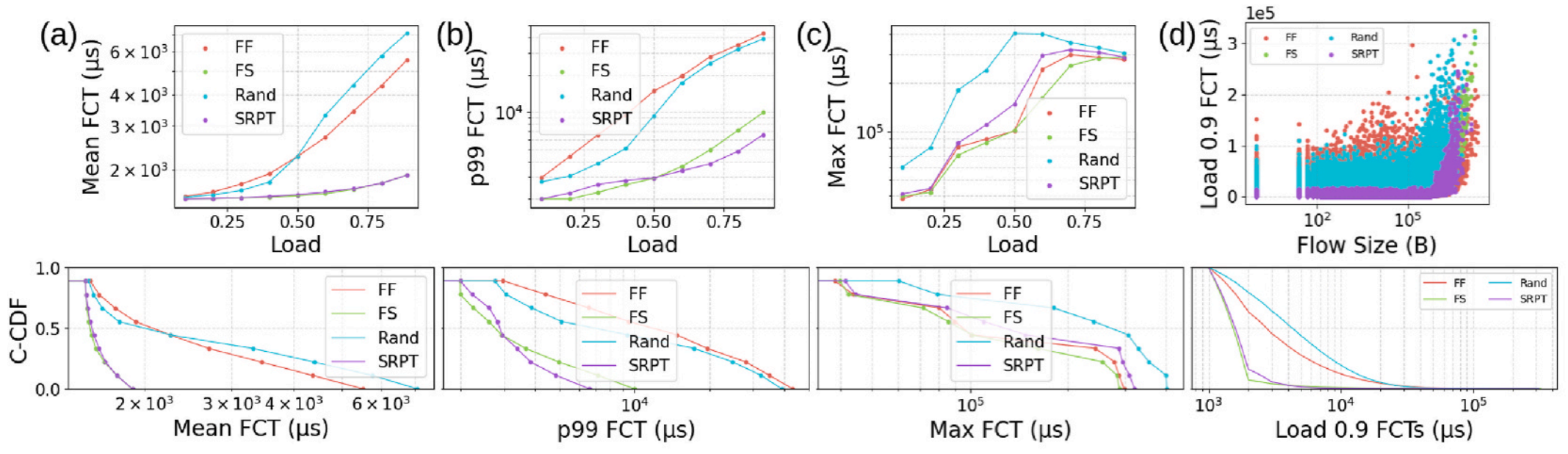

**Fig. F.16.** The schedulers' (a) mean, (b) 99th percentile, and (c) maximum flow completion time metrics for the **uniform node distribution** across loads 0.1–0.9, and (d) a scatter plot of flow completion time as a function of flow size for the same distribution at load 0.9.

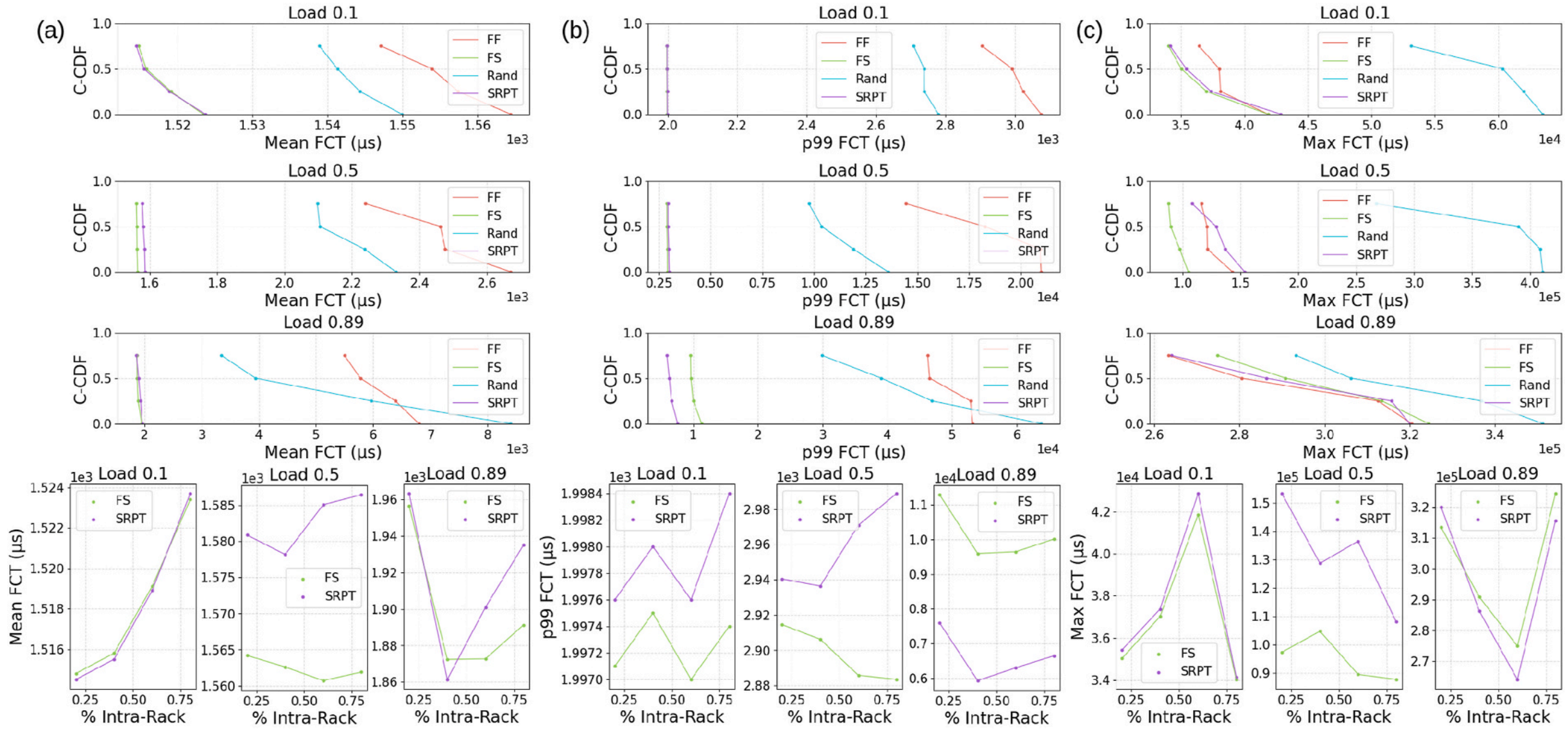

**Fig. F.17.** Sensitivity of the schedulers' (a) mean, (b) 99th percentile, and (c) maximum flow completion times to the changing **intra-rack distribution** for loads 0.1, 0.5, and 0.9. The complementary CDF plots include data for all 4 schedulers, whereas the scatter plots contain the top 2 performing schedulers (SRPT and FS) for clarity.

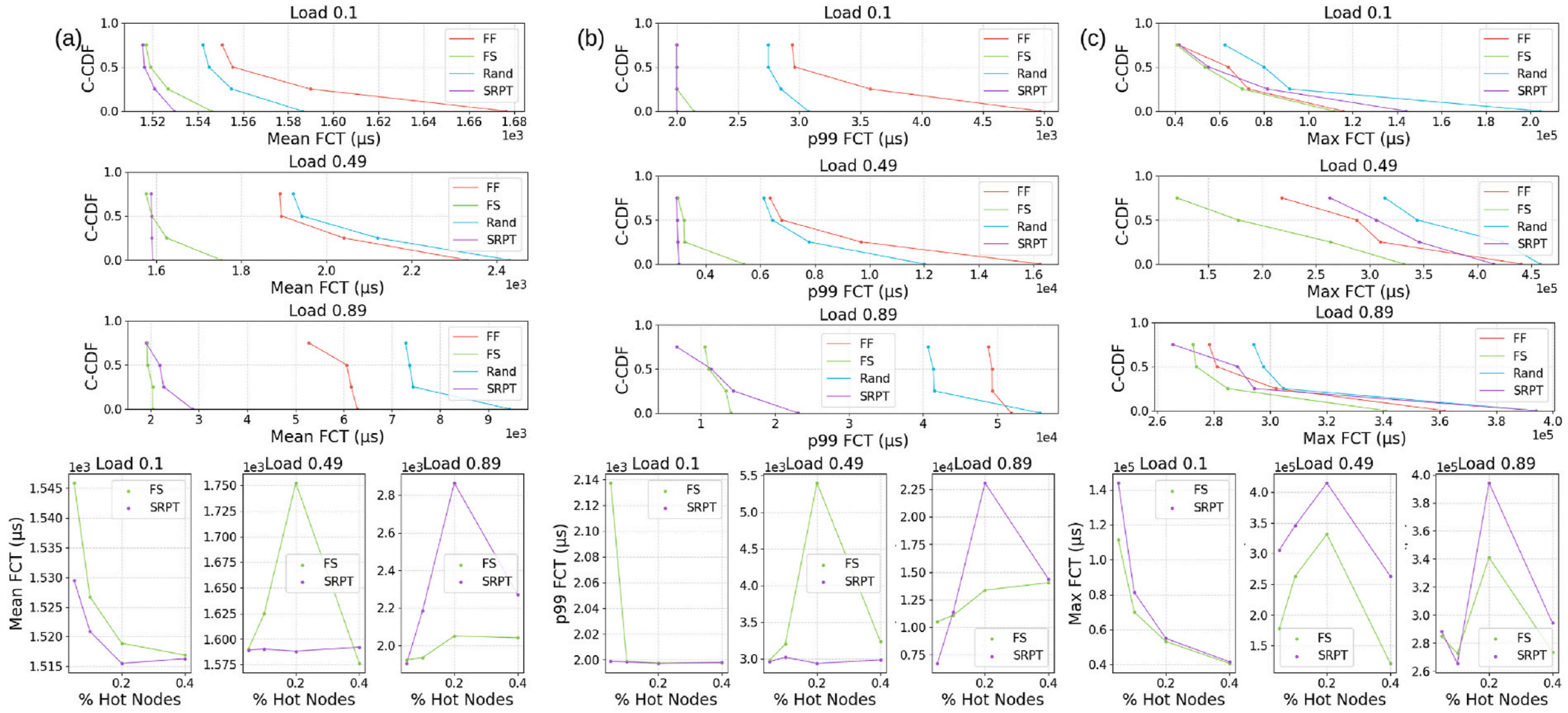

**Fig. F.18.** Sensitivity of the schedulers' (a) mean, (b) 99th percentile, and (c) maximum flow completion times to the changing **skewed nodes distribution** for loads 0.1, 0.5, and 0.9. The complementary CDF plots include data for all 4 schedulers, whereas the scatter plots contain the top 2 performing schedulers (SRPT and FS) for clarity.

### *Appendix F.2. Throughput and Flows Accepted Performance Plots*

Plots showing the schedulers' throughput and accepted flow performances are provided for the realistic DCN (Fig. F.19), uniform (Fig. F.20), extreme rack (Fig. F.21), and extreme nodes (Fig. F.22) traffic traces.

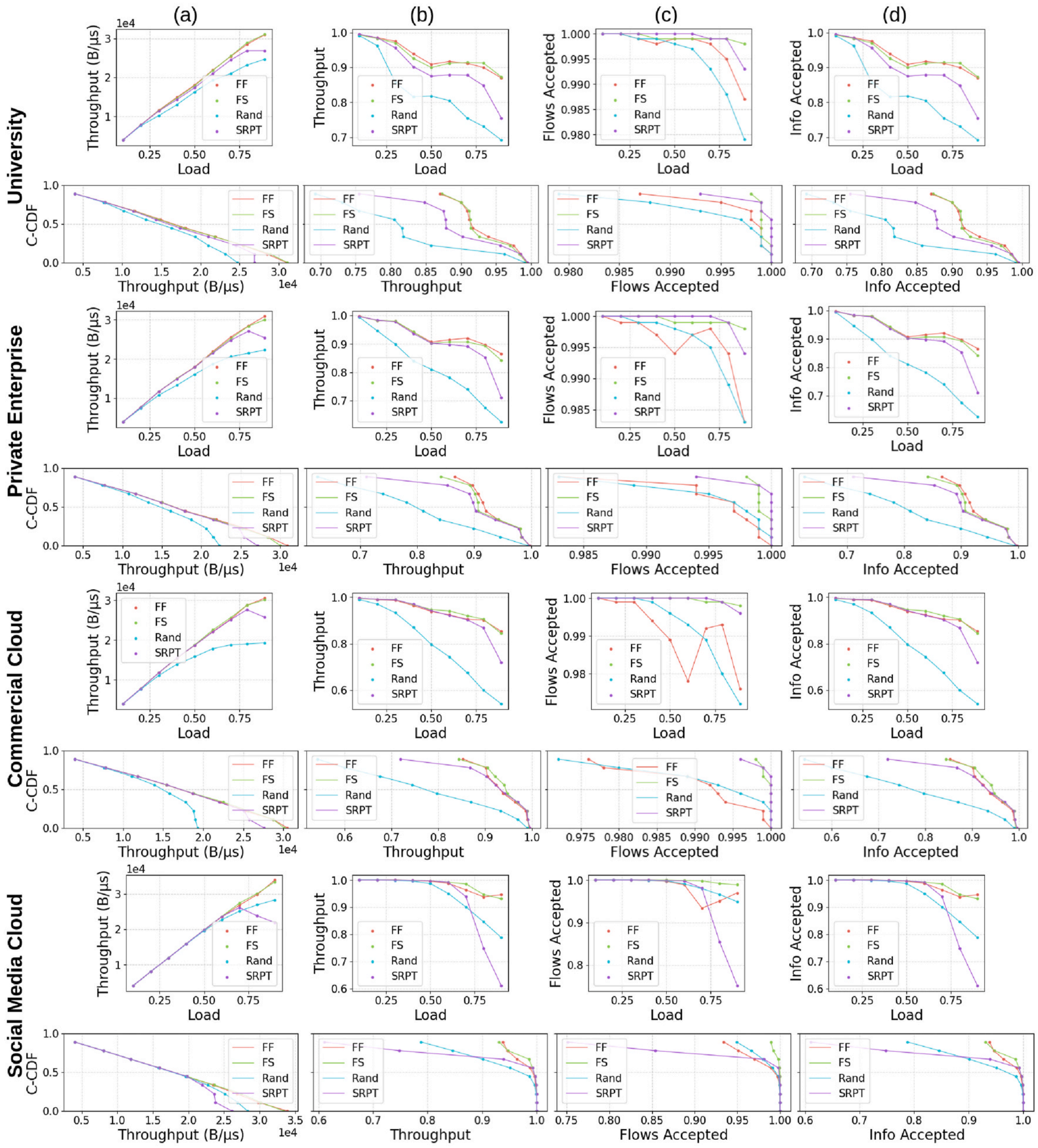

**Fig. F.19.** The schedulers' (a) absolute throughput (information units transported per unit time), (b) relative throughput (fraction of arrived information successfully transported), (c) fraction of arrived flows accepted, and (d) fraction of arrived information accepted metrics for the **DCN benchmark distributions** across loads 0.1–0.9.

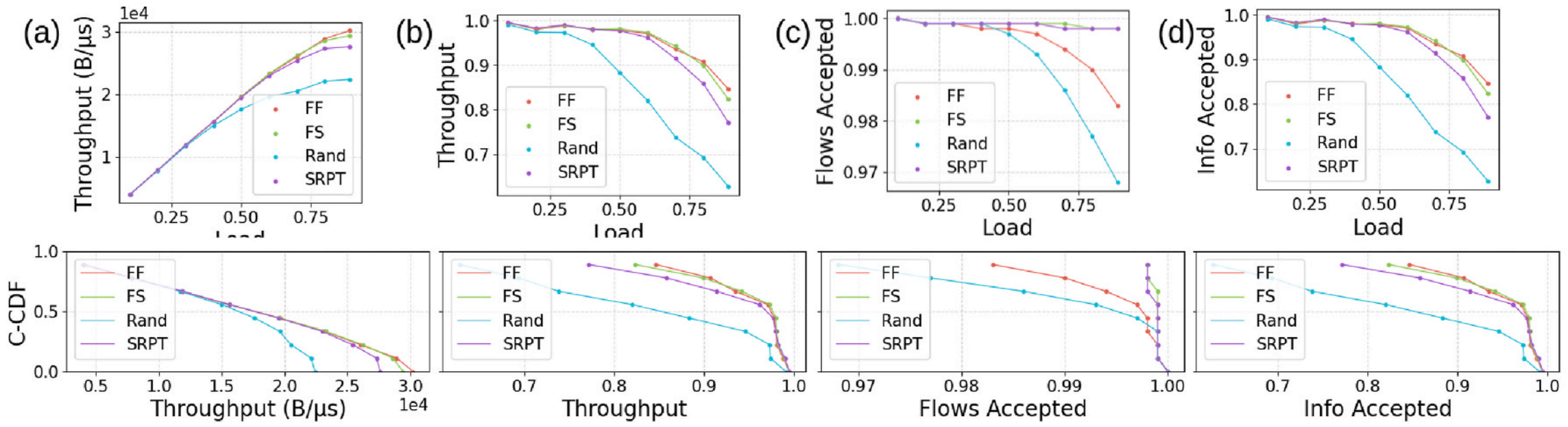

**Fig. F.20.** The schedulers' (a) absolute throughput (information units transported per unit time), (b) relative throughput (fraction of arrived information successfully transported), (c) fraction of arrived flows accepted, and (d) fraction of arrived information accepted metrics for the **uniform node distribution** across loads 0.1–0.9.

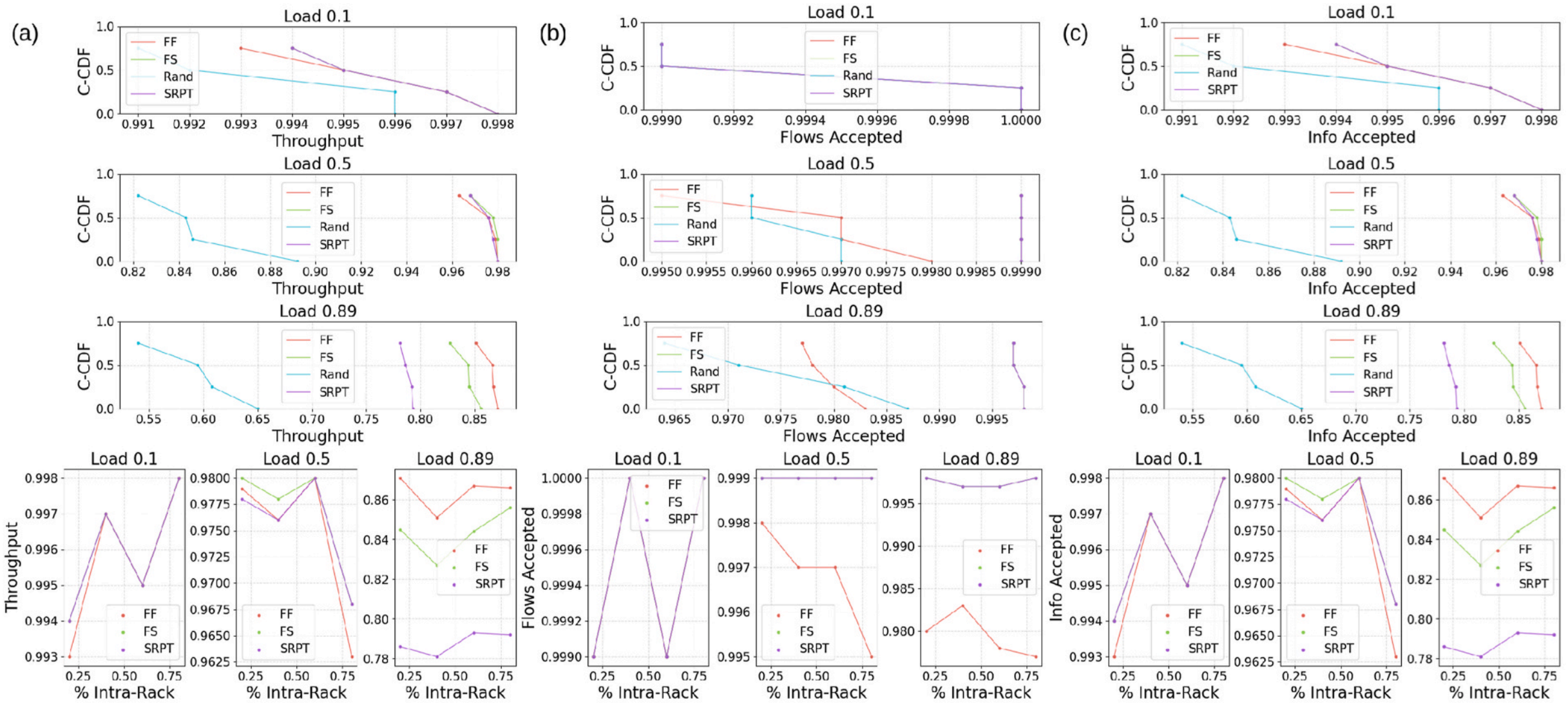

**Fig. F.21.** Sensitivity of the schedulers' (a) relative throughput, (b) fraction of arrived flows accepted, and (c) fraction of arrived information accepted metrics to the changing **intra-rack distribution** for loads 0.1, 0.5, and 0.9. The complementary CDF plots include data for all 4 schedulers, whereas the scatter plots contain the top 3 performing schedulers (SRPT, FS, and FF) for clarity.

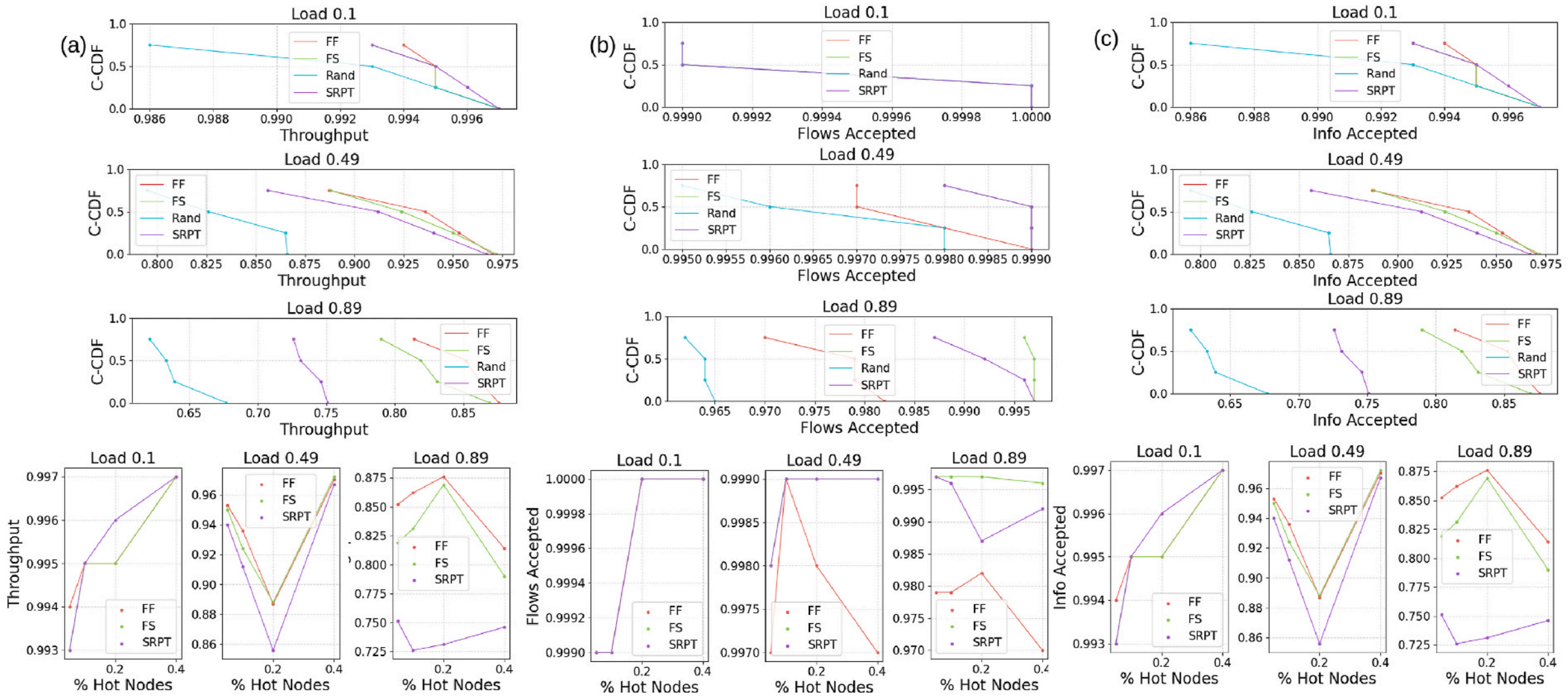

**Fig. F.22.** Sensitivity of the schedulers' (a) relative throughput, (b) fraction of arrived flows accepted, and (c) fraction of arrived information accepted metrics to the changing **skewed nodes distribution** for loads 0.1, 0.5, and 0.9. The complementary CDF plots include data for all 4 schedulers, whereas the scatter plots contain the top 3 performing schedulers (SRPT, FS, and FF) for clarity.

### *Appendix F.3. Performance Metric Tables*

The below performance tables summarise the schedulers' mean performances (averaged across 5 runs, 95% confidence intervals reported) for each $P_{KPI}$, each load, and each benchmark.

#### *Appendix F.3.1. DCN Benchmarks*

**Table F.4**
Scheduler performance summary with 95% confidence intervals for the **University** benchmark.

| Load | Subject | Mean FCT ($\mu s$) | p99 FCT ($\mu s$) | Max FCT ($\mu s$) | Throughput (Frac) | Flows Accepted (Frac) | Info Accepted (Frac) |
|---|---|---|---|---|---|---|---|
| 0.10 | FF | 1557.2 ± 0.19% | 2903.2 ± 0.77% | 44249.8 ± 8.9% | 0.994 ± 0.2% | 1.0 ± 0.0012% | 0.994 ± 0.2% |
| 0.10 | FS | 1521.5 ± 0.028% | 1997.2 ± 0.0059% | 45984.4 ± 11.0% | 0.993 ± 0.24% | 1.0 ± 0.00082% | 0.993 ± 0.24% |
| 0.10 | Rand | 1543.5 ± 0.051% | 2708.2 ± 0.38% | 72316.3 ± 9.1% | 0.991 ± 0.2% | 1.0 ± 0.00078% | 0.991 ± 0.2% |
| 0.10 | SRPT | 1518.8 ± 0.021% | 1996.9 ± 0.0039% | 50036.6 ± 11.0% | 0.995 ± 0.2% | 1.0 ± 0.00025% | 0.995 ± 0.2% |
| 0.20 | FF | 1677.7 ± 1.0% | 5629.1 ± 8.4% | 77986.8 ± 8.3% | 0.985 ± 0.39% | 1.0 ± 0.01% | 0.985 ± 0.39% |
| 0.20 | FS | 1537.6 ± 0.11% | 1999.4 ± 0.0039% | 72962.6 ± 5.9% | 0.983 ± 0.4% | 1.0 ± 0.0019% | 0.983 ± 0.4% |
| 0.20 | Rand | 1600.8 ± 0.18% | 3050.2 ± 1.3% | 182454.6 ± 11.0% | 0.962 ± 0.34% | 1.0 ± 0.0025% | 0.962 ± 0.34% |
| 0.20 | SRPT | 1529.5 ± 0.079% | 2014.7 ± 0.56% | 102306.4 ± 12.0% | 0.985 ± 0.32% | 1.0 ± 0.0019% | 0.985 ± 0.32% |

(continued on next page)

**Table F.4** (*continued*)

| Load | Subject | Mean FCT (μs) | p99 FCT (μs) | Max FCT (μs) | Throughput (Frac) | Flows Accepted (Frac) | Info Accepted (Frac) |
|---|---|---|---|---|---|---|---|
| 0.30 | FF | 1887.8 ± 0.78% | 10474.4 ± 4.9% | 174541.8 ± 16.0% | 0.975 ± 0.17% | 0.999 ± 0.0073% | 0.975 ± 0.17% |
| 0.30 | FS | 1575.3 ± 0.19% | 2630.4 ± 2.8% | 134195.3 ± 3.0% | 0.97 ± 0.12% | 1.0 ± 0.0013% | 0.97 ± 0.12% |
| 0.30 | Rand | 1682.3 ± 0.2% | 3937.4 ± 0.35% | 381073.0 ± 4.0% | 0.857 ± 0.87% | 0.999 ± 0.0063% | 0.857 ± 0.87% |
| 0.30 | SRPT | 1551.2 ± 0.099% | 2500.5 ± 0.29% | 235811.0 ± 5.7% | 0.956 ± 0.29% | 1.0 ± 0.00062% | 0.956 ± 0.29% |
| 0.40 | FF | 2124.1 ± 2.2% | 15235.4 ± 11.0% | 247350.9 ± 7.0% | 0.939 ± 0.38% | 0.998 ± 0.02% | 0.939 ± 0.38% |
| 0.40 | FS | 1643.5 ± 0.12% | 3562.8 ± 4.5% | 230440.4 ± 6.6% | 0.926 ± 0.58% | 0.999 ± 0.0025% | 0.926 ± 0.58% |
| 0.40 | Rand | 1762.5 ± 0.23% | 5081.8 ± 0.67% | 295319.0 ± 1.8% | 0.816 ± 0.75% | 0.999 ± 0.0092% | 0.816 ± 0.75% |
| 0.40 | SRPT | 1561.9 ± 0.08% | 2771.3 ± 0.31% | 221163.5 ± 5.0% | 0.902 ± 0.41% | 1.0 ± 0.0014% | 0.902 ± 0.41% |
| 0.50 | FF | 1902.1 ± 1.1% | 6389.1 ± 2.7% | 391005.8 ± 7.6% | 0.909 ± 0.94% | 0.999 ± 0.0067% | 0.909 ± 0.94% |
| 0.50 | FS | 1740.5 ± 1.2% | 4533.5 ± 12.0% | 344343.1 ± 7.9% | 0.9 ± 1.1% | 0.999 ± 0.0055% | 0.9 ± 1.1% |
| 0.50 | Rand | 1947.7 ± 1.8% | 6365.3 ± 4.5% | 443976.4 ± 11.0% | 0.818 ± 1.2% | 0.998 ± 0.0037% | 0.818 ± 1.2% |
| 0.50 | SRPT | 1582.2 ± 0.16% | 2904.8 ± 0.36% | 363481.8 ± 7.4% | 0.875 ± 0.76% | 1.0 ± 0.0012% | 0.875 ± 0.76% |
| 0.60 | FF | 1989.3 ± 1.0% | 7602.7 ± 4.6% | 335234.2 ± 5.2% | 0.917 ± 0.39% | 0.999 ± 0.0057% | 0.917 ± 0.39% |
| 0.60 | FS | 1677.7 ± 0.53% | 3701.9 ± 1.1% | 314020.0 ± 4.8% | 0.912 ± 0.31% | 0.999 ± 0.0036% | 0.912 ± 0.31% |
| 0.60 | Rand | 2322.4 ± 2.7% | 9921.0 ± 8.2% | 398738.8 ± 2.5% | 0.805 ± 0.48% | 0.997 ± 0.027% | 0.805 ± 0.48% |
| 0.60 | SRPT | 1630.0 ± 0.084% | 3630.4 ± 0.48% | 322416.8 ± 5.0% | 0.879 ± 0.47% | 1.0 ± 0.0022% | 0.879 ± 0.47% |
| 0.70 | FF | 2434.1 ± 1.8% | 12649.6 ± 5.0% | 305610.1 ± 2.9% | 0.912 ± 0.35% | 0.998 ± 0.033% | 0.912 ± 0.35% |
| 0.70 | FS | 1672.2 ± 0.4% | 4415.8 ± 1.9% | 246486.9 ± 2.9% | 0.914 ± 0.3% | 0.999 ± 0.0033% | 0.914 ± 0.3% |
| 0.70 | Rand | 3083.8 ± 1.4% | 19421.0 ± 4.0% | 377667.2 ± 2.1% | 0.755 ± 1.1% | 0.993 ± 0.048% | 0.755 ± 1.1% |
| 0.70 | SRPT | 1712.6 ± 0.28% | 4502.1 ± 1.8% | 280418.9 ± 5.9% | 0.878 ± 0.46% | 0.999 ± 0.008% | 0.878 ± 0.46% |
| 0.79 | FF | 3394.1 ± 2.1% | 23179.1 ± 3.5% | 265525.7 ± 5.5% | 0.9 ± 0.23% | 0.995 ± 0.033% | 0.9 ± 0.23% |
| 0.79 | FS | 1724.5 ± 0.31% | 6302.9 ± 1.9% | 236377.1 ± 3.3% | 0.913 ± 0.28% | 0.999 ± 0.004% | 0.913 ± 0.28% |
| 0.79 | Rand | 3861.5 ± 1.8% | 25389.9 ± 1.9% | 317002.4 ± 2.3% | 0.731 ± 0.83% | 0.988 ± 0.033% | 0.731 ± 0.83% |
| 0.79 | SRPT | 1950.3 ± 1.3% | 7574.3 ± 6.7% | 271794.0 ± 1.7% | 0.848 ± 0.36% | 0.999 ± 0.017% | 0.848 ± 0.36% |
| 0.89 | FF | 5550.1 ± 1.9% | 44869.3 ± 2.5% | 333023.3 ± 11.0% | 0.87 ± 0.62% | 0.987 ± 0.041% | 0.87 ± 0.62% |
| 0.89 | FS | 2015.9 ± 0.54% | 12793.3 ± 1.7% | 254036.6 ± 10.0% | 0.873 ± 0.9% | 0.998 ± 0.013% | 0.873 ± 0.9% |
| 0.89 | Rand | 5718.1 ± 7.5% | 38174.0 ± 8.7% | 346773.2 ± 12.0% | 0.692 ± 0.71% | 0.979 ± 0.045% | 0.692 ± 0.71% |
| 0.89 | SRPT | 2645.0 ± 5.0% | 19839.5 ± 12.0% | 319581.9 ± 11.0% | 0.755 ± 0.43% | 0.993 ± 0.12% | 0.755 ± 0.43% |

**Table F.5**
Scheduler performance summary with 95% confidence intervals for the **Private Enterprise** benchmark.

| Load | Subject | Mean FCT (μs) | p99 FCT (μs) | Max FCT (μs) | Throughput (Frac) | Flows Accepted (Frac) | Info Accepted (Frac) |
|---|---|---|---|---|---|---|---|
| 0.10 | FF | 1576.7 ± 0.34% | 3207.9 ± 3.5% | 50143.5 ± 5.5% | 0.998 ± 0.085% | 1.0 ± 0.00094% | 0.998 ± 0.085% |
| 0.10 | FS | 1522.1 ± 0.021% | 1997.1 ± 0.0079% | 46335.0 ± 4.4% | 0.997 ± 0.095% | 1.0 ± 0.0006% | 0.997 ± 0.095% |
| 0.10 | Rand | 1550.9 ± 0.053% | 2765.2 ± 0.49% | 82610.5 ± 7.8% | 0.994 ± 0.19% | 1.0 ± 0.00074% | 0.994 ± 0.19% |
| 0.10 | SRPT | 1520.3 ± 0.01% | 1997.3 ± 0.0079% | 48062.1 ± 5.8% | 0.997 ± 0.13% | 1.0 ± 0.00047% | 0.997 ± 0.13% |
| 0.20 | FF | 1726.6 ± 1.6% | 6794.6 ± 11.0% | 70833.7 ± 3.0% | 0.983 ± 0.29% | 0.999 ± 0.01% | 0.983 ± 0.29% |
| 0.20 | FS | 1532.2 ± 0.13% | 2048.4 ± 0.76% | 66026.7 ± 2.0% | 0.983 ± 0.22% | 1.0 ± 0.00072% | 0.983 ± 0.22% |
| 0.20 | Rand | 1598.9 ± 0.16% | 3199.8 ± 2.0% | 166233.2 ± 8.3% | 0.946 ± 0.6% | 1.0 ± 0.0044% | 0.946 ± 0.6% |
| 0.20 | SRPT | 1529.5 ± 0.11% | 2214.8 ± 1.6% | 87532.1 ± 7.1% | 0.984 ± 0.22% | 1.0 ± 0.00048% | 0.984 ± 0.22% |
| 0.30 | FF | 2058.9 ± 3.1% | 16033.0 ± 12.0% | 149462.6 ± 8.8% | 0.98 ± 0.19% | 0.999 ± 0.016% | 0.98 ± 0.19% |
| 0.30 | FS | 1549.9 ± 0.13% | 2528.8 ± 1.1% | 121311.0 ± 7.3% | 0.981 ± 0.24% | 1.0 ± 0.001% | 0.981 ± 0.24% |
| 0.30 | Rand | 1684.2 ± 0.39% | 4149.8 ± 1.9% | 285851.7 ± 4.8% | 0.899 ± 0.73% | 0.999 ± 0.0088% | 0.899 ± 0.73% |
| 0.30 | SRPT | 1543.2 ± 0.056% | 2616.2 ± 0.41% | 196424.2 ± 9.0% | 0.978 ± 0.22% | 1.0 ± 0.00089% | 0.978 ± 0.22% |
| 0.40 | FF | 2638.3 ± 4.1% | 30026.6 ± 9.2% | 205182.9 ± 8.2% | 0.942 ± 0.6% | 0.997 ± 0.036% | 0.942 ± 0.6% |
| 0.40 | FS | 1599.4 ± 0.25% | 3333.2 ± 1.9% | 211188.7 ± 4.4% | 0.943 ± 0.21% | 1.0 ± 0.002% | 0.943 ± 0.21% |
| 0.40 | Rand | 1799.1 ± 0.54% | 5653.6 ± 2.3% | 280714.7 ± 3.0% | 0.84 ± 1.1% | 0.999 ± 0.015% | 0.84 ± 1.1% |
| 0.40 | SRPT | 1564.1 ± 0.085% | 2802.8 ± 0.32% | 210192.4 ± 7.6% | 0.937 ± 0.46% | 1.0 ± 0.0017% | 0.937 ± 0.46% |
| 0.50 | FF | 2824.6 ± 5.9% | 34301.5 ± 14.0% | 365468.3 ± 13.0% | 0.907 ± 1.0% | 0.994 ± 0.11% | 0.907 ± 1.0% |
| 0.50 | FS | 1682.6 ± 0.72% | 5048.5 ± 3.6% | 311288.1 ± 9.1% | 0.902 ± 1.2% | 0.999 ± 0.0061% | 0.902 ± 1.2% |
| 0.50 | Rand | 1993.9 ± 1.9% | 7870.4 ± 4.9% | 381296.9 ± 10.0% | 0.811 ± 1.1% | 0.998 ± 0.019% | 0.811 ± 1.1% |
| 0.50 | SRPT | 1582.9 ± 0.26% | 2938.1 ± 0.38% | 332134.3 ± 13.0% | 0.903 ± 0.65% | 1.0 ± 0.0026% | 0.903 ± 0.65% |
| 0.60 | FF | 2230.4 ± 1.3% | 11218.7 ± 5.3% | 339021.9 ± 2.3% | 0.915 ± 0.44% | 0.997 ± 0.065% | 0.915 ± 0.44% |
| 0.60 | FS | 1705.0 ± 0.53% | 5843.2 ± 3.2% | 326252.1 ± 3.6% | 0.907 ± 0.43% | 0.999 ± 0.0044% | 0.907 ± 0.43% |
| 0.60 | Rand | 2282.4 ± 1.6% | 12522.1 ± 6.0% | 412445.3 ± 4.4% | 0.782 ± 1.4% | 0.997 ± 0.029% | 0.782 ± 1.4% |
| 0.60 | SRPT | 1624.0 ± 0.21% | 3425.1 ± 1.3% | 375244.9 ± 5.9% | 0.898 ± 0.38% | 1.0 ± 0.0028% | 0.898 ± 0.38% |
| 0.70 | FF | 2449.3 ± 0.71% | 13110.2 ± 2.3% | 297091.8 ± 4.2% | 0.921 ± 0.26% | 0.998 ± 0.02% | 0.921 ± 0.26% |
| 0.70 | FS | 1696.4 ± 0.49% | 5751.0 ± 4.4% | 283512.5 ± 4.4% | 0.907 ± 0.17% | 0.999 ± 0.003% | 0.907 ± 0.17% |
| 0.70 | Rand | 2636.5 ± 0.7% | 18278.2 ± 2.0% | 363011.5 ± 2.3% | 0.74 ± 1.0% | 0.995 ± 0.029% | 0.74 ± 1.0% |
| 0.70 | SRPT | 1691.0 ± 0.23% | 4085.2 ± 1.5% | 315470.7 ± 7.2% | 0.892 ± 0.36% | 1.0 ± 0.0026% | 0.892 ± 0.36% |
| 0.79 | FF | 3400.0 ± 0.81% | 24127.3 ± 1.5% | 275964.6 ± 3.9% | 0.897 ± 0.37% | 0.994 ± 0.03% | 0.897 ± 0.37% |
| 0.79 | FS | 1732.4 ± 0.24% | 6508.5 ± 1.6% | 258779.8 ± 3.2% | 0.893 ± 0.45% | 0.999 ± 0.0034% | 0.893 ± 0.45% |
| 0.79 | Rand | 3264.4 ± 1.7% | 27586.4 ± 3.0% | 325223.7 ± 2.3% | 0.675 ± 0.73% | 0.989 ± 0.04% | 0.675 ± 0.73% |
| 0.79 | SRPT | 1841.9 ± 0.58% | 5834.0 ± 2.5% | 292946.1 ± 3.7% | 0.853 ± 0.25% | 0.999 ± 0.011% | 0.853 ± 0.25% |
| 0.90 | FF | 5851.8 ± 1.9% | 48861.6 ± 2.7% | 274329.9 ± 2.0% | 0.866 ± 0.64% | 0.983 ± 0.08% | 0.866 ± 0.64% |
| 0.90 | FS | 1940.3 ± 0.35% | 11084.7 ± 2.1% | 268340.1 ± 3.0% | 0.842 ± 0.52% | 0.998 ± 0.0064% | 0.842 ± 0.52% |
| 0.90 | Rand | 4124.7 ± 1.4% | 36647.1 ± 2.2% | 294642.1 ± 0.58% | 0.625 ± 1.5% | 0.983 ± 0.055% | 0.625 ± 1.5% |
| 0.90 | SRPT | 2492.0 ± 5.5% | 16474.5 ± 15.0% | 267699.8 ± 2.0% | 0.711 ± 0.41% | 0.994 ± 0.2% | 0.711 ± 0.41% |

**Table F.6**
Scheduler performance summary with 95% confidence intervals for the **Commercial Cloud** benchmark.

| Load | Subject | Mean FCT ($\mu s$) | p99 FCT ($\mu s$) | Max FCT ($\mu s$) | Throughput (Frac) | Flows Accepted (Frac) | Info Accepted (Frac) |
|---|---|---|---|---|---|---|---|
| 0.10 | FF | 1588.2 ± 0.46% | 3604.1 ± 2.4% | 49490.3 ± 6.8% | 0.996 ± 0.052% | 1.0 ± 0.0019% | 0.996 ± 0.052% |
| 0.10 | FS | 1520.1 ± 0.083% | 1997.1 ± 0.0039% | 42361.1 ± 4.6% | 0.994 ± 0.16% | 1.0 ± 0.00059% | 0.994 ± 0.16% |
| 0.10 | Rand | 1551.4 ± 0.12% | 2816.9 ± 0.47% | 75051.0 ± 13.0% | 0.99 ± 0.2% | 1.0 ± 0.0023% | 0.99 ± 0.2% |
| 0.10 | SRPT | 1519.3 ± 0.077% | 1997.7 ± 0.0059% | 42911.8 ± 5.8% | 0.996 ± 0.08% | 1.0 ± 0.00037% | 0.996 ± 0.08% |
| 0.20 | FF | 1747.8 ± 1.2% | 7437.1 ± 6.9% | 67090.9 ± 3.3% | 0.99 ± 0.29% | 0.999 ± 0.018% | 0.99 ± 0.29% |
| 0.20 | FS | 1524.9 ± 0.14% | 1998.8 ± 0.0059% | 59363.5 ± 6.5% | 0.991 ± 0.3% | 1.0 ± 0.0013% | 0.991 ± 0.3% |
| 0.20 | Rand | 1602.1 ± 0.24% | 3372.3 ± 1.2% | 171058.6 ± 6.9% | 0.97 ± 0.64% | 1.0 ± 0.0033% | 0.97 ± 0.64% |
| 0.20 | SRPT | 1525.8 ± 0.13% | 2276.5 ± 0.57% | 71962.2 ± 7.8% | 0.991 ± 0.26% | 1.0 ± 0.0013% | 0.991 ± 0.26% |
| 0.30 | FF | 2274.3 ± 2.6% | 21086.7 ± 9.0% | 116200.4 ± 8.8% | 0.987 ± 0.06% | 0.999 ± 0.012% | 0.987 ± 0.06% |
| 0.30 | FS | 1538.4 ± 0.061% | 2149.4 ± 0.9% | 85571.7 ± 5.5% | 0.99 ± 0.066% | 1.0 ± 0.00071% | 0.99 ± 0.066% |
| 0.30 | Rand | 1707.2 ± 0.29% | 4544.2 ± 1.7% | 249283.9 ± 9.6% | 0.933 ± 0.58% | 1.0 ± 0.003% | 0.933 ± 0.58% |
| 0.30 | SRPT | 1540.7 ± 0.023% | 2620.5 ± 0.31% | 119981.5 ± 12.0% | 0.989 ± 0.092% | 1.0 ± 0.00056% | 0.989 ± 0.092% |
| 0.40 | FF | 3203.2 ± 3.6% | 39373.6 ± 7.1% | 153040.1 ± 4.9% | 0.964 ± 0.31% | 0.994 ± 0.11% | 0.964 ± 0.31% |
| 0.40 | FS | 1557.2 ± 0.17% | 2559.2 ± 0.57% | 129399.8 ± 9.6% | 0.968 ± 0.36% | 1.0 ± 0.00083% | 0.968 ± 0.36% |
| 0.40 | Rand | 1889.3 ± 0.56% | 6600.9 ± 3.0% | 259317.4 ± 3.9% | 0.87 ± 0.65% | 0.999 ± 0.012% | 0.87 ± 0.65% |
| 0.40 | SRPT | 1564.1 ± 0.13% | 2830.9 ± 0.38% | 190613.2 ± 10.0% | 0.97 ± 0.25% | 1.0 ± 0.00072% | 0.97 ± 0.25% |
| 0.50 | FF | 4495.2 ± 3.4% | 60948.4 ± 4.3% | 255736.7 ± 14.0% | 0.939 ± 0.64% | 0.989 ± 0.18% | 0.939 ± 0.64% |
| 0.50 | FS | 1584.6 ± 0.13% | 2963.7 ± 0.38% | 196875.6 ± 7.7% | 0.947 ± 0.84% | 1.0 ± 0.0039% | 0.947 ± 0.84% |
| 0.50 | Rand | 2324.1 ± 3.3% | 12139.1 ± 11.0% | 353111.3 ± 13.0% | 0.797 ± 0.74% | 0.996 ± 0.027% | 0.797 ± 0.74% |
| 0.50 | SRPT | 1585.3 ± 0.082% | 2962.4 ± 0.21% | 254463.8 ± 8.2% | 0.942 ± 0.56% | 1.0 ± 0.0022% | 0.942 ± 0.56% |
| 0.60 | FF | 4837.1 ± 5.1% | 68328.0 ± 3.3% | 387525.7 ± 2.3% | 0.924 ± 0.23% | 0.978 ± 0.2% | 0.924 ± 0.23% |
| 0.60 | FS | 1639.9 ± 0.14% | 3835.1 ± 0.83% | 268943.4 ± 3.6% | 0.941 ± 0.14% | 1.0 ± 0.0018% | 0.941 ± 0.14% |
| 0.60 | Rand | 3236.8 ± 0.65% | 22198.9 ± 0.66% | 439374.7 ± 1.1% | 0.744 ± 0.42% | 0.993 ± 0.015% | 0.744 ± 0.42% |
| 0.60 | SRPT | 1628.1 ± 0.15% | 3565.0 ± 0.8% | 308435.8 ± 4.8% | 0.922 ± 0.26% | 1.0 ± 0.0026% | 0.922 ± 0.26% |
| 0.70 | FF | 3173.6 ± 0.7% | 22472.4 ± 2.9% | 327840.2 ± 2.7% | 0.905 ± 0.31% | 0.992 ± 0.044% | 0.905 ± 0.31% |
| 0.70 | FS | 1686.9 ± 0.23% | 4915.5 ± 1.0% | 254484.7 ± 1.8% | 0.921 ± 0.44% | 0.999 ± 0.0024% | 0.921 ± 0.44% |
| 0.70 | Rand | 3760.3 ± 0.94% | 31788.5 ± 2.2% | 365861.9 ± 2.2% | 0.675 ± 0.25% | 0.989 ± 0.027% | 0.675 ± 0.25% |
| 0.70 | SRPT | 1715.2 ± 0.24% | 4404.1 ± 1.2% | 264969.5 ± 5.5% | 0.903 ± 0.33% | 1.0 ± 0.004% | 0.903 ± 0.33% |
| 0.79 | FF | 4144.2 ± 2.0% | 30541.3 ± 4.0% | 301349.2 ± 2.6% | 0.902 ± 0.18% | 0.993 ± 0.025% | 0.902 ± 0.18% |
| 0.79 | FS | 1743.5 ± 0.24% | 6572.0 ± 1.6% | 259058.4 ± 2.9% | 0.905 ± 0.18% | 0.999 ± 0.0026% | 0.905 ± 0.18% |
| 0.79 | Rand | 4740.4 ± 0.98% | 46094.7 ± 2.0% | 344636.1 ± 0.65% | 0.6 ± 0.65% | 0.98 ± 0.032% | 0.6 ± 0.65% |
| 0.79 | SRPT | 1889.5 ± 0.74% | 6169.9 ± 3.9% | 292500.7 ± 4.5% | 0.868 ± 0.038% | 0.999 ± 0.0052% | 0.868 ± 0.038% |
| 0.89 | FF | 6856.2 ± 0.89% | 54158.7 ± 2.0% | 272757.7 ± 1.4% | 0.853 ± 0.25% | 0.976 ± 0.14% | 0.853 ± 0.25% |
| 0.89 | FS | 1940.3 ± 0.16% | 10891.2 ± 0.75% | 253250.9 ± 1.6% | 0.844 ± 0.37% | 0.998 ± 0.0061% | 0.844 ± 0.37% |
| 0.89 | Rand | 5320.7 ± 1.0% | 55646.5 ± 1.6% | 300652.9 ± 0.86% | 0.541 ± 0.4% | 0.972 ± 0.051% | 0.541 ± 0.4% |
| 0.89 | SRPT | 2234.9 ± 1.5% | 10623.8 ± 4.9% | 267587.5 ± 2.8% | 0.719 ± 0.62% | 0.996 ± 0.025% | 0.719 ± 0.62% |

**Table F.7**
Scheduler performance summary with 95% confidence intervals for the **Social Media Cloud** benchmark.

| Load | Subject | Mean FCT ($\mu s$) | p99 FCT ($\mu s$) | Max FCT ($\mu s$) | Throughput (Frac) | Flows Accepted (Frac) | Info Accepted (Frac) |
|---|---|---|---|---|---|---|---|
| 0.10 | FF | 1536.7 ± 0.074% | 2766.2 ± 0.3% | 7153.9 ± 2.0% | 1.0 ± 0.0064% | 1.0 ± 0.00041% | 1.0 ± 0.0064% |
| 0.10 | FS | 1513.9 ± 0.062% | 2053.4 ± 0.84% | 6892.3 ± 3.1% | 1.0 ± 0.0062% | 1.0 ± 0.00024% | 1.0 ± 0.0062% |
| 0.10 | Rand | 1536.5 ± 0.054% | 2762.4 ± 0.35% | 13551.4 ± 11.0% | 1.0 ± 0.0079% | 1.0 ± 0.00056% | 1.0 ± 0.0079% |
| 0.10 | SRPT | 1515.2 ± 0.062% | 2189.1 ± 0.38% | 6820.5 ± 3.6% | 1.0 ± 0.0063% | 1.0 ± 0.00032% | 1.0 ± 0.0063% |
| 0.20 | FF | 1591.6 ± 0.11% | 3410.3 ± 1.2% | 12773.2 ± 12.0% | 1.0 ± 0.0045% | 1.0 ± 0.00091% | 1.0 ± 0.0045% |
| 0.20 | FS | 1523.7 ± 0.033% | 2560.6 ± 0.61% | 11206.1 ± 11.0% | 1.0 ± 0.0056% | 1.0 ± 0.00047% | 1.0 ± 0.0056% |
| 0.20 | Rand | 1581.4 ± 0.097% | 3237.3 ± 1.2% | 29019.5 ± 19.0% | 1.0 ± 0.0099% | 1.0 ± 0.0015% | 1.0 ± 0.0099% |
| 0.20 | SRPT | 1532.6 ± 0.054% | 2720.7 ± 0.49% | 11620.8 ± 11.0% | 1.0 ± 0.0052% | 1.0 ± 0.0005% | 1.0 ± 0.0052% |
| 0.30 | FF | 1707.8 ± 0.42% | 4849.3 ± 2.3% | 24735.7 ± 10.0% | 1.0 ± 0.011% | 1.0 ± 0.0032% | 1.0 ± 0.011% |
| 0.30 | FS | 1539.5 ± 0.056% | 2859.5 ± 0.27% | 15729.0 ± 8.7% | 1.0 ± 0.0089% | 1.0 ± 0.0006% | 1.0 ± 0.0089% |
| 0.30 | Rand | 1660.7 ± 0.13% | 4184.2 ± 1.1% | 47524.1 ± 19.0% | 0.999 ± 0.025% | 1.0 ± 0.0033% | 0.999 ± 0.025% |
| 0.30 | SRPT | 1565.8 ± 0.095% | 2972.7 ± 0.14% | 18417.1 ± 12.0% | 1.0 ± 0.0073% | 1.0 ± 0.00098% | 1.0 ± 0.0073% |
| 0.40 | FF | 1924.6 ± 0.8% | 7639.7 ± 3.0% | 39600.9 ± 9.7% | 0.998 ± 0.021% | 0.999 ± 0.0098% | 0.998 ± 0.021% |
| 0.40 | FS | 1563.9 ± 0.11% | 3266.6 ± 1.1% | 17450.8 ± 3.5% | 0.999 ± 0.019% | 1.0 ± 0.0023% | 0.999 ± 0.019% |
| 0.40 | Rand | 1808.3 ± 0.31% | 5802.5 ± 0.92% | 92643.3 ± 23.0% | 0.996 ± 0.042% | 0.999 ± 0.0058% | 0.996 ± 0.042% |
| 0.40 | SRPT | 1622.6 ± 0.19% | 3731.7 ± 0.93% | 23635.4 ± 7.2% | 0.999 ± 0.01% | 1.0 ± 0.0038% | 0.999 ± 0.01% |
| 0.50 | FF | 2646.7 ± 2.9% | 20076.0 ± 7.9% | 117682.7 ± 9.7% | 0.996 ± 0.066% | 0.997 ± 0.052% | 0.996 ± 0.066% |
| 0.50 | FS | 1624.4 ± 0.21% | 4201.7 ± 1.4% | 31567.8 ± 3.4% | 0.997 ± 0.058% | 1.0 ± 0.0047% | 0.997 ± 0.058% |
| 0.50 | Rand | 2218.8 ± 0.77% | 10570.1 ± 3.4% | 207351.1 ± 11.0% | 0.987 ± 0.15% | 0.998 ± 0.019% | 0.987 ± 0.15% |
| 0.50 | SRPT | 1737.3 ± 0.53% | 4829.9 ± 1.8% | 49492.8 ± 6.5% | 0.997 ± 0.045% | 0.999 ± 0.013% | 0.997 ± 0.045% |
| 0.60 | FF | 4495.9 ± 4.4% | 55356.7 ± 7.6% | 237610.0 ± 7.3% | 0.988 ± 0.16% | 0.989 ± 0.053% | 0.988 ± 0.16% |
| 0.60 | FS | 1755.8 ± 0.41% | 6110.1 ± 1.8% | 47599.2 ± 5.6% | 0.992 ± 0.15% | 0.999 ± 0.024% | 0.992 ± 0.15% |
| 0.60 | Rand | 3262.0 ± 1.6% | 24348.0 ± 2.2% | 269243.0 ± 2.0% | 0.951 ± 0.31% | 0.991 ± 0.049% | 0.951 ± 0.31% |
| 0.60 | SRPT | 2034.5 ± 2.0% | 8447.1 ± 8.2% | 193698.4 ± 8.9% | 0.992 ± 0.12% | 0.998 ± 0.071% | 0.992 ± 0.12% |
| 0.69 | FF | 8175.5 ± 2.7% | 121246.2 ± 3.5% | 468538.0 ± 5.4% | 0.964 ± 0.22% | 0.934 ± 0.82% | 0.964 ± 0.22% |
| 0.69 | FS | 2384.6 ± 1.7% | 14253.5 ± 3.9% | 138806.7 ± 6.4% | 0.986 ± 0.14% | 0.998 ± 0.026% | 0.986 ± 0.14% |
| 0.69 | Rand | 6394.4 ± 1.4% | 72096.8 ± 3.3% | 507914.9 ± 2.5% | 0.901 ± 0.29% | 0.98 ± 0.049% | 0.901 ± 0.29% |
| 0.69 | SRPT | 4937.4 ± 9.9% | 64798.0 ± 18.0% | 500125.6 ± 2.8% | 0.939 ± 0.81% | 0.981 ± 0.55% | 0.939 ± 0.81% |
| 0.80 | FF | 7182.3 ± 1.7% | 77566.2 ± 3.8% | 443785.0 ± 2.9% | 0.938 ± 0.13% | 0.951 ± 0.19% | 0.938 ± 0.13% |
| 0.80 | FS | 4026.1 ± 2.1% | 32187.7 ± 2.7% | 243834.5 ± 3.5% | 0.947 ± 0.19% | 0.992 ± 0.034% | 0.947 ± 0.19% |

(*continued on next page*)

**Table F.7** (*continued*)

| Load | Subject | Mean FCT ($\mu s$) | p99 FCT ($\mu s$) | Max FCT ($\mu s$) | Throughput (Frac) | Flows Accepted (Frac) | Info Accepted (Frac) |
|---|---|---|---|---|---|---|---|
| 0.80 | Rand | 8489.0 ± 2.2% | 89488.4 ± 1.5% | 446095.0 ± 1.4% | 0.846 ± 0.23% | 0.966 ± 0.06% | 0.846 ± 0.23% |
| 0.80 | SRPT | 11412.4 ± 4.1% | 154590.0 ± 3.9% | 443708.3 ± 1.7% | 0.748 ± 0.42% | 0.854 ± 0.88% | 0.748 ± 0.42% |
| 0.90 | FF | 8731.6 ± 1.5% | 76236.3 ± 1.8% | 380339.7 ± 2.7% | 0.946 ± 0.13% | 0.97 ± 0.2% | 0.946 ± 0.13% |
| 0.90 | FS | 4809.9 ± 1.4% | 40007.0 ± 2.0% | 228118.7 ± 2.1% | 0.931 ± 0.14% | 0.989 ± 0.038% | 0.931 ± 0.14% |
| 0.90 | Rand | 10800.9 ± 0.96% | 110549.1 ± 0.69% | 407971.9 ± 1.1% | 0.788 ± 0.41% | 0.949 ± 0.15% | 0.788 ± 0.41% |
| 0.90 | SRPT | 18401.3 ± 2.4% | 204251.6 ± 2.0% | 416090.4 ± 0.56% | 0.61 ± 0.78% | 0.751 ± 0.98% | 0.61 ± 0.78% |

### *Appendix F.3.2. Skewed Nodes Distribution Benchmark*

**Table F.8**
Scheduler performance summary with 95% confidence intervals for the **skewed_nodes_sensitivity_uniform** and **rack_sensitivity_uniform** benchmarks.

| Load | Subject | Mean FCT ($\mu s$) | p99 FCT ($\mu s$) | Max FCT ($\mu s$) | Throughput (Frac) | Flows Accepted (Frac) | Info Accepted (Frac) |
|---|---|---|---|---|---|---|---|
| 0.1 | FF | 1554.5 ± 0.15% | 2977.0 ± 0.79% | 38288.3 ± 6.7% | 0.995 ± 0.11% | 1.0 ± 0.024% | 0.995 ± 0.11% |
| 0.1 | FS | 1518.8 ± 0.12% | 1997.5 ± 0.0039% | 39693.2 ± 4.6% | 0.995 ± 0.11% | 1.0 ± 0.024% | 0.995 ± 0.11% |
| 0.1 | Rand | 1544.1 ± 0.11% | 2750.1 ± 0.34% | 60170.8 ± 9.4% | 0.991 ± 0.15% | 1.0 ± 0.024% | 0.991 ± 0.15% |
| 0.1 | SRPT | 1518.3 ± 0.12% | 1998.0 ± 0.0039% | 41190.0 ± 5.2% | 0.995 ± 0.11% | 1.0 ± 0.024% | 0.995 ± 0.11% |
| 0.2 | FF | 1620.8 ± 0.34% | 4398.3 ± 4.0% | 43732.0 ± 4.4% | 0.98 ± 0.3% | 0.999 ± 0.054% | 0.98 ± 0.3% |
| 0.2 | FS | 1524.3 ± 0.1% | 1999.6 ± 0.016% | 42196.8 ± 4.5% | 0.982 ± 0.32% | 0.999 ± 0.055% | 0.982 ± 0.32% |
| 0.2 | Rand | 1579.3 ± 0.18% | 3049.8 ± 1.1% | 79304.0 ± 9.3% | 0.974 ± 0.25% | 0.999 ± 0.057% | 0.974 ± 0.25% |
| 0.2 | SRPT | 1524.9 ± 0.087% | 2234.7 ± 1.0% | 44396.8 ± 4.8% | 0.983 ± 0.28% | 0.999 ± 0.055% | 0.983 ± 0.28% |
| 0.3 | FF | 1744.2 ± 0.55% | 6564.0 ± 2.3% | 80217.5 ± 5.9% | 0.988 ± 0.18% | 0.999 ± 0.069% | 0.988 ± 0.18% |
| 0.3 | FS | 1532.9 ± 0.1% | 2255.9 ± 0.46% | 71447.0 ± 7.3% | 0.989 ± 0.16% | 0.999 ± 0.064% | 0.989 ± 0.16% |
| 0.3 | Rand | 1643.6 ± 0.22% | 3856.5 ± 0.36% | 180283.0 ± 6.6% | 0.973 ± 0.27% | 0.999 ± 0.066% | 0.973 ± 0.27% |
| 0.3 | SRPT | 1537.1 ± 0.071% | 2612.7 ± 0.59% | 84911.1 ± 7.0% | 0.99 ± 0.15% | 0.999 ± 0.064% | 0.99 ± 0.15% |
| 0.4 | FF | 1917.3 ± 0.82% | 9481.8 ± 2.8% | 89676.1 ± 6.4% | 0.981 ± 0.29% | 0.998 ± 0.057% | 0.981 ± 0.29% |
| 0.4 | FS | 1544.5 ± 0.079% | 2602.7 ± 0.7% | 85476.6 ± 5.7% | 0.98 ± 0.3% | 0.999 ± 0.049% | 0.98 ± 0.3% |
| 0.4 | Rand | 1776.3 ± 0.22% | 5093.5 ± 0.91% | 239854.0 ± 7.3% | 0.946 ± 0.38% | 0.999 ± 0.05% | 0.946 ± 0.38% |
| 0.4 | SRPT | 1554.7 ± 0.058% | 2819.0 ± 0.48% | 109885.3 ± 8.2% | 0.98 ± 0.25% | 0.999 ± 0.05% | 0.98 ± 0.25% |
| 0.5 | FF | 2254.6 ± 0.82% | 14792.9 ± 1.9% | 100669.6 ± 4.8% | 0.978 ± 0.24% | 0.998 ± 0.046% | 0.978 ± 0.24% |
| 0.5 | FS | 1563.8 ± 0.16% | 2927.5 ± 0.58% | 101281.5 ± 7.7% | 0.981 ± 0.23% | 0.999 ± 0.042% | 0.981 ± 0.23% |
| 0.5 | Rand | 2259.1 ± 1.4% | 9368.2 ± 3.4% | 403534.7 ± 12.0% | 0.883 ± 0.74% | 0.997 ± 0.041% | 0.883 ± 0.74% |
| 0.5 | SRPT | 1580.4 ± 0.069% | 2948.5 ± 0.13% | 148065.0 ± 4.9% | 0.977 ± 0.25% | 0.999 ± 0.04% | 0.977 ± 0.25% |
| 0.6 | FF | 2696.5 ± 1.4% | 19574.0 ± 3.3% | 242541.4 ± 13.0% | 0.971 ± 0.36% | 0.997 ± 0.051% | 0.971 ± 0.36% |
| 0.6 | FS | 1595.6 ± 0.15% | 3652.1 ± 0.98% | 161242.9 ± 14.0% | 0.973 ± 0.24% | 0.999 ± 0.051% | 0.973 ± 0.24% |
| 0.6 | Rand | 3309.7 ± 1.1% | 17326.4 ± 1.6% | 401082.8 ± 4.4% | 0.82 ± 0.87% | 0.993 ± 0.066% | 0.82 ± 0.87% |
| 0.6 | SRPT | 1620.9 ± 0.077% | 3373.6 ± 0.78% | 294496.9 ± 6.8% | 0.962 ± 0.35% | 0.999 ± 0.051% | 0.962 ± 0.35% |
| 0.7 | FF | 3436.8 ± 1.0% | 27933.1 ± 2.3% | 297748.1 ± 3.1% | 0.935 ± 0.51% | 0.994 ± 0.077% | 0.935 ± 0.51% |
| 0.7 | FS | 1660.9 ± 0.21% | 4953.9 ± 1.2% | 255268.4 ± 4.4% | 0.942 ± 0.36% | 0.999 ± 0.078% | 0.942 ± 0.36% |
| 0.7 | Rand | 4393.5 ± 1.1% | 24778.6 ± 1.8% | 354839.4 ± 2.2% | 0.738 ± 1.1% | 0.986 ± 0.082% | 0.738 ± 1.1% |
| 0.7 | SRPT | 1668.4 ± 0.15% | 3827.3 ± 0.71% | 320957.0 ± 4.4% | 0.914 ± 0.47% | 0.998 ± 0.077% | 0.914 ± 0.47% |
| 0.8 | FF | 4361.4 ± 2.1% | 34817.0 ± 2.7% | 287276.9 ± 3.8% | 0.907 ± 0.59% | 0.99 ± 0.15% | 0.907 ± 0.59% |
| 0.8 | FS | 1758.1 ± 0.5% | 7135.0 ± 1.9% | 283104.7 ± 1.3% | 0.899 ± 0.8% | 0.998 ± 0.1% | 0.899 ± 0.8% |
| 0.8 | Rand | 5762.2 ± 1.5% | 32239.6 ± 1.4% | 329015.7 ± 2.0% | 0.693 ± 1.1% | 0.977 ± 0.15% | 0.693 ± 1.1% |
| 0.8 | SRPT | 1758.1 ± 0.41% | 4842.1 ± 2.4% | 309165.9 ± 3.6% | 0.858 ± 0.63% | 0.998 ± 0.11% | 0.858 ± 0.63% |
| 0.9 | FF | 5520.3 ± 1.7% | 43104.1 ± 2.4% | 278164.1 ± 2.2% | 0.846 ± 0.61% | 0.983 ± 0.061% | 0.846 ± 0.61% |
| 0.9 | FS | 1890.9 ± 0.47% | 9974.2 ± 1.9% | 287700.1 ± 3.1% | 0.823 ± 0.74% | 0.998 ± 0.038% | 0.823 ± 0.74% |
| 0.9 | Rand | 7095.9 ± 1.3% | 39006.4 ± 1.4% | 306075.6 ± 1.6% | 0.627 ± 0.93% | 0.968 ± 0.041% | 0.627 ± 0.93% |
| 0.9 | SRPT | 1890.8 ± 0.89% | 6584.6 ± 4.1% | 287161.6 ± 2.9% | 0.771 ± 0.54% | 0.998 ± 0.036% | 0.771 ± 0.54% |

**Table F.9**
Scheduler performance summary with 95% confidence intervals for the **skewed_nodes_sensitivity_0.05** benchmark.

| Load | Subject | Mean FCT ($\mu s$) | p99 FCT ($\mu s$) | Max FCT ($\mu s$) | Throughput (Frac) | Flows Accepted (Frac) | Info Accepted (Frac) |
|---|---|---|---|---|---|---|---|
| 0.10 | FF | 1676.3 ± 1.3% | 4965.6 ± 5.9% | 115613.1 ± 12.0% | 0.994 ± 0.17% | 0.999 ± 0.042% | 0.994 ± 0.17% |
| 0.10 | FS | 1545.9 ± 0.21% | 2137.2 ± 1.5% | 111455.6 ± 11.0% | 0.993 ± 0.15% | 0.999 ± 0.04% | 0.993 ± 0.15% |
| 0.10 | Rand | 1586.2 ± 0.22% | 3071.9 ± 1.5% | 204371.5 ± 6.6% | 0.986 ± 0.18% | 0.999 ± 0.04% | 0.986 ± 0.18% |
| 0.10 | SRPT | 1529.5 ± 0.14% | 1998.7 ± 0.0078% | 144042.5 ± 11.0% | 0.993 ± 0.16% | 0.999 ± 0.041% | 0.993 ± 0.16% |
| 0.20 | FF | 1769.9 ± 2.2% | 4943.9 ± 12.0% | 281567.2 ± 4.5% | 0.922 ± 0.66% | 0.997 ± 0.086% | 0.922 ± 0.66% |
| 0.20 | FS | 1653.3 ± 0.56% | 3724.2 ± 11.0% | 264636.5 ± 5.0% | 0.896 ± 0.51% | 0.998 ± 0.092% | 0.896 ± 0.51% |
| 0.20 | Rand | 1691.1 ± 0.83% | 4168.9 ± 5.4% | 185373.4 ± 3.9% | 0.901 ± 0.24% | 0.998 ± 0.091% | 0.901 ± 0.24% |
| 0.20 | SRPT | 1547.1 ± 0.16% | 2306.6 ± 1.5% | 165611.6 ± 5.4% | 0.933 ± 0.29% | 0.999 ± 0.093% | 0.933 ± 0.29% |
| 0.30 | FF | 1697.5 ± 0.24% | 4419.9 ± 2.0% | 289568.9 ± 7.6% | 0.949 ± 0.49% | 0.999 ± 0.037% | 0.949 ± 0.49% |
| 0.30 | FS | 1612.6 ± 0.46% | 2501.7 ± 1.3% | 297525.1 ± 2.3% | 0.927 ± 0.69% | 0.999 ± 0.035% | 0.927 ± 0.69% |
| 0.30 | Rand | 1686.6 ± 0.47% | 3854.9 ± 0.88% | 210069.0 ± 5.6% | 0.911 ± 0.71% | 0.999 ± 0.036% | 0.911 ± 0.71% |
| 0.30 | SRPT | 1551.1 ± 0.14% | 2604.4 ± 0.24% | 228406.2 ± 12.0% | 0.943 ± 0.35% | 0.999 ± 0.038% | 0.943 ± 0.35% |
| 0.40 | FF | 1789.6 ± 0.58% | 6066.7 ± 3.7% | 257805.2 ± 5.8% | 0.955 ± 0.24% | 0.998 ± 0.096% | 0.955 ± 0.24% |
| 0.40 | FS | 1584.6 ± 0.29% | 2728.7 ± 0.95% | 201816.6 ± 2.3% | 0.938 ± 0.16% | 0.999 ± 0.088% | 0.938 ± 0.16% |
| 0.40 | Rand | 1783.5 ± 0.32% | 4928.8 ± 0.76% | 275464.8 ± 7.1% | 0.905 ± 0.21% | 0.998 ± 0.092% | 0.905 ± 0.21% |
| 0.40 | SRPT | 1561.7 ± 0.11% | 2830.3 ± 0.32% | 266258.6 ± 11.0% | 0.945 ± 0.21% | 0.999 ± 0.09% | 0.945 ± 0.21% |

(*continued on next page*)

**Table F.9** (*continued*)

| Load | Subject | Mean FCT ($\mu s$) | p99 FCT ($\mu s$) | Max FCT ($\mu s$) | Throughput (Frac) | Flows Accepted (Frac) | Info Accepted (Frac) |
|---|---|---|---|---|---|---|---|
| 0.50 | FF | 2040.0 ± 1.8% | 9688.9 ± 8.6% | 287779.7 ± 19.0% | 0.953 ± 0.52% | 0.997 ± 0.15% | 0.953 ± 0.52% |
| 0.50 | FS | 1589.9 ± 0.25% | 2981.8 ± 0.49% | 177708.6 ± 6.5% | 0.95 ± 0.48% | 0.998 ± 0.13% | 0.95 ± 0.48% |
| 0.50 | Rand | 2120.0 ± 1.7% | 7781.7 ± 3.5% | 314269.3 ± 12.0% | 0.866 ± 0.72% | 0.996 ± 0.14% | 0.866 ± 0.72% |
| 0.50 | SRPT | 1589.3 ± 0.16% | 2963.8 ± 0.28% | 306084.3 ± 13.0% | 0.94 ± 0.52% | 0.998 ± 0.14% | 0.94 ± 0.52% |
| 0.60 | FF | 2468.3 ± 1.2% | 14704.2 ± 3.3% | 311801.5 ± 8.9% | 0.956 ± 0.3% | 0.998 ± 0.042% | 0.956 ± 0.3% |
| 0.60 | FS | 1620.5 ± 0.097% | 3756.5 ± 0.77% | 197184.5 ± 4.0% | 0.954 ± 0.23% | 0.999 ± 0.038% | 0.954 ± 0.23% |
| 0.60 | Rand | 3082.2 ± 1.8% | 15591.2 ± 3.3% | 430919.2 ± 2.4% | 0.815 ± 0.9% | 0.995 ± 0.056% | 0.815 ± 0.9% |
| 0.60 | SRPT | 1633.7 ± 0.19% | 3493.7 ± 1.2% | 337388.2 ± 4.1% | 0.94 ± 0.21% | 0.999 ± 0.038% | 0.94 ± 0.21% |
| 0.70 | FF | 3267.8 ± 2.8% | 23735.1 ± 7.2% | 301004.1 ± 4.5% | 0.939 ± 0.27% | 0.995 ± 0.062% | 0.939 ± 0.27% |
| 0.70 | FS | 1659.3 ± 0.14% | 4784.4 ± 0.97% | 251399.3 ± 4.4% | 0.937 ± 0.32% | 0.999 ± 0.052% | 0.937 ± 0.32% |
| 0.70 | Rand | 4312.8 ± 1.8% | 23854.6 ± 2.8% | 362330.2 ± 2.8% | 0.751 ± 1.2% | 0.988 ± 0.088% | 0.751 ± 1.2% |
| 0.70 | SRPT | 1695.4 ± 0.39% | 4072.7 ± 2.1% | 320406.5 ± 3.1% | 0.918 ± 0.35% | 0.999 ± 0.054% | 0.918 ± 0.35% |
| 0.79 | FF | 4478.6 ± 1.1% | 36615.6 ± 3.5% | 307393.6 ± 1.8% | 0.905 ± 0.4% | 0.989 ± 0.085% | 0.905 ± 0.4% |
| 0.79 | FS | 1763.6 ± 0.15% | 7054.1 ± 1.4% | 269808.6 ± 5.0% | 0.896 ± 0.29% | 0.998 ± 0.067% | 0.896 ± 0.29% |
| 0.79 | Rand | 5939.0 ± 1.5% | 33275.3 ± 2.1% | 332125.7 ± 1.8% | 0.679 ± 1.2% | 0.977 ± 0.087% | 0.679 ± 1.2% |
| 0.79 | SRPT | 1792.6 ± 0.76% | 5219.3 ± 3.7% | 303203.6 ± 3.1% | 0.842 ± 0.32% | 0.998 ± 0.071% | 0.842 ± 0.32% |
| 0.90 | FF | 6062.3 ± 2.5% | 48771.8 ± 3.1% | 278389.8 ± 2.2% | 0.852 ± 0.38% | 0.979 ± 0.075% | 0.852 ± 0.38% |
| 0.90 | FS | 1924.4 ± 0.59% | 10517.0 ± 2.1% | 284887.0 ± 4.4% | 0.819 ± 0.73% | 0.997 ± 0.1% | 0.819 ± 0.73% |
| 0.90 | Rand | 7280.9 ± 2.0% | 40622.7 ± 2.7% | 304640.3 ± 2.2% | 0.621 ± 0.53% | 0.965 ± 0.071% | 0.621 ± 0.53% |
| 0.90 | SRPT | 1905.5 ± 0.72% | 6722.4 ± 2.8% | 288426.4 ± 2.8% | 0.751 ± 0.27% | 0.997 ± 0.095% | 0.751 ± 0.27% |

**Table F.10**
Scheduler performance summary with 95% confidence intervals for the **skewed_nodes_sensitivity_0.1** benchmark.

| Load | Subject | Mean FCT ($\mu s$) | p99 FCT ($\mu s$) | Max FCT ($\mu s$) | Throughput (Frac) | Flows Accepted (Frac) | Info Accepted (Frac) |
|---|---|---|---|---|---|---|---|
| 0.10 | FF | 1589.7 ± 0.2% | 3580.9 ± 0.61% | 72990.8 ± 4.8% | 0.995 ± 0.13% | 0.999 ± 0.044% | 0.995 ± 0.13% |
| 0.10 | FS | 1526.7 ± 0.14% | 1998.8 ± 0.016% | 70198.0 ± 4.2% | 0.995 ± 0.14% | 0.999 ± 0.045% | 0.995 ± 0.14% |
| 0.10 | Rand | 1554.9 ± 0.18% | 2849.2 ± 0.33% | 91598.0 ± 5.9% | 0.995 ± 0.14% | 0.999 ± 0.045% | 0.995 ± 0.14% |
| 0.10 | SRPT | 1520.9 ± 0.11% | 1998.2 ± 0.0098% | 81555.4 ± 5.0% | 0.995 ± 0.13% | 0.999 ± 0.045% | 0.995 ± 0.13% |
| 0.20 | FF | 1904.3 ± 1.3% | 11165.2 ± 7.4% | 170783.1 ± 17.0% | 0.966 ± 0.67% | 0.998 ± 0.044% | 0.966 ± 0.67% |
| 0.20 | FS | 1575.6 ± 0.12% | 2708.5 ± 4.0% | 172624.9 ± 5.3% | 0.957 ± 0.77% | 0.999 ± 0.045% | 0.957 ± 0.77% |
| 0.20 | Rand | 1641.1 ± 0.21% | 3799.1 ± 0.95% | 258243.8 ± 8.6% | 0.901 ± 0.8% | 0.999 ± 0.047% | 0.901 ± 0.8% |
| 0.20 | SRPT | 1542.8 ± 0.11% | 2384.0 ± 1.5% | 237546.1 ± 11.0% | 0.951 ± 0.71% | 0.999 ± 0.046% | 0.951 ± 0.71% |
| 0.30 | FF | 2110.4 ± 5.3% | 13637.6 ± 22.0% | 364074.5 ± 7.9% | 0.922 ± 0.62% | 0.997 ± 0.057% | 0.922 ± 0.62% |
| 0.30 | FS | 1695.1 ± 0.62% | 6015.8 ± 11.0% | 348982.1 ± 3.3% | 0.908 ± 0.47% | 0.999 ± 0.031% | 0.908 ± 0.47% |
| 0.30 | Rand | 1734.6 ± 1.1% | 5030.5 ± 5.8% | 329509.4 ± 5.2% | 0.871 ± 0.9% | 0.999 ± 0.036% | 0.871 ± 0.9% |
| 0.30 | SRPT | 1551.9 ± 0.15% | 2671.0 ± 0.84% | 347195.5 ± 7.9% | 0.911 ± 0.58% | 0.999 ± 0.031% | 0.911 ± 0.58% |
| 0.40 | FF | 1757.4 ± 0.3% | 5007.6 ± 1.5% | 232866.4 ± 6.1% | 0.933 ± 0.32% | 0.998 ± 0.086% | 0.933 ± 0.32% |
| 0.40 | FS | 1640.0 ± 0.54% | 2879.0 ± 1.3% | 290705.3 ± 4.7% | 0.903 ± 0.38% | 0.998 ± 0.086% | 0.903 ± 0.38% |
| 0.40 | Rand | 1738.4 ± 0.63% | 4569.2 ± 1.4% | 253293.4 ± 7.2% | 0.869 ± 0.54% | 0.998 ± 0.093% | 0.869 ± 0.54% |
| 0.40 | SRPT | 1564.4 ± 0.083% | 2821.2 ± 0.2% | 236239.4 ± 6.3% | 0.909 ± 0.35% | 0.999 ± 0.087% | 0.909 ± 0.35% |
| 0.50 | FF | 1890.3 ± 0.82% | 6780.9 ± 4.0% | 309771.9 ± 13.0% | 0.936 ± 0.71% | 0.999 ± 0.036% | 0.936 ± 0.71% |
| 0.50 | FS | 1624.6 ± 0.76% | 3202.1 ± 3.5% | 263314.3 ± 5.5% | 0.924 ± 0.54% | 0.999 ± 0.03% | 0.924 ± 0.54% |
| 0.50 | Rand | 1921.4 ± 0.49% | 6121.4 ± 1.8% | 344062.4 ± 11.0% | 0.865 ± 0.87% | 0.998 ± 0.036% | 0.865 ± 0.87% |
| 0.50 | SRPT | 1590.1 ± 0.1% | 3024.8 ± 0.66% | 345835.9 ± 12.0% | 0.912 ± 0.45% | 0.999 ± 0.031% | 0.912 ± 0.45% |
| 0.60 | FF | 2228.0 ± 1.3% | 11127.2 ± 4.4% | 325509.6 ± 4.4% | 0.941 ± 0.43% | 0.998 ± 0.063% | 0.941 ± 0.43% |
| 0.60 | FS | 1619.8 ± 0.32% | 3642.9 ± 1.6% | 278038.2 ± 4.2% | 0.935 ± 0.29% | 0.999 ± 0.041% | 0.935 ± 0.29% |
| 0.60 | Rand | 2611.0 ± 3.2% | 11568.8 ± 5.7% | 414642.6 ± 4.4% | 0.839 ± 0.76% | 0.996 ± 0.064% | 0.839 ± 0.76% |
| 0.60 | SRPT | 1634.8 ± 0.17% | 3676.2 ± 1.4% | 310853.9 ± 6.2% | 0.915 ± 0.3% | 0.999 ± 0.044% | 0.915 ± 0.3% |
| 0.70 | FF | 2875.4 ± 1.1% | 18523.2 ± 3.0% | 278140.2 ± 6.7% | 0.932 ± 0.48% | 0.996 ± 0.055% | 0.932 ± 0.48% |
| 0.70 | FS | 1653.5 ± 0.21% | 4697.6 ± 0.6% | 230876.0 ± 2.8% | 0.935 ± 0.28% | 0.999 ± 0.042% | 0.935 ± 0.28% |
| 0.70 | Rand | 4114.6 ± 1.3% | 22245.1 ± 2.0% | 368369.0 ± 2.3% | 0.784 ± 0.92% | 0.99 ± 0.13% | 0.784 ± 0.92% |
| 0.70 | SRPT | 1719.6 ± 0.24% | 4455.5 ± 1.8% | 254458.0 ± 2.9% | 0.904 ± 0.36% | 0.999 ± 0.044% | 0.904 ± 0.36% |
| 0.80 | FF | 4161.5 ± 3.0% | 32209.9 ± 5.0% | 290395.7 ± 3.7% | 0.908 ± 0.36% | 0.989 ± 0.055% | 0.908 ± 0.36% |
| 0.80 | FS | 1754.4 ± 0.31% | 7051.4 ± 1.7% | 270181.8 ± 3.0% | 0.9 ± 0.21% | 0.998 ± 0.081% | 0.9 ± 0.21% |
| 0.80 | Rand | 5293.0 ± 2.7% | 29396.4 ± 3.3% | 306156.3 ± 1.7% | 0.719 ± 1.1% | 0.978 ± 0.095% | 0.719 ± 1.1% |
| 0.80 | SRPT | 1862.6 ± 0.57% | 6197.4 ± 2.1% | 296322.5 ± 3.7% | 0.858 ± 0.27% | 0.997 ± 0.088% | 0.858 ± 0.27% |
| 0.89 | FF | 6157.7 ± 1.4% | 49317.2 ± 1.9% | 281000.4 ± 3.5% | 0.862 ± 0.28% | 0.979 ± 0.12% | 0.862 ± 0.28% |
| 0.89 | FS | 1936.4 ± 0.45% | 11082.2 ± 2.2% | 272673.6 ± 2.7% | 0.831 ± 0.44% | 0.997 ± 0.071% | 0.831 ± 0.44% |
| 0.89 | Rand | 7365.4 ± 0.86% | 41371.3 ± 1.0% | 297468.5 ± 1.7% | 0.639 ± 0.88% | 0.964 ± 0.12% | 0.639 ± 0.88% |
| 0.89 | SRPT | 2185.8 ± 3.7% | 11390.2 ± 12.0% | 265498.8 ± 1.9% | 0.726 ± 0.49% | 0.996 ± 0.086% | 0.726 ± 0.49% |

**Table F.11**
Scheduler performance summary with 95% confidence intervals for the **skewed_nodes_sensitivity_0.2** benchmark.

| Load | Subject | Mean FCT ($\mu s$) | p99 FCT ($\mu s$) | Max FCT ($\mu s$) | Throughput (Frac) | Flows Accepted (Frac) | Info Accepted (Frac) |
|---|---|---|---|---|---|---|---|
| 0.10 | FF | 1555.2 ± 0.21% | 2960.3 ± 0.45% | 63958.8 ± 7.2% | 0.995 ± 0.078% | 1.0 ± 0.026% | 0.995 ± 0.078% |
| 0.10 | FS | 1518.9 ± 0.18% | 1997.5 ± 0.0059% | 53307.4 ± 6.3% | 0.995 ± 0.11% | 1.0 ± 0.025% | 0.995 ± 0.11% |
| 0.10 | Rand | 1544.8 ± 0.22% | 2746.4 ± 0.76% | 80003.7 ± 8.1% | 0.993 ± 0.11% | 1.0 ± 0.025% | 0.993 ± 0.11% |

(*continued on next page*)

**Table F.11** (*continued*)

| Load | Subject | Mean FCT ($\mu s$) | p99 FCT ($\mu s$) | Max FCT ($\mu s$) | Throughput (Frac) | Flows Accepted (Frac) | Info Accepted (Frac) |
|---|---|---|---|---|---|---|---|
| 0.10 | SRPT | 1515.5 ± 0.18% | 1997.1 ± 0.0039% | 55035.6 ± 6.0% | 0.996 ± 0.068% | 1.0 ± 0.025% | 0.996 ± 0.068% |
| 0.20 | FF | 1653.6 ± 0.43% | 4948.5 ± 3.1% | 100796.4 ± 7.0% | 0.98 ± 0.39% | 0.999 ± 0.063% | 0.98 ± 0.39% |
| 0.20 | FS | 1538.1 ± 0.21% | 2078.9 ± 1.6% | 86478.5 ± 8.6% | 0.978 ± 0.44% | 0.999 ± 0.059% | 0.978 ± 0.44% |
| 0.20 | Rand | 1604.6 ± 0.33% | 3121.7 ± 1.8% | 210697.6 ± 5.6% | 0.937 ± 0.7% | 0.999 ± 0.06% | 0.937 ± 0.7% |
| 0.20 | SRPT | 1529.4 ± 0.1% | 2199.9 ± 0.86% | 107139.2 ± 10.0% | 0.979 ± 0.48% | 0.999 ± 0.058% | 0.979 ± 0.48% |
| 0.30 | FF | 1879.0 ± 1.4% | 9864.3 ± 5.0% | 219455.7 ± 9.2% | 0.968 ± 0.39% | 0.998 ± 0.059% | 0.968 ± 0.39% |
| 0.30 | FS | 1587.8 ± 0.47% | 2934.7 ± 2.7% | 165814.7 ± 9.3% | 0.965 ± 0.38% | 0.999 ± 0.06% | 0.965 ± 0.38% |
| 0.30 | Rand | 1688.9 ± 0.27% | 4259.1 ± 1.6% | 388493.8 ± 6.6% | 0.811 ± 1.2% | 0.998 ± 0.063% | 0.811 ± 1.2% |
| 0.30 | SRPT | 1555.3 ± 0.14% | 2580.5 ± 0.5% | 318733.6 ± 8.1% | 0.947 ± 0.66% | 0.999 ± 0.061% | 0.947 ± 0.66% |
| 0.40 | FF | 2047.8 ± 1.9% | 12996.0 ± 8.4% | 253351.7 ± 3.1% | 0.901 ± 0.99% | 0.997 ± 0.077% | 0.901 ± 0.99% |
| 0.40 | FS | 1656.3 ± 0.3% | 4469.8 ± 0.9% | 237826.1 ± 6.6% | 0.901 ± 0.88% | 0.999 ± 0.058% | 0.901 ± 0.88% |
| 0.40 | Rand | 1750.5 ± 0.46% | 5216.2 ± 1.9% | 272425.1 ± 5.7% | 0.774 ± 0.66% | 0.997 ± 0.059% | 0.774 ± 0.66% |
| 0.40 | SRPT | 1565.6 ± 0.083% | 2783.8 ± 0.28% | 235162.6 ± 6.6% | 0.88 ± 0.78% | 0.999 ± 0.058% | 0.88 ± 0.78% |
| 0.50 | FF | 1893.6 ± 1.5% | 6355.4 ± 4.8% | 440695.9 ± 8.9% | 0.887 ± 0.41% | 0.998 ± 0.073% | 0.887 ± 0.41% |
| 0.50 | FS | 1752.4 ± 1.1% | 5396.1 ± 12.0% | 331678.2 ± 8.6% | 0.888 ± 1.2% | 0.999 ± 0.052% | 0.888 ± 1.2% |
| 0.50 | Rand | 1941.6 ± 1.1% | 6437.1 ± 3.0% | 458290.7 ± 9.6% | 0.795 ± 0.49% | 0.998 ± 0.056% | 0.795 ± 0.49% |
| 0.50 | SRPT | 1588.2 ± 0.059% | 2940.4 ± 0.42% | 415335.3 ± 7.7% | 0.856 ± 0.52% | 0.999 ± 0.056% | 0.856 ± 0.52% |
| 0.61 | FF | 1981.7 ± 0.88% | 7326.2 ± 3.0% | 372958.5 ± 3.3% | 0.901 ± 0.21% | 0.998 ± 0.043% | 0.901 ± 0.21% |
| 0.61 | FS | 1692.9 ± 0.47% | 3992.5 ± 2.3% | 297476.6 ± 4.3% | 0.898 ± 0.22% | 0.999 ± 0.037% | 0.898 ± 0.22% |
| 0.61 | Rand | 2203.6 ± 1.2% | 8062.8 ± 2.4% | 407016.8 ± 2.1% | 0.801 ± 0.43% | 0.997 ± 0.043% | 0.801 ± 0.43% |
| 0.61 | SRPT | 1638.4 ± 0.15% | 3706.4 ± 0.78% | 327127.1 ± 6.4% | 0.863 ± 0.25% | 0.999 ± 0.039% | 0.863 ± 0.25% |
| 0.70 | FF | 2412.4 ± 0.75% | 12132.2 ± 2.0% | 307320.4 ± 1.3% | 0.897 ± 0.33% | 0.997 ± 0.054% | 0.897 ± 0.33% |
| 0.70 | FS | 1671.6 ± 0.3% | 4565.0 ± 1.8% | 292849.3 ± 3.5% | 0.906 ± 0.18% | 0.999 ± 0.051% | 0.906 ± 0.18% |
| 0.70 | Rand | 3156.7 ± 0.98% | 15098.8 ± 1.5% | 369120.8 ± 2.5% | 0.782 ± 0.31% | 0.993 ± 0.057% | 0.782 ± 0.31% |
| 0.70 | SRPT | 1756.4 ± 0.26% | 5157.5 ± 1.5% | 326751.6 ± 4.6% | 0.862 ± 0.16% | 0.999 ± 0.053% | 0.862 ± 0.16% |
| 0.80 | FF | 3541.7 ± 0.85% | 24415.3 ± 1.9% | 304075.1 ± 3.6% | 0.892 ± 0.17% | 0.993 ± 0.075% | 0.892 ± 0.17% |
| 0.80 | FS | 1731.9 ± 0.23% | 6430.1 ± 1.5% | 234881.5 ± 4.1% | 0.901 ± 0.21% | 0.999 ± 0.061% | 0.901 ± 0.21% |
| 0.80 | Rand | 5311.2 ± 3.2% | 30099.0 ± 3.7% | 329220.0 ± 1.8% | 0.728 ± 1.2% | 0.98 ± 0.12% | 0.728 ± 1.2% |
| 0.80 | SRPT | 2006.7 ± 0.79% | 8444.4 ± 3.0% | 291953.9 ± 5.8% | 0.833 ± 0.37% | 0.998 ± 0.059% | 0.833 ± 0.37% |
| 0.90 | FF | 6282.4 ± 3.1% | 51863.0 ± 4.5% | 361626.4 ± 12.0% | 0.876 ± 0.5% | 0.982 ± 0.17% | 0.876 ± 0.5% |
| 0.90 | FS | 2051.1 ± 0.77% | 13365.0 ± 2.2% | 340927.7 ± 9.7% | 0.869 ± 1.1% | 0.997 ± 0.092% | 0.869 ± 1.1% |
| 0.90 | Rand | 9434.3 ± 8.7% | 55751.7 ± 9.7% | 394053.6 ± 12.0% | 0.677 ± 0.61% | 0.962 ± 0.14% | 0.677 ± 0.61% |
| 0.90 | SRPT | 2864.2 ± 4.1% | 23077.6 ± 10.0% | 394099.5 ± 12.0% | 0.731 ± 1.4% | 0.987 ± 0.23% | 0.731 ± 1.4% |

**Table F.12**
Scheduler performance summary with 95% confidence intervals for the **skewed_nodes_sensitivity_0.4** benchmark.

| Load | Subject | Mean FCT ($\mu s$) | p99 FCT ($\mu s$) | Max FCT ($\mu s$) | Throughput (Frac) | Flows Accepted (Frac) | Info Accepted (Frac) |
|---|---|---|---|---|---|---|---|
| 0.10 | FF | 1550.7 ± 0.14% | 2940.7 ± 0.61% | 41390.8 ± 6.7% | 0.997 ± 0.072% | 1.0 ± 0.018% | 0.997 ± 0.072% |
| 0.10 | FS | 1516.9 ± 0.13% | 1997.5 ± 0.0078% | 40765.3 ± 4.2% | 0.997 ± 0.07% | 1.0 ± 0.018% | 0.997 ± 0.07% |
| 0.10 | Rand | 1542.2 ± 0.16% | 2746.4 ± 0.23% | 62404.4 ± 12.0% | 0.997 ± 0.082% | 1.0 ± 0.017% | 0.997 ± 0.082% |
| 0.10 | SRPT | 1516.3 ± 0.13% | 1997.9 ± 0.0098% | 41765.3 ± 4.4% | 0.997 ± 0.07% | 1.0 ± 0.018% | 0.997 ± 0.07% |
| 0.20 | FF | 1626.3 ± 0.32% | 4422.1 ± 2.7% | 55331.8 ± 5.9% | 0.98 ± 0.11% | 0.999 ± 0.044% | 0.98 ± 0.11% |
| 0.20 | FS | 1527.7 ± 0.11% | 2008.0 ± 0.33% | 46606.0 ± 7.2% | 0.981 ± 0.11% | 0.999 ± 0.045% | 0.981 ± 0.11% |
| 0.20 | Rand | 1582.2 ± 0.082% | 2999.5 ± 0.33% | 98692.2 ± 7.2% | 0.961 ± 0.22% | 0.999 ± 0.044% | 0.961 ± 0.22% |
| 0.20 | SRPT | 1528.6 ± 0.12% | 2280.1 ± 0.98% | 53343.4 ± 7.0% | 0.983 ± 0.22% | 0.999 ± 0.045% | 0.983 ± 0.22% |
| 0.30 | FF | 1748.0 ± 0.97% | 6884.9 ± 6.0% | 70468.9 ± 9.3% | 0.99 ± 0.19% | 0.999 ± 0.084% | 0.99 ± 0.19% |
| 0.30 | FS | 1534.4 ± 0.1% | 2385.5 ± 0.78% | 68968.2 ± 8.7% | 0.991 ± 0.2% | 0.999 ± 0.084% | 0.991 ± 0.2% |
| 0.30 | Rand | 1662.0 ± 0.58% | 3976.1 ± 1.9% | 232725.1 ± 11.0% | 0.972 ± 0.47% | 0.999 ± 0.085% | 0.972 ± 0.47% |
| 0.30 | SRPT | 1538.9 ± 0.064% | 2658.7 ± 0.41% | 85328.2 ± 9.3% | 0.991 ± 0.18% | 0.999 ± 0.084% | 0.991 ± 0.18% |
| 0.40 | FF | 1940.0 ± 0.89% | 9772.0 ± 3.0% | 88904.4 ± 3.9% | 0.981 ± 0.23% | 0.998 ± 0.086% | 0.981 ± 0.23% |
| 0.40 | FS | 1552.0 ± 0.17% | 2718.8 ± 0.66% | 81504.4 ± 5.4% | 0.983 ± 0.25% | 0.999 ± 0.082% | 0.983 ± 0.25% |
| 0.40 | Rand | 1836.8 ± 0.6% | 5756.4 ± 1.2% | 274773.4 ± 3.7% | 0.908 ± 0.39% | 0.998 ± 0.085% | 0.908 ± 0.39% |
| 0.40 | SRPT | 1561.4 ± 0.11% | 2844.9 ± 0.29% | 111871.6 ± 4.0% | 0.981 ± 0.24% | 0.999 ± 0.082% | 0.981 ± 0.24% |
| 0.51 | FF | 2329.1 ± 1.4% | 16228.3 ± 6.2% | 218249.3 ± 18.0% | 0.97 ± 0.62% | 0.997 ± 0.1% | 0.97 ± 0.62% |
| 0.51 | FS | 1576.2 ± 0.19% | 3237.0 ± 1.5% | 120960.8 ± 6.1% | 0.972 ± 0.6% | 0.999 ± 0.078% | 0.972 ± 0.6% |
| 0.51 | Rand | 2429.3 ± 3.3% | 11991.0 ± 7.5% | 422835.2 ± 8.8% | 0.826 ± 0.52% | 0.995 ± 0.089% | 0.826 ± 0.52% |
| 0.51 | SRPT | 1592.1 ± 0.16% | 2987.7 ± 0.52% | 263110.8 ± 12.0% | 0.967 ± 0.66% | 0.999 ± 0.078% | 0.967 ± 0.66% |
| 0.60 | FF | 2939.0 ± 2.1% | 23736.5 ± 4.0% | 343896.8 ± 6.1% | 0.948 ± 0.36% | 0.996 ± 0.045% | 0.948 ± 0.36% |
| 0.60 | FS | 1633.6 ± 0.4% | 4389.9 ± 2.1% | 258643.0 ± 5.9% | 0.959 ± 0.32% | 0.999 ± 0.043% | 0.959 ± 0.32% |
| 0.60 | Rand | 3201.2 ± 2.1% | 19085.7 ± 4.6% | 436718.7 ± 1.5% | 0.766 ± 1.7% | 0.993 ± 0.055% | 0.766 ± 1.7% |
| 0.60 | SRPT | 1632.6 ± 0.094% | 3514.9 ± 1.3% | 323235.8 ± 4.6% | 0.934 ± 0.47% | 0.999 ± 0.044% | 0.934 ± 0.47% |
| 0.71 | FF | 3837.7 ± 1.9% | 34431.1 ± 4.0% | 322903.8 ± 2.5% | 0.911 ± 0.37% | 0.992 ± 0.084% | 0.911 ± 0.37% |
| 0.71 | FS | 1730.6 ± 0.38% | 6601.9 ± 1.2% | 274442.8 ± 2.9% | 0.922 ± 0.39% | 0.999 ± 0.05% | 0.922 ± 0.39% |
| 0.71 | Rand | 3911.6 ± 1.4% | 24538.5 ± 3.2% | 381889.7 ± 2.0% | 0.731 ± 1.2% | 0.989 ± 0.076% | 0.731 ± 1.2% |
| 0.71 | SRPT | 1706.2 ± 0.21% | 4321.1 ± 1.7% | 365187.1 ± 2.6% | 0.886 ± 0.35% | 0.999 ± 0.054% | 0.886 ± 0.35% |
| 0.80 | FF | 4505.2 ± 3.5% | 40048.5 ± 6.2% | 297883.9 ± 2.8% | 0.854 ± 0.38% | 0.985 ± 0.18% | 0.854 ± 0.38% |
| 0.80 | FS | 1843.6 ± 0.83% | 9336.0 ± 3.2% | 284147.5 ± 4.5% | 0.856 ± 0.62% | 0.997 ± 0.095% | 0.856 ± 0.62% |
| 0.80 | Rand | 4761.7 ± 2.7% | 28060.7 ± 2.3% | 315479.6 ± 2.5% | 0.694 ± 0.71% | 0.982 ± 0.12% | 0.694 ± 0.71% |
| 0.80 | SRPT | 1807.6 ± 0.51% | 5691.1 ± 2.9% | 275652.6 ± 5.5% | 0.819 ± 0.61% | 0.998 ± 0.1% | 0.819 ± 0.61% |
| 0.89 | FF | 5277.0 ± 2.7% | 49286.4 ± 4.5% | 301906.9 ± 1.8% | 0.814 ± 0.44% | 0.97 ± 0.43% | 0.814 ± 0.44% |
| 0.89 | FS | 2042.1 ± 0.48% | 14036.0 ± 2.1% | 273754.2 ± 3.0% | 0.79 ± 0.55% | 0.996 ± 0.11% | 0.79 ± 0.55% |
| 0.89 | Rand | 7441.6 ± 3.5% | 41471.0 ± 2.8% | 294162.8 ± 1.2% | 0.633 ± 1.2% | 0.964 ± 0.2% | 0.633 ± 1.2% |
| 0.89 | SRPT | 2271.4 ± 3.0% | 14379.1 ± 11.0% | 294316.1 ± 2.1% | 0.746 ± 0.67% | 0.992 ± 0.29% | 0.746 ± 0.67% |

*Appendix F.3.3. Rack Distribution Benchmark*

**Table F.13**
Scheduler performance summary with 95% confidence intervals for the **rack_sensitivity_0.2** benchmark.

| Load | Subject | Mean FCT ($\mu$s) | p99 FCT ($\mu$s) | Max FCT ($\mu$s) | Throughput (Frac) | Flows Accepted (Frac) | Info Accepted (Frac) |
|---|---|---|---|---|---|---|---|
| 0.1 | FF | 1547.1 ± 0.2% | 2905.6 ± 0.53% | 36420.2 ± 3.8% | 0.993 ± 0.23% | 0.999 ± 0.047% | 0.993 ± 0.23% |
| 0.1 | FS | 1514.8 ± 0.14% | 1997.1 ± 0.0059% | 35026.4 ± 2.6% | 0.994 ± 0.23% | 0.999 ± 0.046% | 0.994 ± 0.23% |
| 0.1 | Rand | 1538.9 ± 0.15% | 2708.2 ± 0.35% | 53118.6 ± 2.9% | 0.991 ± 0.21% | 0.999 ± 0.046% | 0.991 ± 0.21% |
| 0.1 | SRPT | 1514.5 ± 0.14% | 1997.6 ± 0.0078% | 35426.4 ± 2.6% | 0.994 ± 0.22% | 0.999 ± 0.046% | 0.994 ± 0.22% |
| 0.2 | FF | 1613.3 ± 0.16% | 4210.8 ± 1.1% | 43491.3 ± 3.3% | 0.985 ± 0.34% | 0.999 ± 0.032% | 0.985 ± 0.34% |
| 0.2 | FS | 1522.9 ± 0.14% | 1998.9 ± 0.012% | 38988.6 ± 1.6% | 0.986 ± 0.37% | 1.0 ± 0.029% | 0.986 ± 0.37% |
| 0.2 | Rand | 1575.6 ± 0.2% | 3009.0 ± 0.77% | 70182.4 ± 2.6% | 0.978 ± 0.45% | 0.999 ± 0.029% | 0.978 ± 0.45% |
| 0.2 | SRPT | 1524.5 ± 0.14% | 2252.2 ± 1.0% | 41095.6 ± 1.6% | 0.987 ± 0.31% | 1.0 ± 0.028% | 0.987 ± 0.31% |
| 0.3 | FF | 1751.4 ± 0.79% | 6744.9 ± 3.8% | 67480.0 ± 9.5% | 0.989 ± 0.17% | 0.999 ± 0.042% | 0.989 ± 0.17% |
| 0.3 | FS | 1534.7 ± 0.13% | 2247.8 ± 0.56% | 63424.2 ± 7.2% | 0.99 ± 0.17% | 0.999 ± 0.042% | 0.99 ± 0.17% |
| 0.3 | Rand | 1649.2 ± 0.34% | 3891.9 ± 0.76% | 148433.8 ± 8.6% | 0.977 ± 0.19% | 0.999 ± 0.041% | 0.977 ± 0.19% |
| 0.3 | SRPT | 1539.5 ± 0.12% | 2626.4 ± 0.49% | 83252.7 ± 9.4% | 0.989 ± 0.2% | 0.999 ± 0.041% | 0.989 ± 0.2% |
| 0.4 | FF | 1924.1 ± 1.5% | 9755.3 ± 7.2% | 88414.1 ± 9.8% | 0.977 ± 0.23% | 0.998 ± 0.086% | 0.977 ± 0.23% |
| 0.4 | FS | 1541.5 ± 0.092% | 2542.4 ± 0.5% | 74926.1 ± 11.0% | 0.98 ± 0.19% | 0.999 ± 0.085% | 0.98 ± 0.19% |
| 0.4 | Rand | 1795.0 ± 0.49% | 5339.0 ± 1.4% | 216058.0 ± 7.5% | 0.941 ± 0.46% | 0.998 ± 0.089% | 0.941 ± 0.46% |
| 0.4 | SRPT | 1552.2 ± 0.035% | 2802.5 ± 0.34% | 99179.4 ± 14.0% | 0.979 ± 0.2% | 0.999 ± 0.085% | 0.979 ± 0.2% |
| 0.5 | FF | 2239.7 ± 2.0% | 14440.3 ± 7.6% | 120877.0 ± 5.2% | 0.979 ± 0.27% | 0.998 ± 0.048% | 0.979 ± 0.27% |
| 0.5 | FS | 1564.2 ± 0.13% | 2914.5 ± 0.6% | 97264.9 ± 6.6% | 0.98 ± 0.21% | 0.999 ± 0.05% | 0.98 ± 0.21% |
| 0.5 | Rand | 2330.3 ± 1.7% | 9746.8 ± 4.0% | 408828.1 ± 10.0% | 0.892 ± 1.0% | 0.997 ± 0.055% | 0.892 ± 1.0% |
| 0.5 | SRPT | 1580.9 ± 0.082% | 2940.4 ± 0.36% | 153416.6 ± 10.0% | 0.978 ± 0.3% | 0.999 ± 0.051% | 0.978 ± 0.3% |
| 0.6 | FF | 2842.5 ± 2.5% | 22991.2 ± 7.0% | 308474.4 ± 6.6% | 0.967 ± 0.28% | 0.996 ± 0.067% | 0.967 ± 0.28% |
| 0.6 | FS | 1595.6 ± 0.19% | 3658.7 ± 0.66% | 137386.1 ± 5.7% | 0.972 ± 0.28% | 0.999 ± 0.048% | 0.972 ± 0.28% |
| 0.6 | Rand | 3265.3 ± 0.75% | 16613.3 ± 1.2% | 420951.4 ± 3.7% | 0.825 ± 0.81% | 0.994 ± 0.065% | 0.825 ± 0.81% |
| 0.6 | SRPT | 1619.1 ± 0.097% | 3390.5 ± 1.8% | 336922.0 ± 5.1% | 0.961 ± 0.37% | 0.999 ± 0.049% | 0.961 ± 0.37% |
| 0.7 | FF | 3465.2 ± 0.49% | 27554.1 ± 1.9% | 287240.4 ± 6.0% | 0.95 ± 0.29% | 0.994 ± 0.066% | 0.95 ± 0.29% |
| 0.7 | FS | 1648.7 ± 0.15% | 4775.0 ± 1.0% | 210756.3 ± 2.1% | 0.95 ± 0.28% | 0.999 ± 0.063% | 0.95 ± 0.28% |
| 0.7 | Rand | 4658.5 ± 2.2% | 25482.9 ± 2.8% | 345529.5 ± 3.4% | 0.755 ± 0.84% | 0.985 ± 0.055% | 0.755 ± 0.84% |
| 0.7 | SRPT | 1678.0 ± 0.25% | 3916.3 ± 1.2% | 307069.7 ± 3.6% | 0.927 ± 0.29% | 0.999 ± 0.063% | 0.927 ± 0.29% |
| 0.8 | FF | 4604.8 ± 1.5% | 37588.2 ± 2.3% | 287174.0 ± 1.8% | 0.904 ± 0.52% | 0.988 ± 0.11% | 0.904 ± 0.52% |
| 0.8 | FS | 1759.3 ± 0.18% | 7189.5 ± 1.6% | 278549.2 ± 2.2% | 0.886 ± 0.5% | 0.998 ± 0.071% | 0.886 ± 0.5% |
| 0.8 | Rand | 5891.2 ± 0.77% | 32310.3 ± 1.1% | 323761.6 ± 2.1% | 0.694 ± 1.3% | 0.977 ± 0.11% | 0.694 ± 1.3% |
| 0.8 | SRPT | 1757.8 ± 0.7% | 4908.0 ± 3.1% | 307367.1 ± 4.4% | 0.853 ± 0.39% | 0.998 ± 0.073% | 0.853 ± 0.39% |
| 0.9 | FF | 6385.1 ± 2.0% | 52863.8 ± 3.2% | 320436.3 ± 9.5% | 0.871 ± 0.69% | 0.98 ± 0.041% | 0.871 ± 0.69% |
| 0.9 | FS | 1956.4 ± 1.1% | 11288.2 ± 3.1% | 313425.0 ± 11.0% | 0.845 ± 0.79% | 0.998 ± 0.034% | 0.845 ± 0.79% |
| 0.9 | Rand | 8399.6 ± 6.6% | 46907.3 ± 7.9% | 336830.9 ± 12.0% | 0.65 ± 0.81% | 0.964 ± 0.082% | 0.65 ± 0.81% |
| 0.9 | SRPT | 1963.3 ± 1.2% | 7596.5 ± 5.1% | 320009.5 ± 13.0% | 0.786 ± 0.3% | 0.998 ± 0.042% | 0.786 ± 0.3% |

**Table F.14**
Scheduler performance summary with 95% confidence intervals for the **rack_sensitivity_0.4** benchmark.

| Load | Subject | Mean FCT ($\mu$s) | p99 FCT ($\mu$s) | Max FCT ($\mu$s) | Throughput (Frac) | Flows Accepted (Frac) | Info Accepted (Frac) |
|---|---|---|---|---|---|---|---|
| 0.1 | FF | 1553.9 ± 0.15% | 3023.9 ± 1.1% | 38020.6 ± 5.6% | 0.997 ± 0.15% | 1.0 ± 0.029% | 0.997 ± 0.15% |
| 0.1 | FS | 1515.8 ± 0.057% | 1997.5 ± 0.0039% | 37020.6 ± 5.8% | 0.997 ± 0.15% | 1.0 ± 0.026% | 0.997 ± 0.15% |
| 0.1 | Rand | 1541.3 ± 0.058% | 2739.1 ± 0.31% | 61994.1 ± 11.0% | 0.996 ± 0.16% | 1.0 ± 0.026% | 0.996 ± 0.16% |
| 0.1 | SRPT | 1515.5 ± 0.052% | 1998.0 ± 0.0078% | 37373.3 ± 6.5% | 0.997 ± 0.15% | 1.0 ± 0.026% | 0.997 ± 0.15% |
| 0.2 | FF | 1643.8 ± 0.34% | 4775.7 ± 2.4% | 52879.5 ± 3.8% | 0.986 ± 0.2% | 0.999 ± 0.04% | 0.986 ± 0.2% |
| 0.2 | FS | 1525.7 ± 0.11% | 1999.3 ± 0.0059% | 48949.2 ± 3.9% | 0.987 ± 0.18% | 0.999 ± 0.043% | 0.987 ± 0.18% |
| 0.2 | Rand | 1587.8 ± 0.18% | 3035.9 ± 0.46% | 126408.0 ± 8.0% | 0.976 ± 0.14% | 0.999 ± 0.044% | 0.976 ± 0.14% |
| 0.2 | SRPT | 1526.4 ± 0.12% | 2225.3 ± 1.5% | 51165.4 ± 4.0% | 0.988 ± 0.17% | 0.999 ± 0.044% | 0.988 ± 0.17% |
| 0.3 | FF | 1787.0 ± 0.46% | 7619.8 ± 3.9% | 66882.6 ± 8.2% | 0.988 ± 0.16% | 0.999 ± 0.016% | 0.988 ± 0.16% |
| 0.3 | FS | 1532.2 ± 0.15% | 2231.4 ± 0.81% | 57004.1 ± 7.0% | 0.989 ± 0.21% | 1.0 ± 0.015% | 0.989 ± 0.21% |
| 0.3 | Rand | 1671.2 ± 0.57% | 4113.2 ± 2.9% | 256001.0 ± 7.5% | 0.956 ± 0.31% | 0.999 ± 0.015% | 0.956 ± 0.31% |
| 0.3 | SRPT | 1536.4 ± 0.17% | 2610.9 ± 0.51% | 65648.0 ± 10.0% | 0.989 ± 0.16% | 1.0 ± 0.016% | 0.989 ± 0.16% |
| 0.4 | FF | 1997.7 ± 0.57% | 11546.2 ± 2.4% | 78798.6 ± 6.6% | 0.973 ± 0.27% | 0.998 ± 0.054% | 0.973 ± 0.27% |
| 0.4 | FS | 1542.8 ± 0.11% | 2588.6 ± 0.51% | 66608.7 ± 3.7% | 0.976 ± 0.23% | 0.999 ± 0.065% | 0.976 ± 0.23% |
| 0.4 | Rand | 1805.3 ± 0.55% | 6476.5 ± 3.3% | 287594.6 ± 1.8% | 0.882 ± 0.51% | 0.998 ± 0.074% | 0.882 ± 0.51% |
| 0.4 | SRPT | 1553.2 ± 0.061% | 2820.2 ± 0.21% | 85975.8 ± 3.3% | 0.977 ± 0.22% | 0.999 ± 0.066% | 0.977 ± 0.22% |
| 0.5 | FF | 2476.5 ± 1.5% | 20978.2 ± 5.7% | 115951.9 ± 3.9% | 0.976 ± 0.45% | 0.997 ± 0.053% | 0.976 ± 0.45% |
| 0.5 | FS | 1562.6 ± 0.05% | 2906.2 ± 0.44% | 104707.6 ± 4.4% | 0.978 ± 0.38% | 0.999 ± 0.046% | 0.978 ± 0.38% |
| 0.5 | Rand | 2104.8 ± 2.3% | 11901.7 ± 9.0% | 411058.5 ± 11.0% | 0.822 ± 0.74% | 0.996 ± 0.041% | 0.822 ± 0.74% |
| 0.5 | SRPT | 1578.2 ± 0.12% | 2936.4 ± 0.35% | 128711.3 ± 3.1% | 0.976 ± 0.36% | 0.999 ± 0.044% | 0.976 ± 0.36% |
| 0.6 | FF | 2880.0 ± 1.5% | 24414.7 ± 5.0% | 242585.7 ± 3.1% | 0.971 ± 0.31% | 0.997 ± 0.044% | 0.971 ± 0.31% |
| 0.6 | FS | 1592.5 ± 0.094% | 3616.1 ± 0.56% | 131921.2 ± 7.3% | 0.972 ± 0.22% | 0.999 ± 0.031% | 0.972 ± 0.22% |
| 0.6 | Rand | 2420.9 ± 0.64% | 17877.9 ± 2.6% | 417817.1 ± 1.5% | 0.778 ± 1.0% | 0.995 ± 0.043% | 0.778 ± 1.0% |
| 0.6 | SRPT | 1619.6 ± 0.11% | 3401.4 ± 1.4% | 235066.7 ± 3.8% | 0.966 ± 0.35% | 0.999 ± 0.033% | 0.966 ± 0.35% |
| 0.7 | FF | 3534.3 ± 1.3% | 33314.7 ± 4.1% | 311692.8 ± 3.6% | 0.935 ± 0.27% | 0.994 ± 0.052% | 0.935 ± 0.27% |
| 0.7 | FS | 1642.8 ± 0.05% | 4665.5 ± 1.2% | 245550.4 ± 4.2% | 0.937 ± 0.3% | 0.999 ± 0.057% | 0.937 ± 0.3% |

(*continued on next page*)

**Table F.14** (*continued*)

| Load | Subject | Mean FCT (μs) | p99 FCT (μs) | Max FCT (μs) | Throughput (Frac) | Flows Accepted (Frac) | Info Accepted (Frac) |
|---|---|---|---|---|---|---|---|
| 0.7 | Rand | 2652.7 ± 1.1% | 21768.7 ± 1.6% | 375157.8 ± 2.3% | 0.71 ± 0.89% | 0.993 ± 0.071% | 0.71 ± 0.89% |
| 0.7 | SRPT | 1660.0 ± 0.2% | 3780.8 ± 1.1% | 327831.0 ± 2.0% | 0.915 ± 0.3% | 0.999 ± 0.059% | 0.915 ± 0.3% |
| 0.8 | FF | 4311.5 ± 0.84% | 39028.2 ± 2.3% | 294072.3 ± 2.3% | 0.911 ± 0.39% | 0.99 ± 0.056% | 0.911 ± 0.39% |
| 0.8 | FS | 1731.1 ± 0.25% | 6579.1 ± 1.4% | 238819.8 ± 2.4% | 0.904 ± 0.33% | 0.999 ± 0.047% | 0.904 ± 0.33% |
| 0.8 | Rand | 2906.5 ± 0.56% | 24944.4 ± 1.5% | 320552.6 ± 1.3% | 0.665 ± 1.4% | 0.99 ± 0.063% | 0.665 ± 1.4% |
| 0.8 | SRPT | 1747.3 ± 0.58% | 4642.0 ± 2.8% | 280469.5 ± 6.0% | 0.865 ± 0.3% | 0.999 ± 0.05% | 0.865 ± 0.3% |
| 0.9 | FF | 5497.1 ± 2.4% | 46230.6 ± 2.7% | 280463.7 ± 3.8% | 0.851 ± 0.43% | 0.983 ± 0.086% | 0.851 ± 0.43% |
| 0.9 | FS | 1872.4 ± 0.26% | 9593.3 ± 1.6% | 290850.1 ± 1.7% | 0.827 ± 0.43% | 0.997 ± 0.094% | 0.827 ± 0.43% |
| 0.9 | Rand | 3347.1 ± 1.2% | 29914.9 ± 0.63% | 306219.5 ± 1.8% | 0.608 ± 1.4% | 0.987 ± 0.095% | 0.608 ± 1.4% |
| 0.9 | SRPT | 1861.4 ± 0.99% | 5931.5 ± 4.3% | 286319.1 ± 1.6% | 0.781 ± 0.33% | 0.997 ± 0.095% | 0.781 ± 0.33% |

**Table F.15**
Scheduler performance summary with 95% confidence intervals for the **rack_sensitivity_0.6** benchmark.

| Load | Subject | Mean FCT (μs) | p99 FCT (μs) | Max FCT (μs) | Throughput (Frac) | Flows Accepted (Frac) | Info Accepted (Frac) |
|---|---|---|---|---|---|---|---|
| 0.1 | FF | 1557.4 ± 0.18% | 2992.3 ± 1.3% | 41866.0 ± 3.4% | 0.995 ± 0.1% | 0.999 ± 0.045% | 0.995 ± 0.1% |
| 0.1 | FS | 1519.1 ± 0.041% | 1997.0 ± 0.012% | 41866.0 ± 2.5% | 0.995 ± 0.1% | 0.999 ± 0.044% | 0.995 ± 0.1% |
| 0.1 | Rand | 1544.3 ± 0.069% | 2738.7 ± 0.46% | 63511.4 ± 5.0% | 0.992 ± 0.23% | 0.999 ± 0.044% | 0.992 ± 0.23% |
| 0.1 | SRPT | 1518.9 ± 0.044% | 1997.6 ± 0.0098% | 42866.0 ± 3.2% | 0.995 ± 0.1% | 0.999 ± 0.044% | 0.995 ± 0.1% |
| 0.2 | FF | 1639.1 ± 0.39% | 4710.0 ± 2.0% | 48916.2 ± 3.7% | 0.989 ± 0.26% | 0.999 ± 0.039% | 0.989 ± 0.26% |
| 0.2 | FS | 1522.8 ± 0.17% | 1998.9 ± 0.0039% | 47869.5 ± 2.2% | 0.989 ± 0.27% | 0.999 ± 0.043% | 0.989 ± 0.27% |
| 0.2 | Rand | 1582.2 ± 0.16% | 3046.6 ± 0.46% | 105951.9 ± 8.2% | 0.981 ± 0.28% | 0.999 ± 0.043% | 0.981 ± 0.28% |
| 0.2 | SRPT | 1525.6 ± 0.18% | 2338.1 ± 0.83% | 46897.0 ± 4.4% | 0.99 ± 0.23% | 0.999 ± 0.043% | 0.99 ± 0.23% |
| 0.3 | FF | 1786.7 ± 0.96% | 7425.5 ± 4.7% | 64656.1 ± 7.6% | 0.986 ± 0.29% | 0.999 ± 0.047% | 0.986 ± 0.29% |
| 0.3 | FS | 1531.2 ± 0.19% | 2225.5 ± 1.8% | 57743.4 ± 4.0% | 0.987 ± 0.32% | 0.999 ± 0.046% | 0.987 ± 0.32% |
| 0.3 | Rand | 1663.9 ± 0.3% | 4087.7 ± 1.3% | 233890.5 ± 4.2% | 0.967 ± 0.37% | 0.999 ± 0.046% | 0.967 ± 0.37% |
| 0.3 | SRPT | 1538.9 ± 0.17% | 2680.8 ± 0.29% | 64343.4 ± 5.0% | 0.988 ± 0.32% | 0.999 ± 0.046% | 0.988 ± 0.32% |
| 0.4 | FF | 2070.3 ± 1.5% | 11972.7 ± 4.4% | 89212.8 ± 6.2% | 0.98 ± 0.34% | 0.997 ± 0.09% | 0.98 ± 0.34% |
| 0.4 | FS | 1543.2 ± 0.094% | 2555.3 ± 1.1% | 84834.8 ± 6.3% | 0.981 ± 0.34% | 0.999 ± 0.085% | 0.981 ± 0.34% |
| 0.4 | Rand | 1804.3 ± 0.48% | 6112.3 ± 1.9% | 210247.2 ± 7.1% | 0.917 ± 0.59% | 0.998 ± 0.088% | 0.917 ± 0.59% |
| 0.4 | SRPT | 1558.0 ± 0.083% | 2842.9 ± 0.21% | 108634.8 ± 9.3% | 0.981 ± 0.28% | 0.999 ± 0.085% | 0.981 ± 0.28% |
| 0.5 | FF | 2462.2 ± 0.96% | 18251.1 ± 3.1% | 121295.6 ± 5.9% | 0.98 ± 0.15% | 0.997 ± 0.062% | 0.98 ± 0.15% |
| 0.5 | FS | 1560.7 ± 0.05% | 2885.8 ± 0.48% | 89431.3 ± 4.1% | 0.98 ± 0.18% | 0.999 ± 0.048% | 0.98 ± 0.18% |
| 0.5 | Rand | 2236.8 ± 1.9% | 13576.8 ± 6.8% | 390033.5 ± 9.3% | 0.843 ± 0.62% | 0.997 ± 0.048% | 0.843 ± 0.62% |
| 0.5 | SRPT | 1585.1 ± 0.042% | 2970.9 ± 0.11% | 136545.8 ± 11.0% | 0.98 ± 0.17% | 0.999 ± 0.05% | 0.98 ± 0.17% |
| 0.6 | FF | 2956.1 ± 1.6% | 24090.4 ± 4.4% | 242220.9 ± 13.0% | 0.975 ± 0.13% | 0.996 ± 0.065% | 0.975 ± 0.13% |
| 0.6 | FS | 1586.0 ± 0.16% | 3517.5 ± 0.65% | 138093.6 ± 7.7% | 0.979 ± 0.1% | 0.999 ± 0.063% | 0.979 ± 0.1% |
| 0.6 | Rand | 2728.2 ± 0.78% | 22105.7 ± 1.7% | 432399.7 ± 2.9% | 0.771 ± 0.97% | 0.994 ± 0.081% | 0.771 ± 0.97% |
| 0.6 | SRPT | 1624.4 ± 0.043% | 3519.2 ± 0.73% | 284839.9 ± 4.3% | 0.97 ± 0.16% | 0.999 ± 0.064% | 0.97 ± 0.16% |
| 0.7 | FF | 3858.3 ± 1.5% | 35582.9 ± 3.2% | 272734.1 ± 8.6% | 0.951 ± 0.16% | 0.993 ± 0.051% | 0.951 ± 0.16% |
| 0.7 | FS | 1630.8 ± 0.12% | 4456.6 ± 1.0% | 225655.3 ± 3.9% | 0.953 ± 0.13% | 0.999 ± 0.058% | 0.953 ± 0.13% |
| 0.7 | Rand | 3035.8 ± 0.58% | 26654.0 ± 1.4% | 356900.5 ± 1.4% | 0.708 ± 0.38% | 0.991 ± 0.072% | 0.708 ± 0.38% |
| 0.7 | SRPT | 1680.3 ± 0.17% | 3938.0 ± 0.64% | 291331.3 ± 5.2% | 0.931 ± 0.27% | 0.999 ± 0.058% | 0.931 ± 0.27% |
| 0.8 | FF | 4501.2 ± 1.5% | 38457.9 ± 2.2% | 277274.4 ± 2.6% | 0.915 ± 0.42% | 0.988 ± 0.16% | 0.915 ± 0.42% |
| 0.8 | FS | 1713.2 ± 0.17% | 6223.9 ± 1.2% | 259604.2 ± 3.1% | 0.908 ± 0.33% | 0.998 ± 0.095% | 0.908 ± 0.33% |
| 0.8 | Rand | 3484.0 ± 2.5% | 32752.7 ± 3.8% | 321488.9 ± 4.3% | 0.644 ± 1.1% | 0.986 ± 0.087% | 0.644 ± 1.1% |
| 0.8 | SRPT | 1752.6 ± 0.43% | 4662.8 ± 1.9% | 279493.1 ± 6.2% | 0.869 ± 0.31% | 0.998 ± 0.098% | 0.869 ± 0.31% |
| 0.9 | FF | 5773.6 ± 0.51% | 46545.8 ± 0.72% | 263361.6 ± 3.2% | 0.867 ± 0.22% | 0.978 ± 0.14% | 0.867 ± 0.22% |
| 0.9 | FS | 1872.7 ± 0.3% | 9645.7 ± 1.2% | 274889.1 ± 1.4% | 0.844 ± 0.17% | 0.997 ± 0.07% | 0.844 ± 0.17% |
| 0.9 | Rand | 3943.0 ± 1.1% | 39082.1 ± 1.2% | 293317.7 ± 1.1% | 0.595 ± 0.58% | 0.981 ± 0.062% | 0.595 ± 0.58% |
| 0.9 | SRPT | 1900.9 ± 0.69% | 6304.4 ± 3.1% | 264047.5 ± 2.1% | 0.793 ± 0.19% | 0.997 ± 0.066% | 0.793 ± 0.19% |

**Table F.16**
Scheduler performance summary with 95% confidence intervals for the **rack_sensitivity_0.8** benchmark.

| Load | Subject | Mean FCT (μs) | p99 FCT (μs) | Max FCT (μs) | Throughput (Frac) | Flows Accepted (Frac) | Info Accepted (Frac) |
|---|---|---|---|---|---|---|---|
| 0.1 | FF | 1564.4 ± 0.17% | 3075.8 ± 0.78% | 38130.3 ± 3.3% | 0.998 ± 0.055% | 1.0 ± 0.032% | 0.998 ± 0.055% |
| 0.1 | FS | 1523.4 ± 0.13% | 1997.4 ± 0.002% | 34026.7 ± 2.1% | 0.998 ± 0.055% | 1.0 ± 0.031% | 0.998 ± 0.055% |
| 0.1 | Rand | 1549.9 ± 0.14% | 2779.9 ± 0.22% | 60347.0 ± 7.3% | 0.996 ± 0.08% | 1.0 ± 0.032% | 0.996 ± 0.08% |
| 0.1 | SRPT | 1523.7 ± 0.13% | 1998.4 ± 0.0039% | 34147.9 ± 2.8% | 0.998 ± 0.055% | 1.0 ± 0.032% | 0.998 ± 0.055% |
| 0.2 | FF | 1655.6 ± 0.53% | 4886.6 ± 3.3% | 48187.1 ± 7.1% | 0.991 ± 0.17% | 0.998 ± 0.1% | 0.991 ± 0.17% |
| 0.2 | FS | 1525.7 ± 0.11% | 1998.9 ± 0.0078% | 41674.6 ± 5.2% | 0.991 ± 0.16% | 0.999 ± 0.099% | 0.991 ± 0.16% |
| 0.2 | Rand | 1589.2 ± 0.13% | 3175.5 ± 1.1% | 91618.8 ± 7.3% | 0.983 ± 0.13% | 0.999 ± 0.099% | 0.983 ± 0.13% |
| 0.2 | SRPT | 1528.4 ± 0.11% | 2350.2 ± 0.88% | 43538.3 ± 5.8% | 0.992 ± 0.16% | 0.999 ± 0.098% | 0.992 ± 0.16% |
| 0.3 | FF | 1812.5 ± 0.4% | 7816.8 ± 1.9% | 68547.0 ± 5.1% | 0.986 ± 0.17% | 0.999 ± 0.049% | 0.986 ± 0.17% |
| 0.3 | FS | 1532.3 ± 0.1% | 2202.5 ± 0.51% | 64297.6 ± 5.1% | 0.987 ± 0.2% | 0.999 ± 0.053% | 0.987 ± 0.2% |
| 0.3 | Rand | 1657.7 ± 0.13% | 4051.6 ± 0.72% | 227634.8 ± 3.2% | 0.972 ± 0.33% | 0.999 ± 0.051% | 0.972 ± 0.33% |
| 0.3 | SRPT | 1541.1 ± 0.11% | 2701.4 ± 0.27% | 73297.6 ± 7.1% | 0.989 ± 0.14% | 0.999 ± 0.053% | 0.989 ± 0.14% |
| 0.4 | FF | 2211.4 ± 1.3% | 15442.4 ± 5.6% | 107060.7 ± 14.0% | 0.976 ± 0.26% | 0.997 ± 0.065% | 0.976 ± 0.26% |

(*continued on next page*)

**Table F.16** (*continued*)

| Load | Subject | Mean FCT ($\mu s$) | p99 FCT ($\mu s$) | Max FCT ($\mu s$) | Throughput (Frac) | Flows Accepted (Frac) | Info Accepted (Frac) |
|---|---|---|---|---|---|---|---|
| 0.4 | FS | 1546.2 ± 0.12% | 2605.9 ± 0.68% | 76600.8 ± 12.0% | 0.98 ± 0.3% | 0.999 ± 0.046% | 0.98 ± 0.3% |
| 0.4 | Rand | 1823.3 ± 0.45% | 6253.9 ± 1.4% | 256431.0 ± 7.3% | 0.918 ± 0.43% | 0.998 ± 0.047% | 0.918 ± 0.43% |
| 0.4 | SRPT | 1560.4 ± 0.088% | 2854.5 ± 0.31% | 86910.4 ± 10.0% | 0.98 ± 0.22% | 0.999 ± 0.047% | 0.98 ± 0.22% |
| 0.5 | FF | 2670.2 ± 1.5% | 20930.8 ± 5.2% | 142824.2 ± 12.0% | 0.963 ± 0.51% | 0.995 ± 0.094% | 0.963 ± 0.51% |
| 0.5 | FS | 1561.9 ± 0.088% | 2883.4 ± 0.46% | 87631.8 ± 5.5% | 0.968 ± 0.45% | 0.999 ± 0.086% | 0.968 ± 0.45% |
| 0.5 | Rand | 2097.1 ± 1.6% | 10368.7 ± 5.3% | 266946.9 ± 14.0% | 0.846 ± 0.85% | 0.996 ± 0.093% | 0.846 ± 0.85% |
| 0.5 | SRPT | 1586.5 ± 0.14% | 2988.9 ± 0.73% | 108035.3 ± 3.4% | 0.968 ± 0.46% | 0.999 ± 0.087% | 0.968 ± 0.46% |
| 0.6 | FF | 3437.5 ± 0.59% | 30455.8 ± 3.2% | 221359.1 ± 14.0% | 0.971 ± 0.22% | 0.995 ± 0.097% | 0.971 ± 0.22% |
| 0.6 | FS | 1589.9 ± 0.079% | 3541.0 ± 0.89% | 121075.9 ± 9.3% | 0.978 ± 0.15% | 0.999 ± 0.066% | 0.978 ± 0.15% |
| 0.6 | Rand | 3021.9 ± 1.6% | 24451.3 ± 2.3% | 412148.6 ± 1.0% | 0.771 ± 0.67% | 0.993 ± 0.083% | 0.771 ± 0.67% |
| 0.6 | SRPT | 1632.0 ± 0.072% | 3575.4 ± 0.19% | 219688.9 ± 7.6% | 0.97 ± 0.28% | 0.999 ± 0.069% | 0.97 ± 0.28% |
| 0.7 | FF | 4226.4 ± 1.0% | 37246.2 ± 1.9% | 250830.8 ± 3.5% | 0.955 ± 0.43% | 0.992 ± 0.1% | 0.955 ± 0.43% |
| 0.7 | FS | 1630.6 ± 0.12% | 4431.6 ± 0.92% | 200199.1 ± 3.6% | 0.961 ± 0.23% | 0.999 ± 0.077% | 0.961 ± 0.23% |
| 0.7 | Rand | 3899.6 ± 1.9% | 35618.2 ± 3.2% | 367726.5 ± 1.9% | 0.684 ± 0.88% | 0.988 ± 0.14% | 0.684 ± 0.88% |
| 0.7 | SRPT | 1694.3 ± 0.1% | 4009.5 ± 0.56% | 299390.8 ± 5.7% | 0.936 ± 0.3% | 0.999 ± 0.079% | 0.936 ± 0.3% |
| 0.8 | FF | 5264.1 ± 1.5% | 44602.4 ± 1.7% | 284358.6 ± 6.4% | 0.905 ± 0.58% | 0.985 ± 0.066% | 0.905 ± 0.58% |
| 0.8 | FS | 1721.5 ± 0.35% | 6287.8 ± 2.1% | 249298.1 ± 4.5% | 0.907 ± 0.49% | 0.998 ± 0.076% | 0.907 ± 0.49% |
| 0.8 | Rand | 4485.1 ± 1.8% | 44277.7 ± 2.8% | 331280.5 ± 2.2% | 0.59 ± 0.66% | 0.98 ± 0.1% | 0.59 ± 0.66% |
| 0.8 | SRPT | 1772.7 ± 0.17% | 4871.1 ± 0.47% | 308528.4 ± 3.4% | 0.871 ± 0.35% | 0.998 ± 0.073% | 0.871 ± 0.35% |
| 0.9 | FF | 6797.9 ± 2.0% | 53200.8 ± 2.1% | 312515.9 ± 11.0% | 0.866 ± 0.59% | 0.977 ± 0.08% | 0.866 ± 0.59% |
| 0.9 | FS | 1891.3 ± 1.2% | 10007.9 ± 4.7% | 324448.5 ± 7.9% | 0.856 ± 0.93% | 0.998 ± 0.036% | 0.856 ± 0.93% |
| 0.9 | Rand | 5968.6 ± 7.1% | 63779.3 ± 11.0% | 351222.1 ± 11.0% | 0.54 ± 0.64% | 0.971 ± 0.097% | 0.54 ± 0.64% |
| 0.9 | SRPT | 1935.1 ± 0.79% | 6647.8 ± 2.6% | 315660.8 ± 11.0% | 0.792 ± 0.17% | 0.998 ± 0.049% | 0.792 ± 0.17% |

*Appendix F.4. Winner Tables*

The below 'winner tables' summarise the winning schedulers for each load and benchmark with their performance improvement relative to the worst performing baseline for each $P_{KPI}$ averaged across 5 runs. These tables are useful for gaining an overarching view of the multi-faceted performance results which are often difficult to interpret through graphical means alone.

**Table F.17**
The winning schedulers' performances relative to the losing baselines for (from top to bottom) the 0 (uniform), 0.2, 0.4, 0.6, and 0.8 rack sensitivity traces. For brevity, '–' indicates all schedulers' performances were equal.

| Load | Mean FCT | p99 FCT | Max FCT | Throughput | Flows Accepted |
|---|---|---|---|---|---|
| 0.1 | SRPT, −2.3% | FS, −33% | FF, −36% | FF + FS + SRPT, 0.40% | – |
| 0.2 | FS, −6.0% | FS, −55% | FS, −47% | SRPT, 0.92% | – |
| 0.3 | FS, −12% | FS, −66% | FS, −60% | SRPT, 1.7% | – |
| 0.4 | FS, −19% | FS, −73% | FS, −64% | FF, 3.7% | FS + Rand + SRPT, 0.10% |
| 0.5 | FS, −31% | FS, −80% | FF, −75% | FS, 11% | FS + SRPT, 0.21% |
| 0.6 | FS, −52% | SRPT, −83% | FS, −60% | FS, 19% | FS + SRPT, 0.60% |
| 0.7 | FS, −62% | SRPT, −86% | FS, −28% | FS, 28% | FS, 1.3% |
| 0.8 | FS + SRPT, −69% | SRPT, −86% | FS, −14% | FF, 31% | FS + SRPT, 2.1% |
| 0.9 | SRPT, −73% | SRPT, −85% | FF, −9.1% | FF, 35% | FS + SRPT, 3.1% |
| 0.1 | SRPT, −2.107% | FS, −31.27% | FS, −34.06% | FS + SRPT, 0.3027% | – |
| 0.2 | FS, −5.603% | FS, −52.53% | FS, −44.45% | SRPT, 0.9202% | FS + SRPT, 0.1001% |
| 0.3 | FS, −12.37% | FS, −66.67% | FS, −57.27% | FS, 1.331% | – |
| 0.4 | FS, −19.88% | FS, −73.94% | FS, −65.32% | FS, 4.145% | FS + SRPT, 0.1002% |
| 0.5 | FS, −32.88% | FS, −79.82% | FS, −76.21% | FS, 9.865% | FS + SRPT, 0.2006% |
| 0.6 | FS, −51.13% | SRPT, −85.25% | FS, −67.36% | FS, 17.82% | FS + SRPT, 0.503% |
| 0.7 | FS, −64.61% | SRPT, −85.79% | FS, −39.0% | FF + FS, 25.83% | FS + SRPT, 1.421% |
| 0.8 | SRPT, −70.16% | SRPT, −86.94% | FS, −13.96% | FF, 30.26% | FS + SRPT, 2.149% |
| 0.9 | FS, −76.71% | SRPT, −85.63% | FS, −6.949% | FF, 34.0% | FS + SRPT, 3.527% |
| 0.1 | SRPT, −2.471% | FS, −33.94% | FS, −40.28% | FF + FS + SRPT, 0.1004% | – |
| 0.2 | FS, −7.185% | FS, −58.14% | FS, −61.28% | SRPT, 1.23% | – |
| 0.3 | FS, −14.26% | FS, −70.72% | FS, −77.73% | FS + SRPT, 3.452% | FS + SRPT, 0.1001% |
| 0.4 | FS, −22.77% | FS, −77.58% | FS, −76.84% | SRPT, 10.77% | FS + SRPT, 0.1002% |
| 0.5 | FS, −36.9% | FS, −86.15% | FS, −74.53% | FS, 18.98% | FS + SRPT, 0.3012% |
| 0.6 | FS, −44.7% | SRPT, −86.07% | FS, −68.43% | FS, 24.94% | FS + SRPT, 0.402% |
| 0.7 | FS, −53.52% | SRPT, −88.65% | FS, −34.55% | FS, 31.97% | FS + SRPT, 0.6042% |
| 0.8 | FS, −59.85% | SRPT, −88.11% | FS, −25.5% | FF, 36.99% | FS + SRPT, 0.9091% |
| 0.9 | SRPT, −66.14% | SRPT, −87.17% | FF, −8.411% | FF, 39.97% | FS + SRPT, 1.424% |
| 0.1 | SRPT, −2.472% | FS, −33.26% | FF + FS, −34.08% | FF + FS + SRPT, 0.3024% | – |
| 0.2 | FS, −7.095% | FS, −57.56% | SRPT, −55.74% | SRPT, 0.9174% | – |
| 0.3 | FS, −14.3% | FS, −70.03% | FS, −75.31% | SRPT, 2.172% | – |
| 0.4 | FS, −25.46% | FS, −78.66% | FS, −59.65% | FS + SRPT, 6.979% | FS + SRPT, 0.2006% |
| 0.5 | FS, −36.61% | FS, −84.19% | FS, −77.07% | FF + FS + SRPT, 16.25% | FS + SRPT, 0.2006% |
| 0.6 | FS, −46.35% | FS, −85.4% | FS, −68.06% | FS, 26.98% | FS + SRPT, 0.503% |
| 0.7 | FS, −57.73% | SRPT, −88.93% | FS, −36.77% | FS, 34.6% | FS + SRPT, 0.8073% |
| 0.8 | FS, −61.94% | SRPT, −87.88% | FS, −19.25% | FF, 42.08% | FS + SRPT, 1.217% |

(*continued on next page*)

**Table F.17** (*continued*)

| Load | Mean FCT | p99 FCT | Max FCT | Throughput | Flows Accepted |
|---|---|---|---|---|---|
| 0.9 | FS, −67.56% | SRPT, −86.46% | FF, −10.21% | FF, 45.71% | FS + SRPT, 1.943% |
| 0.1 | FS, −2.621% | FS, −35.06% | FS, −43.61% | FF + FS + SRPT, 0.2008% | – |
| 0.2 | FS, −7.846% | FS, −59.09% | FS, −54.51% | SRPT, 0.9156% | FS + Rand + SRPT, 0.1002% |
| 0.3 | FS, −15.46% | FS, −71.82% | FS, −71.75% | SRPT, 1.749% | – |
| 0.4 | FS, −30.08% | FS, −83.13% | FS, −70.13% | FS + SRPT, 6.754% | FS + SRPT, 0.2006% |
| 0.5 | FS, −41.51% | FS, −86.22% | FS, −67.17% | FS + SRPT, 14.42% | FS + SRPT, 0.402% |
| 0.6 | FS, −53.75% | FS, −88.37% | FS, −70.62% | FS, 26.85% | FS + SRPT, 0.6042% |
| 0.7 | FS, −61.42% | SRPT, −89.24% | FS, −45.56% | FS, 40.5% | FS + SRPT, 1.113% |
| 0.8 | FS, −67.3% | SRPT, −89.08% | FS, −24.75% | FS, 53.73% | FS + SRPT, 1.837% |
| 0.9 | FS, −72.18% | SRPT, −89.58% | FF, −11.02% | FF, 60.37% | FS + SRPT, 2.781% |

**Table F.18**
The winning schedulers' performances relative to the losing baselines for (from top to bottom) the 0 (uniform), 0.05, 0.1, 0.2, and 0.4 skewed nodes sensitivity traces. For brevity, '–' indicates all schedulers' performances were equal.

| Load | Mean FCT | p99 FCT | Max FCT | Throughput | Flows Accepted |
|---|---|---|---|---|---|
| 0.1 | SRPT, −2.329% | FS, −32.9% | FF, −36.37% | FF + FS + SRPT, 0.4036% | – |
| 0.2 | FS, −5.954% | FS, −54.54% | FS, −46.79% | SRPT, 0.924% | – |
| 0.3 | FS, −12.11% | FS, −65.63% | FS, −60.37% | SRPT, 1.747% | – |
| 0.4 | FS, −19.44% | FS, −72.55% | FS, −64.36% | FF, 3.7% | FS + Rand + SRPT, 0.1002% |
| 0.5 | FS, −30.78% | FS, −80.21% | FF, −75.05% | FS, 11.1% | FS + SRPT, 0.2006% |
| 0.6 | FS, −51.79% | SRPT, −82.76% | FS, −59.8% | FS, 18.66% | FS + SRPT, 0.6042% |
| 0.7 | FS, −62.2% | SRPT, −86.3% | FS, −28.06% | FS, 27.64% | FS, 1.318% |
| 0.8 | FS + SRPT, −69.49% | SRPT, −86.09% | FS, −13.95% | FF, 30.88% | FS + SRPT, 2.149% |
| 0.9 | SRPT, −73.35% | SRPT, −84.72% | FF, −9.119% | FF, 34.93% | FS + SRPT, 3.099% |
| 0.10 | SRPT, −8.757% | SRPT, −59.75% | FS, −45.46% | FF, 0.8114% | – |
| 0.20 | SRPT, −12.59% | SRPT, −53.34% | SRPT, −41.18% | SRPT, 4.129% | SRPT, 0.2006% |
| 0.30 | SRPT, −8.624% | FS, −43.4% | Rand, −29.39% | FF, 4.171% | – |
| 0.40 | SRPT, −12.73% | FS, −55.02% | FS, −26.74% | FF, 5.525% | FS + SRPT, 0.1002% |
| 0.50 | SRPT, −25.03% | SRPT, −69.41% | FS, −43.45% | FF, 10.05% | FS + SRPT, 0.2008% |
| 0.60 | FS, −47.42% | SRPT, −77.59% | FS, −54.24% | FF, 17.3% | FS + SRPT, 0.402% |
| 0.70 | FS, −61.53% | SRPT, −82.93% | FS, −30.62% | FF, 25.03% | FS + SRPT, 1.113% |
| 0.79 | FS, −70.3% | SRPT, −85.75% | FS, −18.76% | FF, 33.28% | FS + SRPT, 2.149% |
| 0.90 | SRPT, −73.83% | SRPT, −86.22% | FF, −8.617% | FF, 37.2% | FS + SRPT, 3.316% |
| 0.10 | SRPT, −4.328% | SRPT, −44.2% | FS, −23.36% | – | – |
| 0.20 | SRPT, −18.98% | SRPT, −78.65% | FF, −33.87% | FF, 7.214% | FS + Rand + SRPT, 0.1002% |
| 0.30 | SRPT, −26.46% | SRPT, −80.41% | Rand, −9.494% | FF, 5.855% | FS + Rand + SRPT, 0.2006% |
| 0.40 | SRPT, −10.98% | SRPT, −43.66% | FF, −19.9% | FF, 7.365% | SRPT, 0.1002% |
| 0.50 | SRPT, −17.24% | SRPT, −55.39% | FS, −23.86% | FF, 8.208% | FF + FS + SRPT, 0.1002% |
| 0.60 | FS, −37.96% | FS, −68.51% | FS, −32.95% | FF, 12.16% | FS + SRPT, 0.3012% |
| 0.70 | FS, −59.81% | SRPT, −79.97% | FS, −37.32% | FS, 19.26% | FS + SRPT, 0.9091% |
| 0.80 | FS, −66.85% | SRPT, −80.76% | FS, −11.75% | FF, 26.29% | FS, 2.045% |
| 0.89 | FS, −73.71% | FS, −77.53% | SRPT, −10.75% | FF, 34.9% | FS, 3.423% |
| 0.10 | SRPT, −2.553% | SRPT, −32.54% | FS, −33.37% | SRPT, 0.3021% | – |
| 0.20 | SRPT, −7.511% | FS, −57.99% | FS, −58.96% | FF, 4.589% | – |
| 0.30 | SRPT, −17.23% | SRPT, −73.84% | FS, −57.32% | FF, 19.36% | FS + SRPT, 0.1002% |
| 0.40 | SRPT, −23.55% | SRPT, −78.58% | SRPT, −13.68% | FF + FS, 16.41% | FS + SRPT, 0.2006% |
| 0.50 | SRPT, −18.2% | SRPT, −54.32% | FS, −27.63% | FS, 11.7% | FS + SRPT, 0.1002% |
| 0.61 | SRPT, −25.65% | SRPT, −54.03% | FS, −26.91% | FF, 12.48% | FS + SRPT, 0.2006% |
| 0.70 | FS, −47.05% | FS, −69.77% | FS, −20.66% | FS, 15.86% | FS + SRPT, 0.6042% |
| 0.80 | FS, −67.39% | FS, −78.64% | FS, −28.66% | FS, 23.76% | FS, 1.939% |
| 0.90 | FS, −78.26% | FS, −76.03% | FS, −13.49% | FF, 29.39% | FS, 3.638% |
| 0.10 | SRPT, −2.218% | FS, −32.07% | FS, −34.68% | – | – |
| 0.20 | FS, −6.063% | FS, −54.59% | FS, −52.78% | SRPT, 2.289% | – |
| 0.30 | FS, −12.22% | FS, −65.35% | FS, −70.36% | FS + SRPT, 1.955% | – |
| 0.40 | FS, −20.0% | FS, −72.18% | FS, −70.34% | FS, 8.26% | FS + SRPT, 0.1002% |
| 0.51 | FS, −35.12% | SRPT, −81.59% | FS, −71.39% | FS, 17.68% | FS + SRPT, 0.402% |
| 0.60 | SRPT, −49.0% | SRPT, −85.19% | FS, −40.78% | FS, 25.2% | FS + SRPT, 0.6042% |
| 0.71 | SRPT, −56.38% | SRPT, −87.45% | FS, −28.14% | FS, 26.13% | FS + SRPT, 1.011% |
| 0.80 | SRPT, −62.04% | SRPT, −85.79% | SRPT, −12.62% | FS, 23.34% | SRPT, 1.629% |
| 0.89 | FS, −72.56% | FS, −71.52% | FS, −9.325% | FF, 28.59% | FS, 3.32% |

**Table F.19**
The winning schedulers' performances relative to the losing baselines for (from top to bottom) the University, Private Enterprise, Commercial Cloud, and Social Media Cloud DCN traces. For brevity, '–' indicates all schedulers' performances were equal.

| Load | Mean FCT | p99 FCT | Max FCT | Throughput | Flows |
|---|---|---|---|---|---|
| 0.10 | SRPT, −2.466% | SRPT, −31.22% | FF, −38.81% | SRPT, 0.4036% | – |
| 0.20 | SRPT, −8.834% | FS, −64.48% | FS, −60.01% | FF + SRPT, 2.391% | – |
| 0.30 | SRPT, −17.83% | SRPT, −76.13% | FS, −64.78% | FF, 13.77% | FS + SRPT, 0.1001% |
| 0.40 | SRPT, −26.47% | SRPT, −81.81% | SRPT, −25.11% | FF, 15.07% | SRPT, 0.2004% |
| 0.50 | SRPT, −18.77% | SRPT, −54.54% | FS, −22.44% | FF, 11.12% | SRPT, 0.2004% |
| 0.60 | SRPT, −29.81% | SRPT, −63.41% | FS, −21.25% | FF, 13.91% | SRPT, 0.3009% |
| 0.70 | FS, −45.77% | FS, −77.26% | FS, −34.73% | FS, 21.06% | FS + SRPT, 0.6042% |
| 0.79 | FS, −55.34% | FS, −75.18% | FS, −25.43% | FS, 24.9% | FS + SRPT, 1.113% |
| 0.89 | FS, −64.75% | FS, −71.49% | FS, −26.74% | FS, 26.16% | FS, 1.941% |
| 0.10 | SRPT, −3.577% | FS, −37.74% | FS, −43.91% | FF, 0.4024% | – |
| 0.20 | SRPT, −11.42% | FS, −69.85% | FS, −60.28% | SRPT, 4.017% | FS + Rand + SRPT, 0.1001% |
| 0.30 | SRPT, −25.05% | FS, −84.23% | FS, −57.56% | FS, 9.121% | FS + SRPT, 0.1001% |
| 0.40 | SRPT, −40.72% | SRPT, −90.67% | FF, −26.91% | FS, 12.26% | FS + SRPT, 0.3009% |
| 0.50 | SRPT, −43.96% | SRPT, −91.43% | FS, −18.36% | FF, 11.84% | SRPT, 0.6036% |
| 0.60 | SRPT, −28.85% | SRPT, −72.65% | FS, −20.9% | FF, 17.01% | SRPT, 0.3009% |
| 0.70 | SRPT, −35.86% | SRPT, −77.65% | FS, −21.9% | FF, 24.46% | SRPT, 0.5025% |
| 0.79 | FS, −49.05% | SRPT, −78.85% | FS, −20.43% | FF, 32.89% | FS + SRPT, 1.011% |
| 0.90 | FS, −66.84% | FS, −77.31% | SRPT, −9.144% | FF, 38.56% | FS, 1.526% |
| 0.10 | SRPT, −4.338% | FS, −44.59% | FS, −43.56% | FF + SRPT, 0.6061% | – |
| 0.20 | FS, −12.75% | FS, −73.12% | FS, −65.3% | FS + SRPT, 2.165% | FS + Rand + SRPT, 0.1001% |
| 0.30 | FS, −32.36% | FS, −89.81% | FS, −65.67% | FS, 6.109% | FS + Rand + SRPT, 0.1001% |
| 0.40 | FS, −51.39% | FS, −93.5% | FS, −50.1% | SRPT, 11.49% | FS + SRPT, 0.6036% |
| 0.50 | FS, −64.75% | SRPT, −95.14% | FS, −44.25% | FS, 18.82% | FS + SRPT, 1.112% |
| 0.60 | SRPT, −66.34% | SRPT, −94.78% | FS, −38.79% | FS, 26.48% | FS + SRPT, 2.249% |
| 0.70 | FS, −55.14% | SRPT, −86.15% | FS, −30.44% | FS, 36.44% | SRPT, 1.112% |
| 0.79 | FS, −63.22% | SRPT, −86.61% | FS, −24.83% | FS, 50.83% | FS + SRPT, 1.939% |
| 0.89 | FS, −71.7% | SRPT, −80.91% | FS, −15.77% | FF, 57.67% | FS, 2.675% |
| 0.10 | FS, −1.484% | FS, −25.77% | SRPT, −49.67% | – | – |
| 0.20 | FS, −4.266% | FS, −24.92% | FS, −61.38% | – | – |
| 0.30 | FS, −9.855% | FS, −41.03% | FS, −66.9% | FF + FS + SRPT, 0.1001% | – |
| 0.40 | FS, −18.74% | FS, −57.24% | FS, −81.16% | FS + SRPT, 0.3012% | FS + SRPT, 0.1001% |
| 0.50 | FS, −38.63% | FS, −79.07% | FS, −84.78% | FS + SRPT, 1.013% | FS, 0.3009% |
| 0.60 | FS, −60.95% | FS, −88.96% | FS, −82.32% | FS + SRPT, 4.311% | FS, 1.011% |
| 0.69 | FS, −70.83% | FS, −88.24% | FS, −72.67% | FS, 9.434% | FS, 6.852% |
| 0.80 | FS, −64.72% | FS, −79.18% | FS, −45.34% | FS, 26.6% | FS, 16.16% |
| 0.90 | FS, −73.86% | FS, −80.41% | FS, −45.18% | FF, 55.08% | FS, 31.69% |

## Appendix G. A Note on the Flow- vs. Job-Centric Traffic Paradigms

Common DCN jobs include search queries, generating social media feeds, and performing machine learning tasks such as inference and back-propagation. These jobs are directed acyclic graphs composed of *operations* (nodes) and *dependencies* (edges) [61]. The dependencies are either *control dependencies* (where the child operation can only begin once the parent operation has been completed) or *data dependencies* (where $\geq 1$ tensors are output from the parent operation as required input for the child operation). In the context of DCNs, when a job arrives, each operation in the job is placed onto some machine to execute it. These operations might all be placed onto one machine or, as is often the case, distributed across different machines in the network [62]. The DCN is then used to pass the tensors around between machines executing the operations. Job data dependencies whose parent and child operations are placed onto different machines have their tensors become DCN *flows*.

There are therefore two paradigms when considering traffic demand generation in DCNs; the *flow-centric* paradigm, which is agnostic to the overall computation graph being executed in the DCN when servicing an application, and the *job-centric* paradigm, which does consider the computation graph when generating network flows. For this manuscript, we considered the flow-centric paradigm, where a single demand is a *flow*; a task demanding some information be sent from a source node to a destination node in the network. Flow characteristics include *size* (how much information to send), *arrival time* (the time the flow arrives ready to be transported through the network, as derived from the network-level *inter-arrival time* which is the time between a flow's time of arrival and its predecessor's), and *source-destination node pair* (which machine the flow is queued at and where it is requesting to be sent). Together, these characteristics form a network-level *source-destination node pair distribution* ('how much' (as measured by either probability or load) each machine tends to be requested by arriving flows).

In real DCNs, traffic flows can be correlated with one another since they may be part of the same job and therefore share similar characteristics. An interesting area of future work will be to develop TrafPy to support the job-centric paradigm and have this type of inter-flow correlation. However, this is beyond the scope of this manuscript.